\documentclass[a4paper,twocolumn,unpublished]{quantumarticle}
\pdfoutput=1

\usepackage[utf8]{inputenc}
\usepackage[T1]{fontenc}
\usepackage{amsmath,amssymb,amsthm,bm}
\usepackage{booktabs,array,graphicx}
\usepackage{tabularray}
\UseTblrLibrary{booktabs}
\usepackage{enumitem}
\usepackage{xcolor}
\usepackage{tikz}
\usetikzlibrary{arrows.meta,positioning}
\usepackage{placeins}
\usepackage{float}
\usepackage{hyperref}
\usepackage{cleveref}


\theoremstyle{plain}
\newtheorem{theorem}{Theorem}
\newtheorem{proposition}{Proposition}

\newtheorem{lemma}{Lemma}
\theoremstyle{definition}

\theoremstyle{remark}
\newtheorem{remark}{Remark}
\crefname{assumption}{Assumption}{Assumptions}
\Crefname{theorem}{Theorem}{Theorems}
\Crefname{proposition}{Proposition}{Propositions}
\Crefname{lemma}{Lemma}{Lemmas}
\Crefname{corollary}{Corollary}{Corollaries}
\Crefname{definition}{Definition}{Definitions}
\Crefname{remark}{Remark}{Remarks}
\Crefname{result}{Result}{Results}
\crefname{result}{Result}{Results}
\crefname{theorem}{Theorem}{Theorems}
\crefname{proposition}{Proposition}{Propositions}
\crefname{lemma}{Lemma}{Lemmas}
\crefname{corollary}{Corollary}{Corollaries}
\crefname{definition}{Definition}{Definitions}
\crefname{remark}{Remark}{Remarks}

\newcommand{\cH}{\mathcal H}
\newcommand{\cY}{\mathcal Y}
\newcommand{\cZ}{\mathcal Z}
\newcommand{\cA}{\mathcal A}
\newcommand{\Ran}{\operatorname{ran}}
\newcommand{\Ker}{\operatorname{ker}}

\providecommand{\doi}[1]{\href{https://doi.org/#1}{doi:#1}}

\begin{document}

\title{A Constraint-Preserving Preconditioned Quantum Solver for Heterogeneous Brinkman Flow}

\author{Shreya Savadatti}
\affiliation{Carnegie Mellon University, Pittsburgh, PA, USA}

\author{Franz Franchetti}
\affiliation{Carnegie Mellon University, Pittsburgh, PA, USA}

\begin{abstract}
Quantum amplitude estimation can reduce the number of evaluations needed to
estimate an average, but its advantage in PDE uncertainty quantification
depends on the cost of each coherent solve. We study this problem for
incompressible Brinkman flow through a porous material with spatially
varying drag. We first express the velocity in divergence-free coordinates
using a polar Riesz map, obtaining a positive-definite system that preserves
incompressibility in the exact formulation. We then introduce a
Sobolev--Riesz preconditioner and prove that the resulting condition number
is bounded by $\alpha_{\max}/\alpha_{\min}$, independently of the grid
resolution. We construct block encodings and reversible-arithmetic
circuits for this formulation and show how to estimate the mean kinetic
energy without reconstructing the full velocity field. By accounting for
both sampling and solver costs, we establish conditions under which the
method has a gate-complexity advantage over single-level classical
grid-resolving solvers. Numerical experiments verify the conditioning
bound and the accuracy of the circuit components at the tested sizes.
The complexity advantage is conditional on the stated input and
discretization assumptions; no quantum-hardware runtime advantage
is demonstrated.
\end{abstract}

\section{Introduction}
\label{sec:intro}

Porous materials are used to control fluid flow in applications such as
aerodynamic noise reduction and skin-friction control. For example,
porous inserts near the trailing edge of an airfoil can reduce noise
generated by the flow~\cite{jaworski2013}, while porous coatings have
also been studied for controlling skin friction~\cite{bottaro2019}.
At low flow speeds, the motion of a fluid through a porous material
can be described by the Stokes and Darcy--Brinkman
equations~\cite{brinkman1949}. In the Brinkman model, the resistance
offered by the material is represented by a drag coefficient
$\alpha(x)$, which may vary across the domain.

In practice, the permeability of a porous material is not always
known exactly. Manufacturing variations can produce different local
structures, and these differences affect the resulting flow.
To account for this uncertainty, we consider $M$ possible drag
fields $\{\alpha_s\}_{s=0}^{M-1}$, each corresponding to a different
material scenario. Solving the Brinkman equations for scenario $s$
gives a velocity field $u_s$. Our quantity of interest is the
average squared $L^2$ norm of the velocity,

\[
\bar{\mathcal E}
=
\frac{1}{M}\sum_{s=0}^{M-1}\|u_s\|_{L^2}^2,
\]

which we want to estimate to additive accuracy $\epsilon$.
This is an uncertainty quantification problem: rather than
predicting the flow for only one material, we want to estimate
its average response across many possible materials.

A direct classical approach is to solve the PDE for a collection
of scenarios and average the results. Ordinary Monte Carlo
sampling typically requires $\mathcal O(\epsilon^{-2})$
evaluations to obtain statistical error of order $\epsilon$.
When each evaluation involves solving a spatially discretized
PDE, this can be expensive, particularly when the material
properties vary on fine spatial scales.

Quantum amplitude estimation (QAE) offers a way to reduce the
number of evaluations to $\mathcal O(\epsilon^{-1})$
~\cite{brassard2002,montanaro2015}. However, applying QAE to
a fluid problem is not as simple as evaluating a known function.
Each coherent evaluation must represent the material properties,
solve the corresponding PDE, and encode the quantity of interest
in a measurement probability. If the quantum PDE solver is
expensive, the reduction in sampling cost may not lead to an
overall advantage.

This distinction becomes especially important when comparing
against stronger classical methods. Quasi-Monte Carlo (QMC)
can achieve first-order convergence for sufficiently regular
parameter dependence, while multilevel Monte Carlo (MLMC)
can reduce the cost by combining solutions at different
spatial resolutions~\cite{matousek1998,kuo2012,dick2014,cliffe2011,giles2015}.
If a classical method already uses approximately
$\mathcal O(\epsilon^{-1})$ evaluations, QAE alone does not
improve the sampling exponent. The cost of each quantum PDE
solve must therefore be considered alongside the number of
evaluations.

For incompressible flow, constructing such a solver presents
two main difficulties. First, the velocity must satisfy
$\nabla\cdot u=0$. Standard formulations enforce this
constraint through pressure, leading to an indefinite
saddle-point system~\cite{benzi2005}. Second, the condition
number of the discretized viscous operator typically grows
as the grid is refined. This is costly for quantum
linear-system algorithms based on quantum singular value
transformation (QSVT), whose complexity depends on the
conditioning of the operator~\cite{gilyen2019,lintong2020}.
Spatially varying drag adds another complication because it
couples Fourier modes that would otherwise be independent.

Quantum PDE solvers and preconditioning methods have been
studied previously~\cite{montanaro2016fem,deiml2025},
including Fourier-based methods for periodic Stokes
flow~\cite{gimenez2026}. Here we consider the combination
of incompressibility, spatially varying Brinkman drag, and
coherent evaluation over an ensemble of uncertain inputs.
Our goal is to construct a solver for this setting whose
conditioning does not worsen with grid refinement, and to
determine when its cost allows an advantage in estimating
the ensemble average.

We address this problem in two steps. First, we use a
polar Riesz map to express the velocity in divergence-free
coordinates. This replaces the constrained formulation by
a positive-definite system while preserving incompressibility
in the exact mathematical representation. We then rescale
the reduced operator using a constant-drag reference
operator. The resulting Sobolev--Riesz preconditioner gives

\[
\kappa(\widetilde K_s)
\leq
\frac{\alpha_{\max}}{\alpha_{\min}},
\]

for every grid resolution. Thus, for fixed drag contrast,
the condition number remains bounded as the spatial
discretization is refined.

We also construct quantum block encodings and
reversible-arithmetic circuits for the preconditioned
operator, and show how the kinetic energy can be obtained
from a measurement probability without reading out the
velocity field. Combining the solver cost with QAE leads
to a precision--ensemble criterion for quantum advantage.
Under explicit assumptions on coherent input generation,
forcing, discretization accuracy, and the classical solver
class, we obtain a conditional gate-complexity advantage
over single-level grid-resolving classical methods.
Numerical experiments test the preconditioning and circuit
components in two dimensions, together with additional
operator-level checks in three dimensions. The comparison
does not establish an advantage over all classical
algorithms or a runtime speedup on quantum hardware.

The remainder of the paper is organized as follows.
\Cref{sec:related} discusses related work, and
\Cref{sec:background} introduces the Brinkman model and
the quantum tools used in the analysis.
\Cref{sec:reduction} develops the constraint-preserving
formulation, while \Cref{sec:access} describes coherent
evaluation of the ensemble observable.
\Cref{sec:frontier} compares quantum and classical
sampling costs. The preconditioned Brinkman solver and
its gate complexity are presented in
\Cref{sec:sobolev-riesz}, followed by numerical results
in \Cref{sec:experiment}. The appendices provide the
supporting proofs and circuit details.

\section{Related Work}
\label{sec:related}
\label{sec:literature}

Prior work on quantum PDE solvers, operator preconditioning, and
uncertainty quantification informs our comparison of solver and
sampling costs for heterogeneous Brinkman flow.

\subsection{Quantum PDE Solvers and Preconditioning}

Quantum linear-system algorithms make it possible to represent the
solution of a linear system as a quantum state without computing
and storing every entry of the solution. The algorithm introduced
by Harrow, Hassidim, and Lloyd~\cite{hhl2009} established this
approach, and later work improved its precision and condition-number
dependence~\cite{childs2017,gilyen2019,lintong2020,lowsu2026}.
These improvements are important for PDEs, where a fine spatial
discretization can produce a large and poorly conditioned system.

Applying a quantum linear-system algorithm to a PDE does not,
however, guarantee an overall speedup. The input must be prepared,
the system must be solved to sufficient accuracy, and the desired
quantity must be extracted from the solution. Montanaro and
Pallister~\cite{montanaro2016fem} studied these costs for a
finite-element boundary-value problem. They considered the
calculation of a linear functional of the solution and compared
the quantum approach with classical conjugate gradients. Their
analysis provides an important starting point for evaluating
quantum PDE algorithms beyond the cost of the linear-system
routine itself.

Preconditioning is one way to reduce the cost of the solve.
In classical numerical analysis, operator preconditioning uses
an appropriate function-space norm to obtain systems whose
conditioning is less sensitive to mesh refinement
~\cite{benzi2005,mardal2011}. This principle has also been
used in quantum algorithms. Clader, Jacobs, and Sprouse
~\cite{clader2013} explored preconditioned quantum linear-system
methods, while Deiml and Peterseim~\cite{deiml2025} developed a
quantum realization of a finite-element method using a BPX
preconditioner to control the conditioning of an elliptic
problem.

These results show that preconditioning can be an essential
part of a quantum PDE solver. In our setting, the unknown
is a divergence-free vector field rather than a scalar
solution, and the Brinkman operator contains a spatially
varying drag coefficient. We use the same general principle
of operator preconditioning, but work in divergence-free
Fourier coordinates. This allows us to remove the
incompressibility constraint from the reduced system and
bound its condition number independently of the grid.

\subsection{Quantum Algorithms for Fluid Flow}

Quantum algorithms for fluid dynamics have been developed
using several different approaches. Some use Fourier
representations of differential operators, while others
are based on lattice-Boltzmann methods or formulations
for nonlinear flow. Lubasch and coauthors
~\cite{fourierpde2025} showed how the quantum Fourier
transform can simplify circuit constructions for several
structured PDEs. Fourier methods are especially useful
for periodic problems because differential operators
with constant coefficients become diagonal in
wavenumber space.

The closest work to our solver is the quantum spectral
method of Gimenez, Cirak, and Ortiz~\cite{gimenez2026}.
They developed a quantum circuit for steady incompressible
Stokes flow on a two-dimensional periodic domain.
Their method uses the Fourier representation to separate
velocity components parallel and perpendicular to each
wavenumber, allowing incompressibility to be enforced
mode by mode. They also demonstrate the calculation of
a kinetic-energy observable without reconstructing
the full velocity field.

Our work builds on this spectral approach but considers
a different operator. In the constant-coefficient Stokes
problem, the viscous term is diagonal in Fourier space,
so different wavenumbers can be treated separately.
For Brinkman flow with spatially varying drag,
multiplication by $\alpha_s(x)$ couples those modes.
The quantum solver must therefore handle a
variable-coefficient operator while preserving
incompressibility and controlling its condition number.
We also study repeated coherent evaluations for an
ensemble of drag fields, rather than only a fixed
flow problem.

Other quantum fluid algorithms address more general
dynamics. Jennings and coauthors~\cite{jennings2026}
analyzed an end-to-end quantum lattice-Boltzmann
algorithm for nonlinear fluid flow, including the costs
of its implementation and observable estimation.
Ameri and coauthors~\cite{ameri2026} studied quantum
lower bounds for simulating fluid dynamics, including
inviscid flow. These works concern different physical
models, but they illustrate why the cost of preparing
and measuring a quantum fluid simulation must be
included when assessing an algorithm's advantage.

\subsection{Uncertainty Quantification and Sampling}

Our second concern is the cost of averaging over
uncertain material properties. Classical uncertainty
quantification commonly uses Monte Carlo, quasi-Monte
Carlo (QMC), and multilevel Monte Carlo (MLMC).
QMC uses carefully distributed sample points instead
of independent random samples and can achieve faster
convergence when the dependence on uncertain
parameters is sufficiently regular
~\cite{matousek1998,kuo2012,dick2014}.
MLMC reduces computational cost by combining
many inexpensive coarse-grid evaluations with
fewer fine-grid evaluations~\cite{cliffe2011,giles2015}.
These methods provide stronger classical comparisons
than ordinary Monte Carlo alone.

On the quantum side, amplitude estimation
~\cite{brassard2002} reduces the number of coherent
evaluations needed for general mean estimation,
and Montanaro~\cite{montanaro2015} showed how this
can accelerate Monte Carlo calculations.
Quantum algorithms have also been proposed for
PDEs with uncertain inputs. For example, Golse,
Jin, and Liu~\cite{golse2022} developed
transformations that reformulate certain PDE
uncertainty problems in higher-dimensional spaces,
allowing the uncertain inputs to be treated through
a different computational representation.

An important limitation is that quantum amplitude
estimation does not always improve the asymptotic
sampling rate over classical QMC. Recchia and
coauthors~\cite{recchia2026} studied quantum QMC
using coherently prepared low-discrepancy points.
They identified a possible pre-asymptotic advantage,
while showing why their construction does not
generally improve the asymptotic rate over
classical QMC.

This distinction matters for our problem because
every evaluation includes a Brinkman solve.
Even when QAE reduces the number of evaluations,
the quantum computation is advantageous only if
the cost of those evaluations is sufficiently low.
Our precision--ensemble criterion accounts for
both costs, and our preconditioned solver addresses
the cost of each coherent PDE evaluation.

\subsection{Circuit Implementation and Complexity Models}

The gate cost depends on how the drag field is evaluated and the
Fourier-mode rotations are synthesized. The construction uses
standard controlled operations, reversible arithmetic, multiplexed
rotations, and rotation synthesis~\cite{barenco1995,cuccaro2004,
shende2006,rossselinger2016,haner2018}. The gate analysis includes
the cost and approximation errors of these operations.

Information-based complexity studies how many observations of an
unknown function are needed to compute a quantity
of interest~\cite{traub1988,novak1988,heinrich1994}.
Quantum query bounds for integration and elliptic
PDE problems have been studied by Novak and
Heinrich~\cite{novak2001,heinrich2002,heinrich2006}.
Our gate analysis assumes compact descriptions of the drag fields
that can be evaluated coherently, and the comparison is with
specified classical grid-resolving solvers. Access to unknown drag
through point queries defines a separate information model.

\section{Preliminaries}
\label{sec:background}

We consider a periodic Brinkman problem with uncertain material properties.
This section defines the fluid model and the ensemble quantity we wish to
estimate. We then introduce the Fourier representation and the quantum
operations used later in the paper.

\subsection{Periodic Brinkman Flow}
\label{sec:bg-fluids}

Let $d\geq 2$ be the spatial dimension. We work on the periodic domain
$\mathbb T^d=(\mathbb R/2\pi\mathbb Z)^d$, which can be viewed as the cell
$[0,2\pi)^d$ with opposite faces identified. This allows us to study a
representative region of porous material without prescribing boundaries
for a complete device or airfoil.

For each material scenario $s\in\{0,\ldots,M-1\}$, let
$u_s:\mathbb T^d\to\mathbb R^d$ denote the fluid velocity and
$p_s:\mathbb T^d\to\mathbb R$ its pressure. The material is described by
a positive drag coefficient $\alpha_s(x)$: a larger value means greater
resistance to the flow. In the usual Brinkman scaling, the drag is the
viscosity divided by the local permeability. We write $\nu>0$ for the
viscosity parameter and $f_s$ for the applied body force.

Without porous resistance, the low-Reynolds-number momentum equation is
the Stokes equation $-\nu\Delta u_s+\nabla p_s=f_s$. The Brinkman model
adds the drag force $\alpha_s(x)u_s$~\cite{brinkman1949}. We use the
following periodic, zero-mean formulation:
\begin{equation*}
\begin{aligned}
-\nu\Delta u_s+\alpha_s(x)u_s+\nabla p_s+\lambda_s&=f_s,\\
\nabla\cdot u_s&=0,\\
\int_{\mathbb T^d}u_s\,dx&=0,\\
\int_{\mathbb T^d}p_s\,dx&=0.
\end{aligned}
\end{equation*}
Here $\Delta$ is the spatial Laplacian and $\nabla p_s$ is the pressure
gradient. The condition $\nabla\cdot u_s=0$ expresses incompressibility:
the velocity has zero divergence at every point. The pressure enforces
this local constraint. We also require the average velocity to vanish.
The constant vector $\lambda_s\in\mathbb R^d$ is the multiplier that
enforces this mean-flow constraint; it can be interpreted as a uniform
imposed pressure gradient. The final condition fixes the otherwise
arbitrary additive constant in the periodic pressure. The role of
$\lambda_s$ is discussed further in \Cref{sec:sobolev-riesz}.

\subsection{Fourier Representation}

The periodic domain is useful because the differential operators have
simple expressions in Fourier space. Write the velocity as
\begin{equation*}
 u_s(x)=\sum_{k\in\mathbb Z^d}\widehat u_s(k)e^{ik\cdot x},
\end{equation*}
where $k=(k_1,\ldots,k_d)$ is an integer wavenumber and
$\widehat u_s(k)\in\mathbb C^d$ is its Fourier coefficient. Differentiation
with respect to $x_j$ multiplies a Fourier coefficient by $ik_j$.
Consequently, the zero-mean condition and incompressibility become
\begin{equation*}
 \widehat u_s(0)=0,
 \qquad
 k\cdot\widehat u_s(k)=0\quad(k\neq0).
\end{equation*}
Thus the constant Fourier mode is absent, and at each nonzero
wavenumber the velocity coefficient must be perpendicular to $k$.
This observation is the starting point for the constraint-preserving
coordinates developed in \Cref{sec:reduction}.

Fourier space also makes the viscous operator easy to represent.
For a constant reference drag $\alpha_0>0$, define
$A_0:=-\nu\Delta+\alpha_0 I$, where $I$ denotes the identity.
Its action on mode $k$ is multiplication by
$\nu|k|^2+\alpha_0$. By contrast, multiplication by a spatially
varying coefficient $\alpha_s(x)$ is simple in position space but
couples Fourier modes. Our quantum construction moves between these
two representations using quantum Fourier transforms
(\Cref{sec:sobolev-riesz}).

\subsection{Ensemble Observable and Classical Sampling}
\label{sec:bg-uq}

The uncertain input consists of $M$ possible pairs $(\alpha_s,f_s)$.
Each pair determines a velocity $u_s$, and hence a kinetic-energy
observable $\|u_s\|_{L^2}^2$. The $L^2$ norm uses the normalized spatial
measure $d\mu(x)=(2\pi)^{-d}dx$ on $\mathbb T^d$; in particular,
\begin{equation*}
 \|u_s\|_{L^2}^2
 =\int_{\mathbb T^d}|u_s(x)|^2\,d\mu(x).
\end{equation*}
The quantity we ultimately estimate is an ensemble average. For the
sampling analysis, it is convenient to work with a normalized output
$g_s\in[0,1]$, defined precisely in \Cref{eq:target}. Its mean is
\begin{equation*}
 \mu=\frac1M\sum_{s=0}^{M-1}g_s.
\end{equation*}
Here $\mu$ denotes the scalar ensemble mean, whereas $d\mu(x)$ denotes
the normalized spatial measure. Our target is an estimate of $\mu$ with
additive error at most $\epsilon>0$.

The ensemble size $M$ may be fixed. When the scenarios approximate a
continuous probability distribution, we also allow it to depend on the
target accuracy, writing $M(\epsilon)=\Theta(\epsilon^{-m})$ for some
$m\geq0$. The exponent $m$ describes how quickly the number of
scenarios grows as more accuracy is requested.

To compare sampling methods, we describe a classical rule by its
convergence order $p>0$. Specifically, we assume that it needs
$\Theta(\epsilon^{-1/p})$ scenario evaluations to reach error of order
$\epsilon$ in the regime where the ensemble is large enough. Since all
$M$ scenarios can also be evaluated directly, the corresponding cost
law is $\Theta(\min\{M,\epsilon^{-1/p}\})$ evaluations. Ordinary Monte
Carlo has $p=1/2$. Quasi-Monte Carlo rules based on Sobol' points can
reach $p=1$, up to logarithmic factors, under sufficient regularity of
the map from uncertain parameters to the observable
~\cite{matousek1998,kuo2012}. Nonsmooth inputs may give a lower
order~\cite{he2019}. Multilevel Monte Carlo uses solves at different
resolutions to change the overall cost, so it is a separate comparison
from this single-level sampling law. The assumptions used for our
complexity results appear in \Cref{sec:norm-and-conditional}.

\subsection{Quantum Tools}
\label{sec:bg-quantum}

Our algorithm uses four established quantum operations. The quantum
Fourier transform changes between the two representations of the PDE;
block encoding represents a nonunitary operator within a unitary
circuit; quantum singular value transformation approximately inverts
that operator; and amplitude estimation computes the final average.
We recall only the properties used later.

\paragraph{Quantum registers and Fourier transforms.}
An $n$-qubit register has $2^n$ computational basis states. For
simplicity, take $N=2^n$ grid points along each spatial axis and
$M=2^{n_M}$ material scenarios. The scenario register uses $\log_2M$
qubits, while a $d$-dimensional grid index uses $d\log_2N$ qubits.
The $d$-dimensional quantum Fourier transform $\mathcal F_d$ converts
position basis states $|x\rangle$ into superpositions of wavenumber
states $|k\rangle$, and vice versa with $\mathcal F_d^\dagger$. Its
standard circuit uses $\mathcal O(d\log^2N)$ gates. This bound counts
the transform itself, not the cost of loading an arbitrary drag field
or source vector into the register.

\paragraph{Block encoding.}
Quantum gates are unitary, whereas a PDE operator is generally not.
A block encoding represents a scaled operator as one block of a larger
unitary. Let $X$ be an operator on the system register. A unitary
$U_{\rm BE}$ is an $(a,\gamma,\eta)$-block encoding of $X$ if, with
$a$ auxiliary (ancilla) qubits,
\begin{equation*}
\begin{gathered}
 B=(\langle0^a|\otimes I)U_{\rm BE}
   (|0^a\rangle\otimes I),\\
 \|B-X/\gamma\|_2\leq\eta,
 \qquad \gamma\geq\|X\|_2.
\end{gathered}
\end{equation*}
The number $\gamma$ is the normalization, and $\eta$ is the allowed
operator-norm error. Later quantum operations act on the encoded block
$X/\gamma$, rather than applying $X$ directly. For our solver it is
important that the required normalization does not increase as the
grid is refined.

\paragraph{Quantum singular value transformation.}
Quantum singular value transformation (QSVT) applies a polynomial to
an operator through its block encoding~\cite{gilyen2019,lintong2020}.
For a Hermitian operator $X/\gamma$, QSVT can implement
$P(X/\gamma)$ for suitable real polynomials $P$ of degree
$d_{\rm inv}$ satisfying $|P(t)|\leq 1$ on $[-1,1]$, subject
to the usual parity conditions. This takes
$\mathcal O(d_{\rm inv})$ calls to $U_{\rm BE}$ and related controlled
operations. If the spectrum
of $X$ is positive and bounded away from zero, we can choose $P$ to
approximate a scaled reciprocal. This gives an approximate inverse of
$X$, which is the operation needed to solve the reduced PDE. The
required degree depends on the condition number $\kappa$ (the ratio of
the largest to smallest eigenvalue) and the target approximation
error $\eta$; a standard bound is
$d_{\rm inv}=\mathcal O(\kappa\log(1/\eta))$ under the usual
block-encoding normalization assumptions. The specific polynomial and
its error are analyzed later for our preconditioned operator.

\paragraph{Quantum amplitude estimation.}
Finally, suppose a unitary $\cA$ prepares a state in which the
probability of observing a designated ``good'' flag is $\mu$:
\begin{equation*}
 \cA|0\rangle
 =\sqrt{\mu}\,|\Psi_{\rm good}\rangle
 +\sqrt{1-\mu}\,|\Psi_{\rm bad}\rangle.
\end{equation*}
The two states lie in orthogonal good and bad subspaces, with the good
subspace selected by a projector $\Pi_{\rm good}$. Amplitude estimation
uses $\cA$, its inverse, and reflections about these subspaces to
estimate $\mu$. For example, the associated Grover iterate can be
written as
\begin{equation*}
\begin{aligned}
 \mathcal G={}&\cA(2|0\rangle\langle0|-I)\cA^\dagger\\
 &\times(I-2\Pi_{\rm good}).
\end{aligned}
\end{equation*}
Writing $\mu=\sin^2\theta$, the iterate acts as a rotation by
$2\theta$ on the two-dimensional subspace relevant to amplitude
estimation. Phase-estimation or likelihood-based procedures then
estimate $\mu$ to additive error $\epsilon$ using
$\mathcal O(\epsilon^{-1}\log(1/\delta))$ coherent calls, with success
probability at least $1-\delta$~\cite{brassard2002,suzuki2020}.
In our setting, the good-flag probability represents the normalized
ensemble energy. We therefore need not measure the velocity field
itself.

\subsection{Notation and Conventions}
\label{sec:bg-notation}

For a vector, $\|\cdot\|_2$ is its Euclidean norm; for an operator,
it is the induced spectral norm. We write $A^*$ for the conjugate
transpose, $A^\dagger$ for the Moore--Penrose pseudoinverse, and
$A\preceq B$ when the Hermitian matrix $B-A$ is positive
semidefinite. The $L^2$ inner product and norm on $\mathbb T^d$ use
the normalized spatial measure defined above. With this convention,
Parseval's identity is
\begin{equation*}
 \|w\|_{L^2}^2
 =\sum_{k\in\mathbb Z^d}|\widehat w(k)|_2^2.
\end{equation*}
The space $\dot L^2_{\rm div}(\mathbb T^d)$ consists of square-integrable,
mean-zero, divergence-free vector fields. We use $H^j$ for Sobolev
spaces with $j$ square-integrable weak derivatives and $W^{j,\infty}$
when these derivatives are bounded. We also use the periodic
H\"older space $C^{r_0,\beta}$, with $r=r_0+\beta$,
$r_0\in\mathbb N_0$, and $0<\beta\leq1$; at $r=1$ this denotes the
Lipschitz class $W^{1,\infty}$. As $\epsilon\to0^+$, the symbols
$\mathcal O$, $\Omega$, and $\Theta$ have their standard asymptotic
meanings, while $\widetilde{\mathcal O}$ suppresses polylogarithmic
factors in $M,N,d,\widetilde\kappa$, and $1/\epsilon$.
A list of the principal symbols is given in \Cref{app:notation}.

\section{Eliminating the Incompressibility Constraint}
\label{sec:reduction}

The first step in our solver is to represent only velocities that
satisfy the incompressibility constraint. We begin with a general
finite-dimensional formulation, so the construction does not depend
on a particular choice of Fourier grid. We then show that the
resulting reduced system is equivalent to the original constrained
problem.

\subsection{Divergence-Free Coordinates}

Let $\cH$ be a finite-dimensional complex vector space of discrete
velocities, equipped with an inner product. Let $\cY$ be the space of
constraint values, and write
\begin{equation*}
 D:\cH\longrightarrow\cY
\end{equation*}
for the linear constraint operator. In a discretization of
incompressible flow, $D$ includes the discrete divergence and, when
needed, the mean-flow constraint. The admissible velocities are
exactly those satisfying $Du=0$, namely the nullspace $\Ker D$.

One way to enforce the constraint is to choose another coordinate
space $\cZ$ and a linear map $C:\cZ\to\cH$ whose range is exactly
$\Ker D$. Every velocity of the form $u=C\phi$ is then admissible,
because $DC=0$. Examples are a discrete curl or a Fourier basis
whose vectors are perpendicular to their wavenumbers.

This reduction removes the constraint, but it can introduce a new
problem. The columns of $C$ need not have equal lengths or be
orthogonal. A coordinate vector may therefore be strongly stretched
in some directions and compressed in others. For example, a curl
multiplies Fourier coefficients by their wavenumbers. Using $C$
directly can worsen the conditioning of the reduced system
$C^*A_sC$, even when the physical operator $A_s$ is well behaved on
$\Ker D$. We want to retain the admissible velocities without
introducing this additional scaling.

\subsection{The Polar Riesz Map}

We remove the scaling of $C$ by replacing it with the norm-preserving
part of its polar decomposition. If $C$ has redundant coordinates,
those coordinates lie in $\Ker C$ and do not change the velocity.
We therefore define the nonredundant coordinate space
$\cZ_0=(\Ker C)^\perp$. Let
\begin{equation}
 A_C=C^*C,
 \qquad
 R=C A_C^{\dagger/2},
 \label{eq:riesz}
\end{equation}
where $A_C^{\dagger/2}$ acts as the inverse square root of $A_C$
on $\cZ_0$ and is zero on $\Ker C$. The map $R$ has the same range as
$C$, but its restriction to $\cZ_0$ preserves Euclidean norms.

\begin{proposition}[Constraint-preserving Riesz map]
\label{prop:riesz}
Let $C:\cZ\to\cH$ satisfy $\Ran C=\Ker D$, and let $R$ be defined
by \eqref{eq:riesz}. Then
\begin{equation}
\begin{aligned}
 DR&=0,\\
 R^*R&=P_{\cZ_0},\\
 RR^*&=P_{\Ker D},
\end{aligned}
\label{eq:riesz-identities}
\end{equation}
where $P_{\cZ_0}$ and $P_{\Ker D}$ are the orthogonal projections
onto the indicated spaces. Consequently, $R\xi$ satisfies the
constraint for every $\xi\in\cZ$, and
\begin{equation*}
 \|R(\widetilde\xi-\xi)\|_2
 \leq\|\widetilde\xi-\xi\|_2,
\end{equation*}
with equality when both coordinate vectors lie in $\cZ_0$.
\end{proposition}

\begin{proof}
Let $\rho=\operatorname{rank}(C)$, and take a compact singular-value
decomposition
\begin{equation*}
 C=U_\rho\Sigma_\rho V_\rho^*.
\end{equation*}
Here $\Sigma_\rho$ is a $\rho\times\rho$ diagonal matrix with
strictly positive entries. The columns of $U_\rho$ form an
orthonormal basis for $\Ran C=\Ker D$, and the columns of $V_\rho$
form an orthonormal basis for $\cZ_0$. It follows that
\begin{equation*}
\begin{aligned}
 A_C&=V_\rho\Sigma_\rho^2V_\rho^*,\\
 A_C^{\dagger/2}&=V_\rho\Sigma_\rho^{-1}V_\rho^*.
\end{aligned}
\end{equation*}
Substituting these expressions into \eqref{eq:riesz} cancels the
singular values:
\begin{equation*}
 R=U_\rho V_\rho^*.
\end{equation*}
This is the essential point: $R$ retains the two orthonormal
coordinate systems, but not the unequal scales in $\Sigma_\rho$.
Therefore
\begin{equation*}
\begin{aligned}
 R^*R&=V_\rho V_\rho^*=P_{\cZ_0},\\
 RR^*&=U_\rho U_\rho^*=P_{\Ker D}.
\end{aligned}
\end{equation*}
Also, $DR=0$ because the range of $R$ is contained in $\Ker D$.
Finally, for any $\eta\in\cZ$,
\begin{equation*}
 \|R\eta\|_2^2
 =\langle\eta,P_{\cZ_0}\eta\rangle
 \leq\|\eta\|_2^2,
\end{equation*}
with equality for $\eta\in\cZ_0$. Taking
$\eta=\widetilde\xi-\xi$ completes the proof.
\end{proof}

The proposition separates two kinds of numerical error. If the
coordinates $\xi$ are approximated but the map $R$ is applied
exactly, the reconstructed velocity remains divergence-free: the
coordinate error changes the velocity within $\Ker D$. A real
circuit also approximates $R$. Writing its implementation as
$\widetilde R=R+\Delta R$ gives
$D\widetilde R=D\Delta R$, which need not be zero. The angle
synthesis of \Cref{prop:be-cost} controls
$\|\Delta R\|_2=\mathcal O(2^{-b_{\rm arith}})$, where
$b_{\rm arith}$ is the arithmetic precision in bits. This is an
operator-norm guarantee, not a grid-independent physical divergence
bound, because $\|D\|_2$ may itself grow with grid resolution. The
later analysis uses the operator-norm error to bound the error in
the measured observable (\Cref{lem:bias}).

\subsection{The Reduced Positive-Definite System}

We next use $R$ to remove the constraint from the linear equations.
For scenario $s$, let $A_s:\cH\to\cH$ be a Hermitian operator that is
coercive on $\Ker D$. By coercivity we mean that some constant
$\gamma_s>0$ satisfies
\begin{equation*}
 \langle v,A_sv\rangle
 \geq\gamma_s\|v\|_2^2
 \quad(v\in\Ker D).
\end{equation*}
In particular, the physical operator has a strictly positive
quadratic form on every nonzero admissible velocity. Let
$f_s\in\cH$ be the discrete body force and define its reduced form
$b_s=R^*f_s\in\cZ_0$. We assume the source has been normalized so
that $\|b_s\|_2\leq1$ for all scenarios. Since $RR^*=P_{\Ker D}$,
the vector $Rb_s$ is the projection of the forcing onto the
admissible velocity space.

The constrained problem is to find $u_s\in\Ker D$ such that
\begin{equation}
 \langle v,A_su_s\rangle
 =\langle v,Rb_s\rangle
 \quad\text{for all }v\in\Ker D.
 \label{eq:weak}
\end{equation}
This is a variational statement: the equation must hold when tested
against every admissible velocity $v$. To express it in reduced
coordinates, write $u_s=R\xi_s$ and $v=R\eta$ for
$\xi_s,\eta\in\cZ_0$. The isometry $R^*R=I_{\cZ_0}$ then changes
\eqref{eq:weak} into
\begin{equation*}
 \langle\eta,(R^*A_sR)\xi_s\rangle
 =\langle\eta,b_s\rangle
 \quad(\eta\in\cZ_0).
\end{equation*}
Thus, defining the reduced operator
\begin{equation*}
 K_s:=R^*A_sR\big|_{\cZ_0},
\end{equation*}
we obtain the ordinary linear system $K_s\xi_s=b_s$. The next
proposition shows that this change of coordinates loses neither
the physical velocity nor the multiplier associated with the
constraint.

\begin{proposition}[Reduced--saddle equivalence]
\label{prop:reduced}
For each scenario $s$, $K_s$ is Hermitian positive definite on
$\cZ_0$. The unique solution of \eqref{eq:weak} is
\begin{equation*}
 \xi_s=K_s^{-1}b_s,
 \qquad
 u_s=R\xi_s\in\Ker D,
\end{equation*}
and $\|u_s\|_2=\|\xi_s\|_2$. If $D:\cH\to\cY$ has full row rank,
then there is also a unique multiplier $p_s\in\cY$ such that
$(u_s,p_s)$ solves
\begin{equation}
 \begin{bmatrix}A_s&D^*\\D&0\end{bmatrix}
 \begin{bmatrix}u_s\\p_s\end{bmatrix}
 =\begin{bmatrix}Rb_s\\0\end{bmatrix}.
 \label{eq:saddle}
\end{equation}
\end{proposition}

To see why the reduced operator is positive definite, take any
nonzero $\xi\in\cZ_0$. Since $R$ preserves its norm,
$R\xi\neq0$, and coercivity gives
\begin{equation*}
 \langle\xi,K_s\xi\rangle
 =\langle R\xi,A_sR\xi\rangle
 \geq\gamma_s\|\xi\|_2^2>0.
\end{equation*}
The equivalence follows from the coordinate change above. The
multiplier can be recovered from the residual
$Rb_s-A_su_s$: this residual is orthogonal to $\Ker D$, so it lies
in $\Ran D^*$; full row rank makes its representation as $D^*p_s$
unique. The complete proof is given in \Cref{app:proofs-early}.

This result lets us solve a positive-definite system for $\xi_s$
instead of an indefinite saddle-point system for $(u_s,p_s)$. The
physical velocity is still obtained by applying $R$, and therefore
satisfies the constraint in exact arithmetic. The next stage of the
algorithm will precondition $K_s$ to control its condition number.

\subsection{Normalizing the Observable}

Our eventual goal is not to measure the full velocity vector, but
to estimate an average of a bounded scalar response. Let $O_s$ be a
Hermitian observable operator satisfying
$0\preceq O_s\preceq I_{\cH}$. For total kinetic energy,
$O_s=I_{\cH}$; a projection onto a spatial region can also be used
to measure energy in that region. Choose a constant $\beta$ such
that
\begin{equation*}
 \beta\geq\max_{0\leq s<M}\|K_s^{-1}\|_2.
\end{equation*}
Because $\|b_s\|_2\leq1$ and $R$ is an isometry on $\cZ_0$, we
have $\|u_s\|_2\leq\beta$. We can therefore define
\begin{equation}
\begin{aligned}
 g_s&=\frac{\langle u_s,O_su_s\rangle}{\beta^2}
 \in[0,1],\\
 \mu&=\frac1M\sum_{s=0}^{M-1}g_s\in[0,1].
\end{aligned}
\label{eq:target}
\end{equation}
For $O_s=I_{\cH}$, the unnormalized mean kinetic energy is
$\bar{\mathcal E}=\beta^2\mu$. This normalization makes the desired
average suitable for amplitude estimation.

\section{Coherent Evaluation of the Ensemble Mean}
\label{sec:access}

The reduced system $K_s\xi_s=b_s$ gives the velocity
$u_s=R\xi_s$.
The circuit encodes the ensemble average $\mu$ from \eqref{eq:target}
as the probability of a designated measurement outcome. For kinetic
energy, this avoids reconstructing the velocity field.

\subsection{Encoding the Observable in a Probability}

Recall that $O_s$ is a Hermitian operator satisfying
$0\preceq O_s\preceq I_{\cH}$ and that $\beta$ is chosen so that
$\beta\geq\max_s\|K_s^{-1}\|_2$. For kinetic energy we take
$O_s=I_{\cH}$. More generally, because $O_s$ is positive semidefinite,
we can write $O_s=E_s^*E_s$, for example with $E_s=O_s^{1/2}$.
The bound $O_s\preceq I_{\cH}$ implies $\|E_s\|_2\leq1$.

Define the scaled inverse
\begin{equation*}
 F_s:=\frac{K_s^{-1}}{\beta}.
\end{equation*}
By the choice of $\beta$, $\|F_s\|_2\leq1$. The Riesz map also satisfies
$\|R\|_2\leq1$. Thus $F_s$, $R$, and $E_s$ are all
\emph{contractions}: applying any one of them cannot increase the norm
of a vector. Using $u_s=RK_s^{-1}b_s$, the normalized observable becomes
\begin{equation}
 g_s=\|E_sRF_sb_s\|_2^2.
 \label{eq:coherent-chain}
\end{equation}
This identity gives a way to obtain $g_s$ from a quantum circuit.
Starting with the source $b_s$, we apply the scaled inverse $F_s$,
reconstruct the velocity with $R$, and apply $E_s$ to select the
observable. The squared norm of the resulting vector is $g_s$.

The operators in this sequence are generally not unitary. In the
circuit, each contraction is therefore represented as a block of a
larger unitary, with an ancilla register recording the corresponding
successful branch. No intermediate measurement is needed. The joint
``good'' branch has squared norm $g_s$, while the other branches account
for the remaining probability. This is the connection between the
mathematical expression in \eqref{eq:coherent-chain} and the flag measured
by amplitude estimation.

We must also apply the procedure to all scenarios coherently. We first
prepare the uniform scenario register
\begin{equation*}
 |\mathrm{scen}\rangle
 =\frac{1}{\sqrt M}\sum_{s=0}^{M-1}|s\rangle,
\end{equation*}
and then perform the source preparation, inverse, reconstruction, and
observable operations controlled by $s$. The scenario basis states are
orthogonal. Therefore the probability of the good flag, without
measuring the scenario register, is
\begin{equation*}
 \frac1M\sum_{s=0}^{M-1}g_s=\mu.
\end{equation*}
Applying quantum amplitude estimation to this circuit and its inverse
then gives an estimate of $\mu$. For kinetic energy, the mean of the
unscaled observable can be recovered by multiplying by $\beta^2$.

\subsection{Effect of Circuit Approximation}

The identity above assumes exact operations. In practice, the source
preparation, QSVT inverse, basis rotations, and observable encoding
are implemented with finite precision. We write a tilde to distinguish
an implemented quantity from its ideal counterpart and assign an error
tolerance to each stage:
\begin{enumerate}[label=(\arabic*),leftmargin=2.2em]
\item \emph{Source preparation.} The prepared source vector
$\widetilde b_s$ satisfies
$\|\widetilde b_s-b_s\|_2\leq\eta_b$ and
$\|\widetilde b_s\|_2\leq1$.
\item \emph{Inverse.} The QSVT circuit implements a contraction
$\widetilde F_s$ with
$\|\widetilde F_s-F_s\|_2\leq\eta_F$.
\item \emph{Velocity reconstruction.} The implemented Riesz map
$\widetilde R$ is a contraction with
$\|\widetilde R-R\|_2\leq\eta_R$.
\item \emph{Observable encoding.} The implemented operator
$\widetilde E_s$ is a contraction with
$\|\widetilde E_s-E_s\|_2\leq\eta_E$.
\end{enumerate}
The four errors are assumed to hold uniformly over the scenarios.
The bounds for the approximate operators include the errors of their
block encodings and finite-precision rotations.

For scenario $s$, the actual circuit has good-flag probability
\begin{equation*}
 \widetilde g_s
 =\|\widetilde E_s\widetilde R
        \widetilde F_s\widetilde b_s\|_2^2.
\end{equation*}
We denote the mean probability prepared by the implemented circuit
$\cA$ by
\begin{equation*}
 \widetilde\mu_{\rm cir}
 =\frac1M\sum_{s=0}^{M-1}\widetilde g_s.
\end{equation*}
Amplitude estimation will estimate $\widetilde\mu_{\rm cir}$,
not the ideal value $\mu$. We therefore need to bound the difference
between the two.

Let
\begin{equation*}
 \eta_{\rm impl}:=\eta_b+\eta_F+\eta_R+\eta_E.
\end{equation*}
Since every stage is a contraction, changing one stage does not
amplify errors introduced by another. This gives the following bound.

\begin{lemma}[Observable bias of a contraction chain]
\label{lem:bias}
Suppose $b_s,\widetilde b_s$ have norm at most one, and
$F_s,\widetilde F_s,R,\widetilde R,E_s,\widetilde E_s$ are contractions
satisfying the four error bounds above. Then, for every scenario $s$,
\begin{equation}
 |\widetilde g_s-g_s|
 \leq 2\eta_{\rm impl}+\eta_{\rm impl}^2.
 \label{eq:bias}
\end{equation}
The ensemble mean satisfies the same bound:
\begin{equation*}
 |\widetilde\mu_{\rm cir}-\mu|
 \leq 2\eta_{\rm impl}+\eta_{\rm impl}^2.
\end{equation*}
\end{lemma}

To understand the bound, write
$y_s=E_sRF_sb_s$ and
$\widetilde y_s=\widetilde E_s\widetilde R
\widetilde F_s\widetilde b_s$.
Subtracting the two products one stage at a time and using the
contraction bounds gives
\begin{equation*}
 \|\widetilde y_s-y_s\|_2\leq\eta_{\rm impl}.
\end{equation*}
The difference between their squared norms is therefore at most
$2\eta_{\rm impl}+\eta_{\rm impl}^2$. Averaging this inequality over
all $M$ scenarios gives the stated bound for the ensemble mean.
The full telescoping argument is given in \Cref{app:proofs-early}.

\subsection{Accuracy and Cost of the Coherent Evaluator}

The estimate has two distinct sources of error. The first is the
\emph{implementation bias} just bounded: the circuit encodes
$\widetilde\mu_{\rm cir}$ rather than $\mu$. The second is the
\emph{statistical estimation error} of amplitude estimation, which
approximates the probability encoded by the circuit. We choose their
tolerances so that the total error does not exceed $\epsilon$.

For the cost statement, let $\kappa(K_s)$ denote the condition
number of the positive-definite reduced operator, and set
$\kappa:=\max_s\kappa(K_s)$. Let $\Gamma(\kappa,\beta,\eta)$
denote a bound on the gate cost of one coherent use of the implemented
state-preparation circuit $\cA$ or its inverse, including the
reflections required by amplitude estimation, when the
implementation tolerance is $\eta$. This quantity includes the cost
of preparing the inputs and solving the PDE. We do not yet assume it
is small.

\begin{theorem}[Correctness and cost of one coherent evaluator]
\label{thm:quantum}
Let $0<\epsilon\leq1$ and $0<\delta<1$. Suppose the circuit errors
satisfy $\eta_{\rm impl}\leq\epsilon/8$. Apply quantum amplitude
estimation to $\cA$ with statistical tolerance
$\epsilon_{\rm stat}\leq\epsilon/2$, and repeat the estimation as
needed to achieve confidence $1-\delta$. The resulting estimate
$\widetilde\mu$ satisfies
\begin{equation*}
 \Pr\bigl[|\widetilde\mu-\mu|\leq\epsilon\bigr]
 \geq1-\delta
\end{equation*}
with logical gate complexity
\begin{equation}
 T_Q=\mathcal O\!\left(
  \frac{\Gamma(\kappa,\beta,\epsilon/8)}{\epsilon}
  \log\frac1\delta\right).
 \label{eq:qcost}
\end{equation}
\end{theorem}

Indeed, \Cref{lem:bias} and $\epsilon\leq1$ give
\begin{equation*}
 \begin{aligned}
 |\widetilde\mu_{\rm cir}-\mu|
 &\leq\frac\epsilon4+\frac{\epsilon^2}{64}\\
 &<\frac\epsilon2.
 \end{aligned}
\end{equation*}
Amplitude estimation approximates $\widetilde\mu_{\rm cir}$
within $\epsilon/2$ with a constant success probability. Taking the
median of $\mathcal O(\log(1/\delta))$ independent runs increases
the success probability to at least $1-\delta$. On that event, the
triangle inequality gives total error at most $\epsilon$.
Each run uses $\mathcal O(\epsilon^{-1})$ coherent evaluations,
which gives \eqref{eq:qcost}. A detailed proof is provided in
\Cref{app:proofs-early}.

This theorem separates the sampling cost from the PDE-solver cost.
It does not yet show that the quantum algorithm is efficient:
all of the work needed for one coherent solve is included in
$\Gamma$. In \Cref{sec:sobolev-riesz} we construct a preconditioned
Brinkman solver and bound this quantity explicitly.

\section{When Faster Sampling Pays Off}
\label{sec:frontier}
\label{sec:advantage}

The cost comparison depends on the sampling rate, the size of the
ensemble, and the cost of each PDE evaluation. The budget below
includes classical sampling rules such as quasi-Monte Carlo.

\subsection{The Precision--Ensemble Budget}
\label{sec:budget}

We study how the cost changes as the target error $\epsilon$ decreases.
The number of scenarios may grow with accuracy, and resolving the fluid
solution may require a finer spatial grid. We assume three power-law
relationships:
\begin{enumerate}[label=(\arabic*),leftmargin=2.2em]
\item The ensemble contains
$M(\epsilon)=\Theta(\epsilon^{-m})$ scenarios, where $m\geq0$.
The case $m=0$ describes a fixed-size ensemble; $m>0$ allows the
number of scenarios to grow as $\epsilon$ decreases.
\item One coherent quantum evaluation, including its PDE solve, costs
$\Gamma_Q(\epsilon)=\Theta(\epsilon^{-g_Q})$ gates, where $g_Q\geq0$.
\item One classical PDE solve costs
$G_C(\epsilon)=\Theta(\epsilon^{-g_C})$ operations,
where $g_C\geq0$.
\end{enumerate}
The exponents $g_Q$ and $g_C$ describe the cost of one evaluation,
not the number of evaluations. The theorem below assumes exact
power-law cost expressions. In the later solver application, logarithmic
factors are suppressed; they do not change a strict power-law advantage
when the exponents differ.

As in \Cref{sec:bg-uq}, let $p>0$ be the convergence order of the
chosen classical sampling rule. Under its assumed cost law, obtaining
error of order $\epsilon$ requires
$\Theta(\epsilon^{-1/p})$ scenario evaluations when the ensemble is
large. For a finite ensemble, evaluating all $M$ scenarios is also an
option. We therefore model the number of classical solves by
\begin{equation*}
 n_C(\epsilon)
 =\Theta\!\left(\min\{M(\epsilon),\epsilon^{-1/p}\}\right).
\end{equation*}
Multiplying by the cost of one solve gives
\begin{equation*}
 T_C(\epsilon)
 =\Theta\!\left(G_C(\epsilon)
 \min\{M(\epsilon),\epsilon^{-1/p}\}\right).
\end{equation*}

For the QAE-based procedure of the previous section, we use the
usual worst-case mean-estimation scaling of
$\Theta(\epsilon^{-1})$ coherent evaluations. Each evaluation
encodes the response in an amplitude; it does not return the value
$g_s$ as an exactly readable classical number. The total quantum
cost in this model is therefore
\begin{equation*}
 T_Q(\epsilon)
 =\Theta\!\left(\Gamma_Q(\epsilon)\epsilon^{-1}\right).
\end{equation*}
This comparison concerns the specified quantum and classical
procedures. It does not claim that these are optimal for every
possible way of representing the input.

\begin{theorem}[Precision--ensemble advantage budget]
\label{thm:budget}
Under the power-law cost assumptions above,
\begin{equation}
 \frac{T_Q(\epsilon)}{T_C(\epsilon)}
 =\Theta\!\left(
 \epsilon^{-\left[g_Q-g_C+1-\min\{m,1/p\}\right]}
 \right).
\end{equation}
A strict power-law quantum advantage, meaning that
$T_Q/T_C=\mathcal O(\epsilon^\alpha)$ for some $\alpha>0$,
holds if and only if
\begin{equation}
 \begin{aligned}
 \boxed{g_Q-g_C<B(p,m)},\\
 B(p,m):=\min\{m,1/p\}-1.
 \end{aligned}
 \label{eq:budget-frontier}
\end{equation}
\end{theorem}

\begin{proof}
Since $M(\epsilon)=\Theta(\epsilon^{-m})$,
\begin{equation*}
 \begin{aligned}
 T_C(\epsilon)
 &=\Theta\!\left(
 \epsilon^{-g_C-\min\{m,1/p\}}\right),\\
 T_Q(\epsilon)
 &=\Theta\!\left(\epsilon^{-g_Q-1}\right).
 \end{aligned}
\end{equation*}
Dividing the second expression by the first gives the claimed
ratio. The ratio decreases as a positive power of $\epsilon$ if
and only if
$g_C+\min\{m,1/p\}-g_Q-1>0$.
Rearranging this inequality gives
$g_Q-g_C<\min\{m,1/p\}-1$, as required.
\end{proof}

The quantity $B(p,m)$ is the \emph{precision--ensemble budget}.
It tells us how much more expensive one quantum evaluation may be,
relative to one classical solve, while still allowing a power-law
advantage. A negative budget means that the quantum evaluation must
have a lower cost exponent than the classical solve. Three cases
illustrate the result.

\paragraph{Ordinary Monte Carlo.}
For $p=1/2$ and an ensemble with $m\geq2$,
$B(1/2,m)=1$. The QAE procedure can therefore retain a power-law
advantage even when the quantum cost per evaluation grows faster than
the classical cost, provided $g_Q-g_C<1$. When $m<1$, the budget
becomes negative because the complete ensemble is smaller than the
number of evaluations assumed for QAE.

\paragraph{First-order quasi-Monte Carlo.}
For $p=1$,
\begin{equation*}
 B(1,m)=\min\{m,1\}-1\leq0.
\end{equation*}
If $m\geq1$, an advantage requires $g_Q<g_C$.
If $m<1$, the stronger condition $g_Q-g_C<m-1$ is needed.
Thus QAE's sampling rate alone does not provide an asymptotic
improvement over first-order QMC. The cost of the PDE evaluation
must also be reduced.

\paragraph{The preconditioned Brinkman solver.}
The Sobolev--Riesz preconditioner in \Cref{sec:sobolev-riesz} keeps
the inverse cost from acquiring a power of $N$ through conditioning.
Under the compact-input and low-wavenumber-forcing assumptions
stated there, with fixed numbers of drag modes and source terms,
the gate-cost bound per evaluation is polylogarithmic in
$1/\epsilon$, so we take $g_Q=0$ in the power-law comparison.
For the specified classical grid-resolving solvers, assume
$N=\Theta(\epsilon^{-1/r})$ grid points per spatial direction,
where $r>0$ is the assumed spatial convergence exponent. A solver
that processes all $N^d$ unknowns has cost exponent $g_C=d/r$,
up to logarithmic factors, where $d$ is the spatial dimension.
In this case $g_Q-g_C=-d/r$, and the advantage condition becomes
\begin{equation*}
 \frac{d}{r}>1-\min\{m,1/p\}.
\end{equation*}
This is a conditional comparison with that classical solver class,
not a lower bound on all classical PDE algorithms. It also depends on
the efficient coherent input preparation proved or assumed in the
solver analysis.

\begin{figure*}[tp]
\centering
\includegraphics[width=0.98\linewidth]{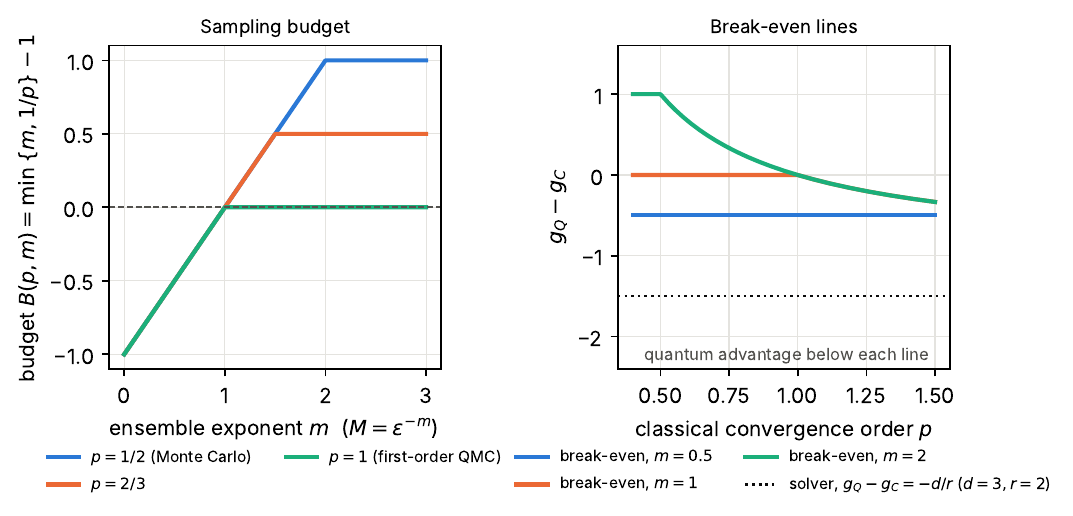}
\caption{The precision--ensemble budget
$B(p,m)=\min\{m,1/p\}-1$.
Left: the budget as a function of the ensemble exponent $m$ for
classical sampling orders between ordinary Monte Carlo ($p=1/2$)
and first-order QMC ($p=1$).
Right: the boundary $g_Q-g_C=B(p,m)$ for three values of $m$.
A power-law quantum advantage is obtained below the boundary.
The dotted line shows the conditional solver exponent
$g_Q-g_C=-d/r$ for $d=3$ and $r=2$.
The curves are evaluations of the cost model, not measured runtimes.}
\label{fig:accuracy-budget}
\end{figure*}

\subsection{A Query Benchmark in Stokes Flow}
\label{sec:stokes-benchmark}

The preceding theorem compares cost laws, but it does not establish a
lower bound for the Brinkman solver. To see the usual sampling query
rates in a fluid setting, consider a deliberately simple Stokes
problem in which the unknown input is a list of binary values.
The fluid solve is elementary; the difficulty is determining the mean
of the hidden list.

Let $z=(z_0,\ldots,z_{M-1})\in\{0,1\}^M$ be a binary string.
For scenario $s$, its bit $z_s$ determines the strength of the body
force. An input query returns $z_s$ for a chosen index $s$; a
quantum query may access the same oracle coherently. We use the
periodic two-dimensional domain $\mathbb T^2$ with normalized
spatial measure $(2\pi)^{-2}dx\,dy$, and the shear profile
\begin{equation*}
 v(x,y)=\sqrt2\,(0,\sin x)^T.
\end{equation*}
The vector field $v$ is divergence-free, has zero mean, satisfies
$-\Delta v=v$, and has squared $L^2$ norm equal to one. These
properties make the response to the body force $z_sv$ explicit.

\begin{theorem}[Optimal query separation inside periodic Stokes flow]
\label{thm:optimal-stokes}
For the hidden string $z\in\{0,1\}^M$, consider the $M$ steady
periodic Stokes systems
\begin{equation*}
 \begin{aligned}
 -\Delta u_s+\nabla p_s&=z_sv,\\
 \nabla\cdot u_s&=0,\\
 \int_{\mathbb T^2}u_s\,dx\,dy&=0,\\
 \int_{\mathbb T^2}p_s\,dx\,dy&=0.
 \end{aligned}
\end{equation*}
Let $g_s=\|u_s\|_{L^2(\mathbb T^2)}^2$, and suppose the hidden
bits are accessed through the forcing oracle
\begin{equation*}
 O_z|s\rangle|b\rangle
 =|s\rangle|b\oplus z_s\rangle,
\end{equation*}
where $b\in\{0,1\}$ and $\oplus$ denotes addition modulo two.
For $0<\epsilon\leq1/16$, estimating
$\mu=M^{-1}\sum_{s=0}^{M-1}g_s$ to additive error $\epsilon$
with success probability at least $2/3$ has worst-case query
complexities
\begin{equation*}
 \begin{aligned}
 Q_C&=\Theta\!\left(\min\{M,\epsilon^{-2}\}\right),\\
 Q_Q&=\Theta\!\left(\min\{M,\epsilon^{-1}\}\right).
 \end{aligned}
\end{equation*}
\end{theorem}

To see why, $u_s=z_sv$ and $p_s=0$ solve the Stokes equations.
Since $z_s$ is either zero or one and $\|v\|_{L^2}^2=1$,
\begin{equation*}
 g_s=\|z_sv\|_{L^2}^2=z_s.
\end{equation*}
Thus the mean fluid energy is exactly the fraction of ones in the
hidden string. Classically, this mean can be estimated by sampling
bits or reading all $M$ entries. Quantumly, it can be estimated by
amplitude estimation or by reading the complete list. The matching
query lower bounds follow from the finite-list argument in
\Cref{thm:classical} and quantum approximate counting
~\cite{nayak1999}; details are in \Cref{app:proofs-early}.

These are query counts, not gate counts or fluid-solver runtimes.
In this example the velocity is a known Fourier mode multiplied by
one hidden bit, so solving the Stokes equation adds no substantial
computational difficulty. The benchmark illustrates the sampling
part of our problem. The additional challenge for heterogeneous
Brinkman flow is to implement each coherent PDE solve efficiently,
which we address in \Cref{sec:sobolev-riesz}.

\section{A Preconditioned Quantum Solver for Brinkman Flow}
\label{sec:sobolev-riesz}

The previous section showed that faster sampling does not guarantee a
faster uncertainty-quantification calculation. In particular, when a
classical method achieves first-order quasi-Monte Carlo convergence
($p=1$), the precision--ensemble budget satisfies
$B(1,m)=\min\{m,1\}-1\leq0$. A strict power-law advantage therefore
requires the quantum PDE evaluation to have a smaller cost exponent
than a classical solve:
\begin{equation}
 g_Q-g_C<0.
 \label{eq:negative-relative-exponent}
\end{equation}
For a smaller ensemble, the required inequality is stronger, as shown
in \Cref{thm:budget}.

The cost of a quantum linear-system algorithm depends in part on the
condition number of the matrix it must invert. The divergence-free
reduction in \Cref{sec:reduction} removes the pressure constraint, but
it does not remove the large eigenvalues associated with the viscous
term. On a grid with $N$ points in each spatial direction, the reduced
Brinkman operator has condition number $\Theta(N^2)$ under the fixed
viscosity and drag bounds considered here. This makes a direct QSVT
solve increasingly expensive as the grid is refined. The penalty may
reduce or eliminate an advantage, although it need not do so for every
classical comparison. Our goal is to remove this grid dependence
before constructing the quantum circuit.

\subsection{Brinkman Operator and the Mean-Flow Constraint}
\label{sec:sobolev-riesz-sub}

We return to the periodic Brinkman problem defined in
\Cref{sec:bg-fluids}. For each scenario $s$, the velocity $u_s$, pressure
$p_s$, and constant mean-flow multiplier $\lambda_s$ satisfy
\begin{equation}
\begin{aligned}
 -\nu\Delta u_s+\alpha_s(x)u_s+\nabla p_s+\lambda_s&=f_s,\\
 \nabla\cdot u_s&=0,\\
 \int_{\mathbb T^d}u_s\,dx&=0,\\
 \int_{\mathbb T^d}p_s\,dx&=0.
\end{aligned}
\label{eq:brinkman-pde}
\end{equation}
Here $\nu>0$ is the viscosity parameter, $f_s$ is the body force,
and $\alpha_s(x)$ is the drag coefficient of scenario $s$. We assume
that the drag is bounded above and below by constants common to all
scenarios:
\begin{equation}
\begin{gathered}
 0<\alpha_{\min}\leq\alpha_s(x)\leq\alpha_{\max}<\infty,\\
 x\in\mathbb T^d,\quad s=0,\ldots,M-1.
\end{gathered}
\label{eq:alpha-bounds}
\end{equation}
The positive lower bound makes the drag contribution coercive; the
upper bound will enter the condition-number estimate. Both bounds are
assumed independent of the grid size.

For one class of structured inputs, the spatially varying drag has an
$L$-term Fourier expansion,
\begin{equation}
\begin{aligned}
 \alpha_s(x)&=\alpha_0+\sum_{\ell=1}^{L}
     c_\ell y_{s,\ell}e^{iq_\ell\cdot x},\\
 \alpha_0&=\frac{\alpha_{\min}+\alpha_{\max}}{2},\\
 y_{s,\ell}&\in[-1,1].
\end{aligned}
\label{eq:kl-expansion}
\end{equation}
Here $q_\ell\in\mathbb Z^d\setminus\{0\}$ is a spatial wavenumber,
$c_\ell\in\mathbb C$ is its fixed coefficient, and $y_{s,\ell}$
describes how that coefficient changes with the scenario. The number
$L$ is the expansion length, while $\alpha_0$ is a constant reference
drag. The field must be real-valued, so the terms occur in conjugate
pairs: a term at $q_\ell$ is accompanied by the complex-conjugate term
at $-q_\ell$. One sufficient condition for
\eqref{eq:alpha-bounds} is
$\sum_{\ell=1}^{L}|c_\ell|\leq
(\alpha_{\max}-\alpha_{\min})/2$.
This Fourier representation is useful for one circuit construction;
a second construction will work directly with a position-space
evaluator of $\alpha_s(x)$.

The constant multiplier $\lambda_s$ in \eqref{eq:brinkman-pde} is
important when the drag varies in space. For constant drag, a
zero-mean forcing produces a zero-mean solution. With variable drag,
the product $\alpha_s(x)u_s(x)$ can have a nonzero spatial mean even
when $u_s$ has none. Integrating the momentum equation over the
periodic cell removes the Laplacian and pressure-gradient terms and
gives
\begin{equation*}
 \lambda_s=\frac{1}{(2\pi)^d}
 \int_{\mathbb T^d}(f_s-\alpha_su_s)\,dx.
\end{equation*}
Thus $\lambda_s$ balances the mean force while the condition
$\int u_s\,dx=0$ is enforced. Without it, the zero-mean constraint
and the momentum equation can be incompatible. Equivalently, the
velocity equation can be projected onto the space of mean-zero,
divergence-free fields. The pressure and $\lambda_s$ can then be
recovered from the residual, as shown in
\Cref{lem:app-matched-pde-oracle}. In the notation of
\Cref{sec:reduction}, the combined constraint operator is
$D_{\rm tot}u=(\nabla\cdot u,(2\pi)^{-d}\int_{\mathbb T^d}u\,dx)$.
The version in which the mean velocity is allowed to be nonzero is
discussed in \Cref{rem:unconstrained}.

We use a Fourier--Galerkin discretization with $N$ grid points per
axis. Its retained wavenumbers are
$k\in\{-N/2,\ldots,N/2-1\}^d$, excluding the zero mode because the
velocity has zero mean. For each $k\neq0$, the velocity coefficient
is perpendicular to $k$; these are exactly the directions represented
by the Riesz map $R$ from \Cref{sec:reduction}. The resulting
finite-dimensional Brinkman operator is denoted by $A_s$, and its
reduced operator is $K_s=R^*A_sR$ on the nonredundant coordinate
space $\cZ_0$.

\subsection{Sobolev--Riesz Preconditioning}
\label{sec:sobolev-riesz-map}

The polar Riesz map removes the constraint without stretching the
remaining coordinates. It does not, however, change the fact that
high-wavenumber modes have a large viscous contribution. On mode $k$,
the operator $-\nu\Delta$ acts as multiplication by $\nu|k|^2$.
As the grid is refined, the largest retained $|k|$ grows with $N$.
We therefore rescale the reduced system by a reference operator that
contains this same wavenumber dependence.

Choose the constant-drag operator
\begin{equation*}
 A_0:=-\nu\Delta+\alpha_0 I,
\end{equation*}
and define its restriction to divergence-free coordinates by
\begin{equation*}
 K_0:=R^*A_0R\big|_{\cZ_0}.
\end{equation*}
Because $A_0$ has constant coefficients, $K_0$ is diagonal in the
Fourier basis. On each nonzero mode its eigenvalue is
$\nu|k|^2+\alpha_0>0$. We can therefore apply its inverse square
root mode by mode. Define the \emph{Sobolev--Riesz map}
\begin{equation}
 R_{\mathcal E}:=RK_0^{-1/2}:\cZ_0\longrightarrow\Ker D.
 \label{eq:sobolev-riesz}
\end{equation}
The factor $K_0^{-1/2}$ rescales a mode $k$ by
$(\nu|k|^2+\alpha_0)^{-1/2}$, while $R$ maps the coordinates back
to a divergence-free velocity. Hence the change of coordinates
preserves the incompressibility constraint in exact arithmetic.

We measure velocities in the reference energy norm
\begin{equation*}
 \|u\|_{\mathcal E}^2:=\langle u,A_0u\rangle.
\end{equation*}
In this norm, the new coordinates have a useful property: applying
$R_{\mathcal E}$ preserves their length. More importantly, the
reference part of the Brinkman operator becomes the identity.
Only the difference between the actual and reference drag remains.

The precise result is as follows. The relation between
$R_{\mathcal E}$ and the equivalent construction
$C(C^*A_0C)^{\dagger/2}$ for a general divergence-free generator
$C$ is given in \Cref{app:generator}.

\begin{proposition}[Sobolev--Riesz isometric preconditioning]
\label{prop:sobolev-riesz}
Let $R_{\mathcal E}$ be defined by \eqref{eq:sobolev-riesz}. Then
$DR_{\mathcal E}=0$, and for every $\zeta\in\cZ_0$,
\begin{equation}
 \|R_{\mathcal E}\zeta\|_{\mathcal E}=\|\zeta\|_2.
 \label{eq:sobolev-isometry}
\end{equation}
Writing $u_s=R_{\mathcal E}\zeta_s$ and defining the
preconditioned source $\widetilde b_s:=K_0^{-1/2}b_s$, the constrained
Brinkman problem is equivalent to
\begin{equation}
\begin{aligned}
 \widetilde K_s\zeta_s&=\widetilde b_s,\\
 \widetilde K_s&:=R_{\mathcal E}^*A_sR_{\mathcal E}\\
 &=I_{\cZ_0}+K_0^{-1/2}R^*\\
 &\qquad\cdot(\alpha_s(x)-\alpha_0I)RK_0^{-1/2}.
\end{aligned}
\label{eq:precond-operator}
\end{equation}
For every spatial dimension $d\geq2$, grid size $N$, and scenario
$s$, the preconditioned operator satisfies
\begin{equation}
\begin{gathered}
 \frac{\alpha_{\min}}{\alpha_0}I_{\cZ_0}
 \preceq\widetilde K_s\preceq
 \frac{\alpha_{\max}}{\alpha_0}I_{\cZ_0},\\
 \kappa(\widetilde K_s)\leq\widetilde\kappa,
 \qquad\widetilde\kappa:=\frac{\alpha_{\max}}{\alpha_{\min}}.
\end{gathered}
\label{eq:mesh-independent-kappa}
\end{equation}
In particular, the condition number is bounded independently of
$N$ when the drag contrast $\widetilde\kappa$ is fixed.
\end{proposition}

\begin{proof}
We first show that the change of coordinates is an isometry. Since
$K_0$ is positive definite on $\cZ_0$, its inverse square root is
well defined. Using $R^*A_0R=K_0$ on this space gives
\begin{equation*}
\begin{aligned}
 \|R_{\mathcal E}\zeta\|_{\mathcal E}^2
 &=\langle RK_0^{-1/2}\zeta,
       A_0RK_0^{-1/2}\zeta\rangle\\
 &=\langle\zeta,K_0^{-1/2}K_0K_0^{-1/2}\zeta\rangle\\
 &=\|\zeta\|_2^2.
\end{aligned}
\end{equation*}
Also $DR_{\mathcal E}=0$, because $DR=0$.
Testing the original variational equation \eqref{eq:weak} with
$v=R_{\mathcal E}\phi$, for any $\phi\in\cZ_0$, gives
$\widetilde K_s\zeta_s=\widetilde b_s$. Substituting
$A_s=A_0+(\alpha_s-\alpha_0)I$ then gives
\eqref{eq:precond-operator}.

It remains to bound the spectrum. The pointwise drag bounds
\eqref{eq:alpha-bounds}, together with $-\nu\Delta\succeq0$, imply
\begin{equation*}
 \frac{\alpha_{\min}}{\alpha_0}A_0
 \preceq A_s\preceq
 \frac{\alpha_{\max}}{\alpha_0}A_0.
\end{equation*}
To see this, the viscous term is nonnegative and
$\alpha_{\min}/\alpha_0\leq1\leq\alpha_{\max}/\alpha_0$;
therefore the same bounds hold for the sum of the viscous and drag
terms. Now take $u=R_{\mathcal E}\zeta$. By the isometry proved
above, $\langle u,A_0u\rangle=\|\zeta\|_2^2$, while
$\langle u,A_su\rangle=
\langle\zeta,\widetilde K_s\zeta\rangle$. Thus
\begin{equation*}
\begin{aligned}
 \frac{\alpha_{\min}}{\alpha_0}\|\zeta\|_2^2
 &\leq\langle\zeta,\widetilde K_s\zeta\rangle\\
 &\leq\frac{\alpha_{\max}}{\alpha_0}\|\zeta\|_2^2.
\end{aligned}
\end{equation*}
These inequalities bound the smallest and largest eigenvalues of
$\widetilde K_s$. Their ratio is at most
$\alpha_{\max}/\alpha_{\min}$, which proves
\eqref{eq:mesh-independent-kappa}.
\end{proof}

This bound is the main mathematical reason for using the
preconditioner. The number of Fourier modes can grow with grid
refinement, but the condition-number dependence of the quantum
inverse no longer grows with it. This does not by itself prove that
the full quantum algorithm is cheap. The next step is to implement
$\widetilde K_s$ without introducing a new grid-dependent cost.

\subsection{Quantum Circuit Construction}
\label{sec:be-architecture}

The preconditioned operator has the form
$\widetilde K_s=I_{\cZ_0}+H_s$, where
\begin{equation*}
\begin{aligned}
 H_s&:=\widetilde K_s-I_{\cZ_0}\\
 &=K_0^{-1/2}R^*(\alpha_s(x)-\alpha_0I)\\
 &\qquad\cdot RK_0^{-1/2}.
\end{aligned}
\end{equation*}
The constant-drag reference operator has become the identity, which
needs no separate Laplacian block encoding. The remaining operator
$H_s$ contains only the difference between the actual and reference
drag. Define
\begin{equation*}
 r_{\max}:=\frac{\alpha_{\max}-\alpha_{\min}}
 {\alpha_{\max}+\alpha_{\min}}
 =\frac{\widetilde\kappa-1}{\widetilde\kappa+1}<1.
\end{equation*}
The drag bounds imply $\|H_s\|_2\leq r_{\max}$. This is useful
because the part of the operator that the circuit must represent is
bounded independently of the grid.

To describe its implementation, consider a nonzero wavenumber $k$.
The matrix $R(k):\mathbb C^{d-1}\to\ker(k^T)$ maps a transverse
polarization vector to a velocity vector perpendicular to $k$.
It is the Fourier-mode form of the Riesz map. We also define the
weight
\begin{equation*}
 w(k):=\sqrt{\frac{\alpha_0}{\nu|k|^2+\alpha_0}},
 \qquad w(0):=0.
\end{equation*}
The weight lies between zero and one and implements the scaling
introduced by $K_0^{-1/2}$, apart from a constant factor.
The circuit combines these operations with a Fourier transform and
an evaluation of the spatially varying drag.

\begin{figure*}[tp]
\centering
\includegraphics[width=0.92\linewidth]{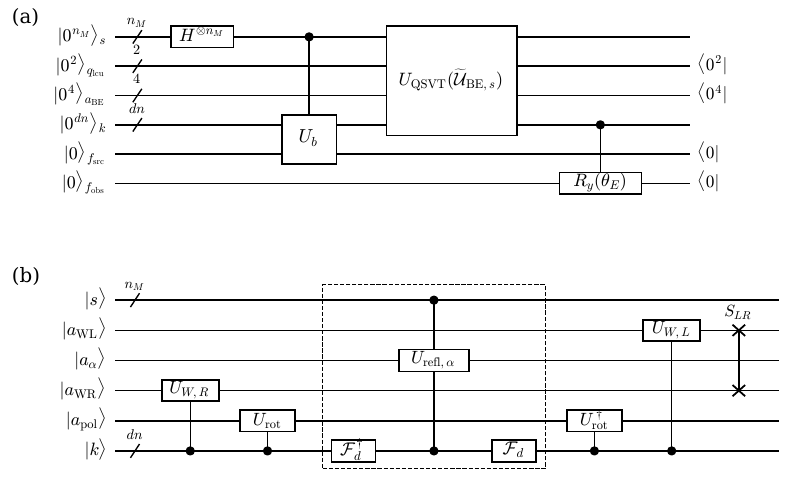}
\caption{Circuit structure of the preconditioned Brinkman solver.
(a)~State-preparation circuit $\cA$: uniform scenario superposition
$H^{\otimes n_M}$, preconditioned source loading $U_b$, QSVT
polynomial inverse $U_{\rm QSVT}(\widetilde{\mathcal U}_{{\rm BE},s})$,
and observable rotation $R_y(\theta_E)$. The good-flag subspace
$|0^2\rangle_{q_{\rm lcu}}|0^4\rangle_{a_{\rm BE}}|0\rangle_{f_{\rm src}}|0\rangle_{f_{\rm obs}}$
has probability $p_{\rm good}=\bar{\mathcal E}/B_{\rm norm}$.
(b)~Symmetric Hermitian block encoding
$\widetilde{\mathcal U}_{{\rm BE},s}
=S_{LR}\,U_{W,L}\,U_{\rm rot}^\dagger\,\mathcal F_d\,U_{{\rm refl},\alpha}\,
\mathcal F_d^\dagger\,U_{\rm rot}\,U_{W,R}$ in Construction~B
(\Cref{app:registers}); the dashed box groups the position-space
stage between $\mathcal F_d^\dagger$ and $\mathcal F_d$, and
$S_{LR}=\operatorname{SWAP}(a_{\rm WL},a_{\rm WR})$ enforces Hermiticity.}
\label{fig:circuit-architecture}
\end{figure*}

One coherent evaluation uses the following stages
(\Cref{fig:circuit-architecture}). All temporary
registers are uncomputed so that the procedure can be used together
with its inverse in amplitude estimation.
\begin{enumerate}[label=(\arabic*),leftmargin=2.2em]
\item \emph{Prepare the source.} Load the preconditioned right-hand
side $\widetilde b_s=K_0^{-1/2}b_s$, scaled by the known bound
$\widetilde b_{\max}$ from \Cref{lem:norm-chain}(i), on the
wavenumber and polarization registers.
\item \emph{Apply the reference weight.} Rotate an ancilla so that
the successful branch of mode $k$ is multiplied by $w(k)$.
\item \emph{Enforce the velocity directions.} Apply $R(k)$ to map
the transverse polarization coordinates to divergence-free velocity
components.
\item \emph{Move to position space.} Apply the inverse quantum
Fourier transform $\mathcal F_d^\dagger$. The drag acts by
multiplication in this representation.
\item \emph{Encode the drag.} Evaluate $\alpha_s(x)$ and rotate an
ancilla to represent the bounded factor
$(\alpha_s(x)-\alpha_0)/(r_{\rm be}\alpha_0)$, where
$r_{\rm be}\in(r_{\max},1)$ is the chosen block-encoding scale.
\item \emph{Return to Fourier coordinates.} Undo the transform,
frame rotation, and weight operations, using the ancilla layout of
\Cref{app:registers}. This gives the Hermitian signal block needed
by the inverse construction.
\item \emph{Apply the inverse polynomial.} Use QSVT to implement a
polynomial approximation to a scaled inverse of
$\widetilde K_s$. The degree $d_{\rm inv}$ counts how many
applications of the encoded operator are needed.
\item \emph{Encode the energy.} Apply the observable rotation so
that the probability of a designated flag is proportional to
$\|u_s\|_{L^2}^2$. The normalization is derived in
\Cref{lem:norm-chain}.
\end{enumerate}

The drag stage can be built in two ways. \emph{Construction A} uses
the Fourier expansion \eqref{eq:kl-expansion}: multiplication by
each Fourier mode becomes a wavenumber shift, and the shifts are
combined through block encodings. Its normalization is
$\alpha_0^{-1}\sum_{\ell}|c_\ell y_{s,\ell}|$. Thus this construction
needs control of the sum of the Fourier coefficients; cancellations
between modes do not reduce that normalization. \emph{Construction B}
evaluates the drag directly at the position-grid point $x$. Its
normalization is at most $r_{\max}$ and needs only the pointwise
bounds \eqref{eq:alpha-bounds}. Both normalizations are independent
of the grid size under their respective assumptions
(\Cref{lem:lcu-subnorm}).

We use two circuit models to check the implementation. \emph{Model I}
loads the required rotation angles through Gray-code multiplexors
\cite{shende2006}. It is convenient for small gate-by-gate
simulations, but its angle table can require $\Theta(MN^d)$ gates.
\emph{Model II} computes the angles reversibly from the scenario and
grid registers using $b_{\rm arith}$ bits of fixed-point precision.
It avoids the full angle table, provided the drag coefficients have
a compact coherent generator. The arithmetic analysis proves a
block-encoding error of order $2^{-b_{\rm arith}}$; the half-angle
construction controls the drag-angle error even at high contrast
(\Cref{prop:be-cost,prop:global-arith-error}). For fixed numbers
of drag terms and sufficiently compact inputs, its gate cost grows
polylogarithmically with $N$ and $M$.

The complete register layout and arithmetic circuits are given in
\Cref{app:registers,app:circuit}. The explicit gate-level frame
construction is two-dimensional. For $d\geq3$ the corresponding
gate bound assumes an efficient reversible frame-synthesis routine,
as stated in \Cref{rem:general-d}; the mathematical preconditioning
bound itself does not need that assumption.

\subsection{End-to-End Gate Complexity}
\label{sec:norm-and-conditional}

So far we have controlled the condition number and the cost of
encoding the operator. To obtain an end-to-end estimate, we must
also account for preparing the source and measuring the kinetic
energy. An unfavorable normalization could make the success
probability very small, forcing amplitude estimation to use many
more evaluations. We therefore keep track of the factors introduced
at each step.

Let $b_s=R^*\widehat f_s$ be the reduced Fourier forcing and let
$\widetilde b_s=K_0^{-1/2}b_s$ denote its \emph{preconditioned} form.
The inverse polynomial applies a scaled version of
$\widetilde K_s^{-1}$. To recover the physical kinetic energy, the
observable stage applies the contraction
\begin{equation*}
 E_s:=\sqrt{\nu+\alpha_0}\,K_0^{-1/2}.
\end{equation*}
The normalization calculation in \Cref{lem:norm-chain} gives
\begin{equation*}
 p_{{\rm good},s}
 =\frac{\mathcal E_s}{B_{\rm norm}},
 \qquad\mathcal E_s:=\|u_s\|_{L^2}^2,
\end{equation*}
where $p_{{\rm good},s}$ is the probability of the designated good
flag and $B_{\rm norm}$ is a known positive scaling factor.
Under the normalization choices of that lemma,
$B_{\rm norm}=\Theta(\widetilde\kappa^2)$.
For forcing supported on fixed low wavenumbers
$0<|k|\leq k_0$, with its norm bounded above and below by positive
constants, the good-flag probability also has a positive lower
bound independent of $N$. Thus refining the grid does not by itself
make the energy flag increasingly unlikely.

The complexity comparison requires assumptions about the inputs and
the classical method. We state them explicitly.
\begin{enumerate}[label=(A\arabic*),leftmargin=2.4em]
\item\label{as:A1} \textbf{Efficient coherent input generation.}
Let the drag expansion length and the number of nonzero forcing
coefficients obey
$L(\epsilon)=\Theta(\epsilon^{-\ell})$ and
$S(\epsilon)=\Theta(\epsilon^{-\sigma})$, where
$\ell,\sigma\geq0$. We assume that the scenario coefficients
$y_{s,\ell}$ can be computed reversibly from their indices using
$\mathcal O(\operatorname{polylog}(M,1/\epsilon))$ gates per term,
for example with the scrambled Sobol' generator studied in the
appendix. Construction B can instead use a similarly efficient
position-space evaluator for $\alpha_s(x)$. We also assume that
the $S$-sparse source can be prepared in
$\mathcal O(S(\epsilon)d\log N)$ gates. These are assumptions on
how the data are supplied. An arbitrary table of all $M$ scenarios
need not have such an efficient loading circuit.
\item\label{as:A2} \textbf{Low-wavenumber forcing.}
The forcing is supported on modes $0<|k|\leq k_0$, where $k_0$ is
fixed, and its reduced norm obeys
$0<b_{\min}\leq\|b_s\|_2\leq b_{\max}$ uniformly in $s$.
These bounds are the conditions used in \Cref{lem:norm-chain} to
control the source normalization and the energy-flag probability.
\item\label{as:A3} \textbf{Discretization accuracy and classical solver class.}
We assume that $N=\Theta(\epsilon^{-1/r})$ grid points per axis
suffice to reduce the error in the observable to
$\mathcal O(\epsilon)$, where $r>0$ is the assumed spatial
convergence exponent. This includes any truncation and aliasing
error of the chosen Fourier--Galerkin or pseudospectral scheme; the
rate is not inferred from our tests on smooth drag fields.
The classical competitor is a single-level method that explicitly
processes the $N^d$ grid unknowns, such as a grid-resolving
FFT-based conjugate-gradient or multigrid solver. Its per-scenario
cost is at least $\Omega(N^d)$ operations. For the specified
near-linear implementations, we also take the upper bound
$\widetilde{\mathcal O}(N^d)$ when a two-sided cost law is needed.
\end{enumerate}

The assumptions separate the two possible sources of quantum cost.
The number of drag and forcing terms is controlled by $L$ and $S$;
the grid size enters through the Fourier transforms and arithmetic
register widths. The preconditioner prevents the inverse polynomial
from acquiring a further power of $N$ through poor conditioning.
The next theorem combines these costs with the sampling comparison
of \Cref{thm:budget}.

\begin{theorem}[Conditional end-to-end gate complexity]
\label{thm:sobolev-advantage}
Assume \ref{as:A1}--\ref{as:A3} and fixed drag contrast
$\widetilde\kappa=\alpha_{\max}/\alpha_{\min}=\mathcal O(1)$.
\begin{enumerate}[label=(\roman*),leftmargin=2.2em]
\item \textbf{Cost of one coherent evaluation.}
For observable bias $\mathcal O(\epsilon)$, one evaluation uses
$\log_2 M+d\log_2 N+\lceil\log_2d\rceil+7$ system qubits and
$\mathcal O(d\log N+b_{\rm arith}^2)$ reusable workspace qubits,
where $b_{\rm arith}$ is the fixed-point arithmetic width. Its
logical Clifford$+T$ gate cost satisfies
\begin{equation}
\begin{aligned}
 \Gamma_Q(\epsilon)
 &=\mathcal O\!\Big(
 \widetilde\kappa(L(\epsilon)+S(\epsilon))d\\
 &\qquad\cdot\operatorname{polylog}
 \!\left(\frac{d\widetilde\kappa M}{\epsilon}\right)\Big)\\
 &=\widetilde{\mathcal O}\!\left(
 \epsilon^{-\max\{\ell,\sigma\}}\right).
\end{aligned}
\label{eq:gamma-q-sobolev}
\end{equation}
Thus the quantum cost exponent satisfies
$g_Q\leq\max\{\ell,\sigma\}$. The explicit frame circuit and
gate count are established for $d=2$; for $d\geq3$, the gate
count assumes the frame-synthesis hypothesis in
\Cref{rem:general-d}. Under the normalization choices of
\Cref{lem:norm-chain}, converting the flag probability into the
physical energy introduces a factor
$B_{\rm norm}=\Theta(\widetilde\kappa^2)$ in the required
amplitude-estimation precision.
\item \textbf{Sufficient condition for advantage.}
Suppose the classical solver of \ref{as:A3} is combined with a
sampling rule of order $p$ and the ensemble has
$M=\Theta(\epsilon^{-m})$ scenarios. If
\begin{equation}
\boxed{
 \max\{\ell,\sigma\}-\frac dr
 < B(p,m)=\min\{m,1/p\}-1
}
\label{eq:sobolev-frontier-exact}
\end{equation}
then the quantum-to-classical cost ratio satisfies
\begin{equation*}
 \frac{T_Q}{T_C}
 =\widetilde{\mathcal O}\!\left(
 \epsilon^{B(p,m)+d/r-\max\{\ell,\sigma\}}\right)
 \longrightarrow0.
\end{equation*}
This is a sufficient condition from the quantum upper bound and the
classical grid-processing lower bound. It is also necessary only
if the corresponding costs obey matching two-sided asymptotic laws.
In the special case $\ell=\sigma=0$, the bound becomes
\begin{equation*}
 \frac{T_Q}{T_C}
 =\widetilde{\mathcal O}\!\left(
 \epsilon^{d/r+\min\{m,1/p\}-1}\right).
\end{equation*}
For $m\geq1$ and $p\leq1$, this gives an advantage exponent of
at least $d/r$. For $m<1$, the exponent is $d/r+m-1$, which is
positive when $d/r>1-m$.
\end{enumerate}
\end{theorem}

\paragraph{Proof sketch.}
The preconditioning proposition bounds
$\kappa(\widetilde K_s)$ independently of the grid, and
\Cref{lem:lcu-subnorm} bounds the block-encoding normalization.
Consequently, the QSVT inverse needs a polynomial of degree
$\mathcal O(\widetilde\kappa
\log(\widetilde\kappa/\epsilon))$, rather than a degree
that grows polynomially with $N$. The uniform polynomial
approximation is proved in \Cref{lem:qsvt-degree}. Each call to
the encoded operator is implemented with the reversible circuits
analyzed in \Cref{prop:be-cost}. The energy normalization is
controlled by \Cref{lem:norm-chain}. Combining these bounds gives
\eqref{eq:gamma-q-sobolev}. Finally, a grid-resolving classical
solve costs at least $\Omega(N^d)=\Omega(\epsilon^{-d/r})$.
Substituting these quantum and classical cost bounds into
\Cref{thm:budget} proves part (ii). The detailed circuit and error
accounting is given in \Cref{app:circuit}.

\begin{remark}[Scope of the gate-complexity comparison]
\label{rem:classical-scope}
The theorem compares specified quantum and classical procedures.
It does not give a lower bound against every classical algorithm.
Reduced-basis methods, surrogate models, or homogenization may be
cheaper for material families with additional structure
(\Cref{rem:reduced-basis}). In particular, the analytic three-mode
drag used in the numerical conditioning test can be solved by
classical Fourier methods with much faster spatial convergence than
the finite-regularity law assumed in \ref{as:A3}; it is a test of
the preconditioner, not evidence for the asymptotic speedup.
\end{remark}

\section{Numerical Results}
\label{sec:experiment}

We now test whether the main parts of the construction behave as
predicted at finite sizes. There are three distinct questions:
whether preconditioning controls the spectrum, whether the circuit
components produce the correct kinetic-energy observable, and how
the circuit resources change with grid and ensemble size. We also
check the operator and mean-flow constraint in three dimensions.

These tests provide different kinds of evidence. Some use classical
matrix computations or PDE solves. Model I is simulated gate by
gate as a quantum statevector on classical hardware. Model II is
examined using emitted circuits, arithmetic subcircuits, and
truth-table checks, rather than a full large statevector simulation.
The cost curves obtained by substituting exponents into our
complexity formulas are analytic illustrations, not measured
runtimes. No experiments were run on quantum hardware.
\Cref{tab:validation-map} records the scope of each test; full
numerical tables are provided in \Cref{app:numerics}.

\begin{table*}[tp]
\centering
\caption{Numerical and circuit checks. ``Numerics'' denotes classical
matrix or PDE calculations; ``simulated'' denotes gate-by-gate
statevector simulation; and ``audited'' denotes circuit and
arithmetic checks without full statevector execution.}
\label{tab:validation-map}
\small
\begin{tblr}{width=\textwidth, colspec={Q[wd=2.7cm,l] Q[wd=1.5cm,l] X[l] X[l] X[l]}}
\toprule
Claim tested & Kind & Test & Range & Not tested \\
\midrule
\Cref{prop:riesz,prop:reduced}, \Cref{lem:bias} & Numerics &
Dense linear algebra; 200 random contraction chains &
$\dim\cH\leq16$ &
Large-scale behavior \\
\Cref{thm:optimal-stokes} & Numerics &
Spectral Stokes solves for 257 hidden-bit scenarios &
$N=64$ &
Quantum circuit for this family \\
QAE scaling & Numerics &
Grover law on the reduced statevector; MLQAE versus sampling, 180 trials &
$N=4$, $M=16$ &
Advantage at these sizes \\
\Cref{prop:sobolev-riesz} & Numerics &
Spectra of the preconditioned operator &
$d=2$, $N\leq32$; $d=3$, $N_3\leq16$ &
The limit $N\to\infty$ \\
Circuit correctness (Model~I) & Simulated &
Gate-by-gate statevector simulation &
16 and 18 qubits &
Larger grids \\
Reversible arithmetic (Model~II) & Audited &
Subcircuit extraction, truth tables, ancilla checks &
$N\leq1024$, $M\leq2^{20}$ &
Fault-tolerant execution \\
Mean-flow multiplier & Numerics &
Brinkman solves with and without $\lambda_s$ &
$d=3$, $N_3=32$ &
Rough drag fields \\
\bottomrule
\end{tblr}
\end{table*}

\subsection{Conditioning Under Grid Refinement}
\label{sec:exp-brinkman-sv}

We first check the main theoretical bound
\eqref{eq:mesh-independent-kappa}. The test uses a two-dimensional
periodic Brinkman operator with viscosity $\nu=0.1$ and drag bounds
$\alpha_{\min}=1$, $\alpha_{\max}=3$. Hence the theorem predicts
$\kappa(\widetilde K_s)\leq3$ for every grid, although the actual
condition number may be smaller. We construct the reduced operator
$K_s$ and preconditioned operator $\widetilde K_s$ on grids with
$N=8,12,16,24,32$ points per axis.

We also check that the map $R_{\mathcal E}$ has the properties
required by the proof. The divergence residual
$\|DR_{\mathcal E}\|_2$ measures the violation of the discrete
constraint, while the energy-isometry defect measures how closely
$R_{\mathcal E}^*A_0R_{\mathcal E}$ agrees with the identity.
Both quantities stay below $7\times10^{-16}$ in the tests, consistent
with floating-point roundoff.

\begin{figure*}[!tp]
\centering
\includegraphics[width=0.95\linewidth]{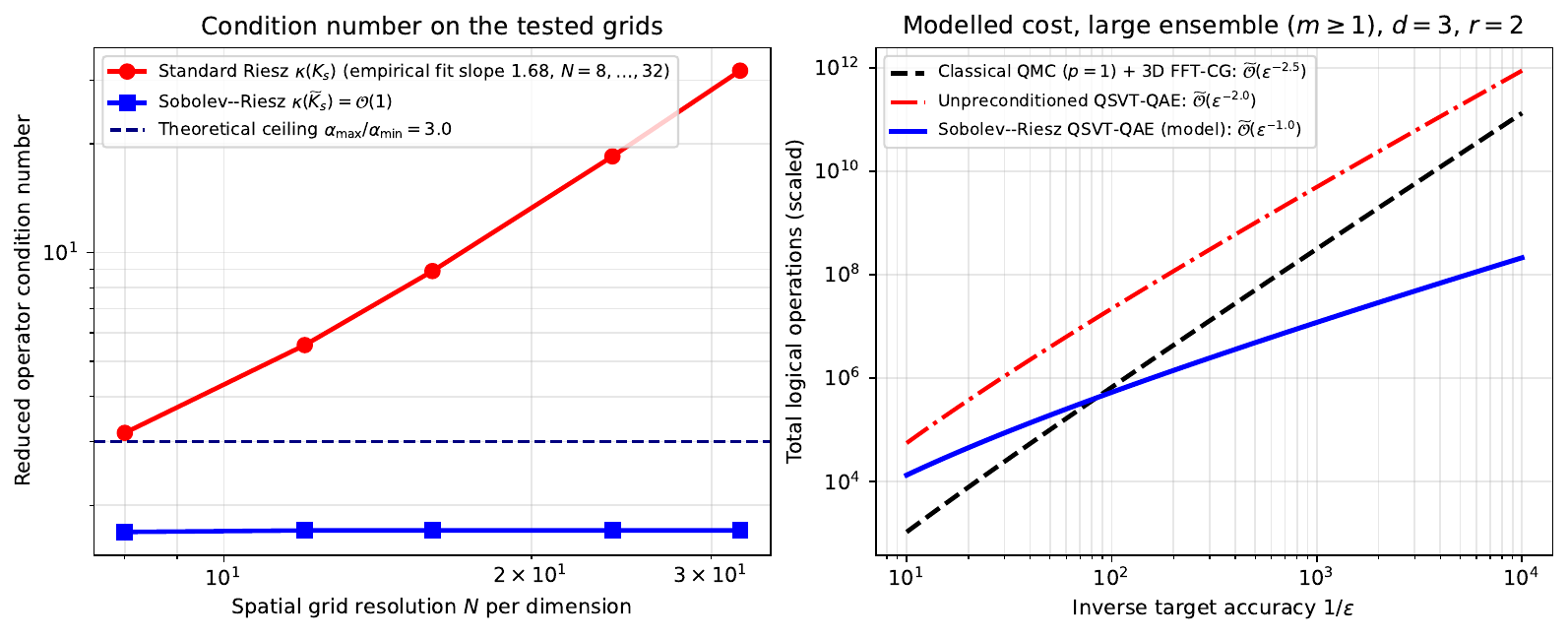}
\caption{Left: the unpreconditioned condition number grows over the
tested grids, whereas the preconditioned value remains near $1.70$.
The slope $1.68$ is a least-squares fit over $N=8,\ldots,32$, not a
proof of the asymptotic rate; the $\Theta(N^2)$ behavior is an
independent analytic property of the unpreconditioned operator.
Right: modelled total-cost exponents for 3D Brinkman flow with
$d=3$, $r=2$, and first-order classical QMC ($p=1$). The curves
use chosen prefactors $1$, $50$, and $120$ for the classical,
unpreconditioned quantum, and preconditioned quantum procedures,
respectively. They illustrate the cost laws and are not measured
runtime comparisons. In this model, preconditioning improves the
quantum exponent even though the unpreconditioned method already has
a better exponent than the specified classical solver.}
\label{fig:sobolev-advantage}
\end{figure*}

\begin{figure*}[!tp]
\centering
\includegraphics[width=0.95\linewidth]{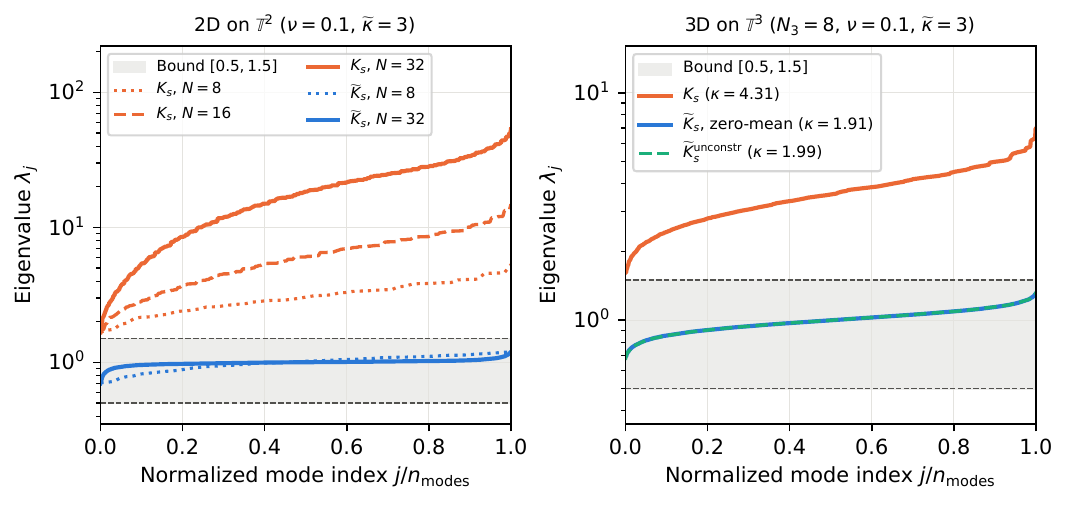}
\caption{Sorted eigenvalues of the reduced Brinkman operator $K_s$
and the Sobolev--Riesz preconditioned operator $\widetilde K_s$
($\nu=0.1$, $\alpha_{\min}=1$, $\alpha_{\max}=3$, $\alpha_0=2$).
The shaded band marks the theoretical spectral interval
$[\alpha_{\min}/\alpha_0,\alpha_{\max}/\alpha_0]=[0.5,1.5]$ of
\Cref{prop:sobolev-riesz}.
Left: two-dimensional operators on $\mathbb T^2$ for $N=8,16,32$.
Right: three-dimensional operators on $\mathbb T^3$ at $N_3=8$,
comparing $K_s$ ($1{,}022$ mean-zero modes, $\kappa=4.31$) with
$\widetilde K_s$ ($\kappa=1.91$) and the unconstrained-mean variant
$\widetilde K_s^{\rm unconstr}$ ($1{,}025$ modes, $\kappa=1.99$;
\Cref{rem:unconstrained}).}
\label{fig:spectral-clustering}
\end{figure*}

The condition numbers show the effect of preconditioning clearly.
The unpreconditioned value grows from $3.17$ at $N=8$ to $31.86$
at $N=32$. By contrast, the preconditioned value stays near $1.70$
for the same grids, well below the bound of $3$. A degree-14 QSVT
inverse polynomial also gives errors of order $10^{-10}$ for the
reported observable calculation. That number measures the inverse
approximation, not the error of a complete uncertainty-quantification
run. The results are listed in \Cref{tab:sobolev-brinkman} and shown
in the left panel of \Cref{fig:sobolev-advantage}; the eigenvalue
spectra are plotted in \Cref{fig:spectral-clustering}.

\begin{table*}[tp]
\centering
\caption{Conditioning and inverse-approximation checks for the
Sobolev--Riesz formulation on $\mathbb T^2$ with $\nu=0.1$ and
drag contrast $\widetilde\kappa=3$. The last column reports the
inverse-polynomial error, not the error of a full QAE estimate.}
\label{tab:sobolev-brinkman}
\begin{tabular}{rrrrrrr}
\toprule
$N$ & active modes & $\|DR_{\mathcal E}\|_2$ & isometry gap
& $\kappa(K_s)$ & $\kappa(\widetilde K_s)$ & QSVT ($d=14$) error \\
\midrule
8  & 63   & $1.11\times10^{-16}$ & $3.33\times10^{-16}$ & 3.17  & 1.687 & $5.90\times10^{-11}$ \\
12 & 143  & $2.22\times10^{-16}$ & $4.44\times10^{-16}$ & 5.55  & 1.704 & $1.20\times10^{-10}$ \\
16 & 255  & $2.22\times10^{-16}$ & $6.66\times10^{-16}$ & 8.89  & 1.704 & $1.01\times10^{-10}$ \\
24 & 575  & $4.44\times10^{-16}$ & $6.66\times10^{-16}$ & 18.46 & 1.704 & $1.01\times10^{-10}$ \\
32 & 1023 & $4.44\times10^{-16}$ & $6.66\times10^{-16}$ & 31.86 & 1.704 & $1.01\times10^{-10}$ \\
\bottomrule
\end{tabular}
\end{table*}

\subsection{Accuracy of the Circuit Construction}
\label{sec:exp-stokes}

Next we check whether the circuit computes the same energy as the
classical PDE solution. We begin with small dense-matrix tests of the
Riesz reduction, reduced-system equivalence, and error bound in
\Cref{prop:riesz,prop:reduced,lem:bias}. These agree with the
corresponding algebraic identities to numerical precision.

For a more direct circuit test, we simulate Model I gate by gate with
$N=8$ and $M=16$. The circuit uses 18 qubits and $48{,}737$ gates.
The computed ensemble kinetic energy is approximately $0.341$ and
differs from the classical PDE result by about
$1.4\times10^{-8}$. Its Grover iterations also reproduce the
expected probability law $\sin^2((2j+1)\theta)$, where
$\sin^2\theta$ is the good-flag probability and $j$ counts Grover
iterations. The maximum reported deviation is about
$1.3\times10^{-12}$. These checks test both the energy encoding
and the amplitude-estimation state preparation at this small size.

\begin{figure*}[!tp]
\centering
\includegraphics[width=0.95\linewidth]{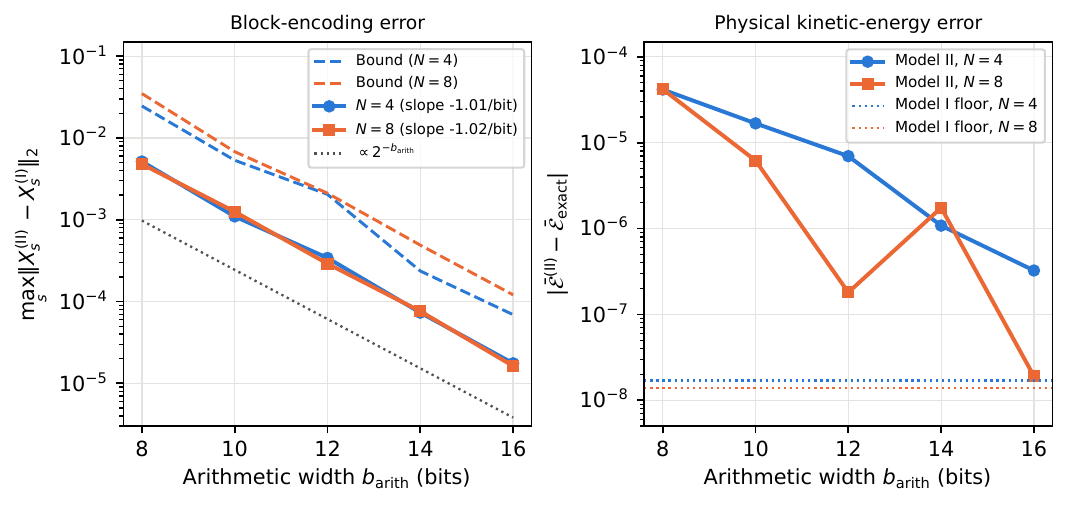}
\caption{Finite-precision convergence of the emitted Model~II circuit
with arithmetic width $b_{\rm arith}\in\{8,10,12,14,16\}$ for $N=4$
and $N=8$ ($M=16$, $\nu=0.1$, $\alpha_0=2$; \Cref{tab:model2-accuracy}).
Left: worst-case block-encoding discrepancy
$\max_s\|X_s^{({\rm II})}-X_s^{({\rm I})}\|_2$ together with the
analytic bound \eqref{eq:be-telescoping-bound} (dashed) and a
$2^{-b_{\rm arith}}$ reference slope.
Right: physical ensemble kinetic-energy error relative to the exact
classical PDE solution; dotted horizontal lines indicate the Model~I
degree-$14$ QSVT floor ($1.69\times10^{-8}$ at $N=4$ and
$1.39\times10^{-8}$ at $N=8$).}
\label{fig:model2-precision}
\end{figure*}

Model II replaces the stored angles with reversible fixed-point
arithmetic. At arithmetic width $b_{\rm arith}=16$, its encoded
energy agrees with the reference to about $1.9\times10^{-8}$,
consistent with the expected finite-precision error. This is not a
full statevector run of the large Model II circuit. That circuit
would contain approximately $3.5\times10^7$ gates on 416 qubits,
including arithmetic workspace. Instead, we validate its netlist
components through subcircuit extraction, truth-table checks of the
arithmetic units at the tested widths, ancilla cleanup, and amplitude
comparisons against Model I. The convergence with $b_{\rm arith}$ is
shown in \Cref{fig:model2-precision}, and the full accuracy and
resource tables are given in \Cref{app:numerics}.

\begin{figure*}[!tp]
\centering
\includegraphics[width=0.95\linewidth]{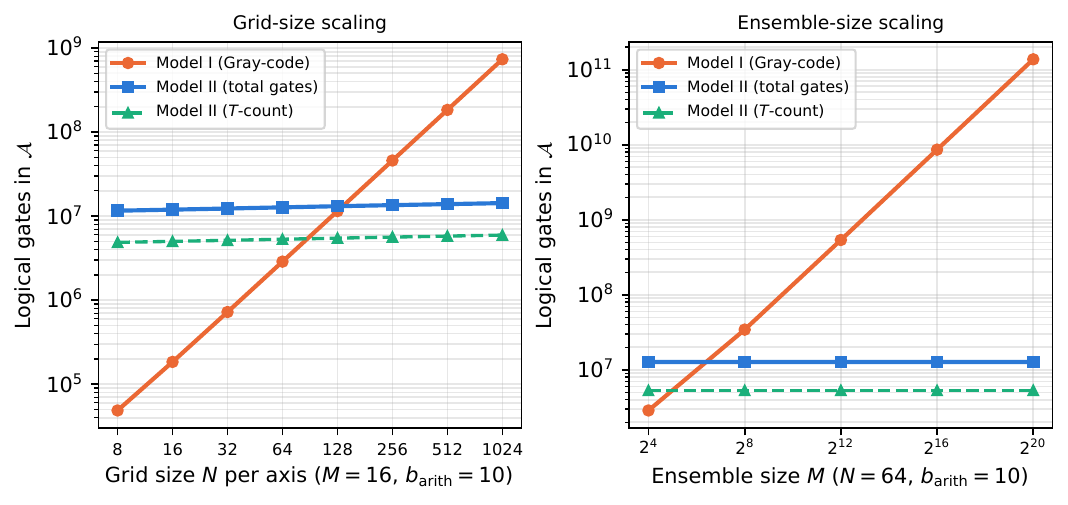}
\caption{Logical gate counts of the state-preparation circuit $\cA$
comparing Model~I (Gray-code angle tables) with Model~II (reversible
arithmetic, total gates and $T$-count) at $b_{\rm arith}=10$
(\Cref{tab:polylog-gate-audit-b,tab:scenario-m-scaling-b}).
Left: scaling with grid resolution $N=8,\ldots,1024$ at $M=16$;
Model~II falls below Model~I at $N=256$ and uses $51\times$ fewer
gates at $N=1024$.
Right: scaling with ensemble size $M=2^4,\ldots,2^{20}$ at $N=64$;
Model~II total gates increase by $0.11\%$ while its $T$-count stays
constant at $5.31\times10^6$.
Both panels compare two quantum compilations, not a classical solver.}
\label{fig:circuit-resource-scaling}
\end{figure*}

\subsection{Circuit Resources and Sampling Error}
\label{sec:exp-model2}

We next examine how the compiled circuit size changes when the grid
or ensemble grows. Model I must load a table of rotation angles,
whereas Model II evaluates the angles from the input registers.
Across the tested grids $N=8$ to $N=1024$, the audited Model II
gate count follows a quadratic trend in $\log_2N$. At $N=1024$,
its gate count is 51 times smaller than the corresponding Model I
count. When the ensemble size increases from $M=16$ to $M=2^{20}$,
the Model II gate count changes by only $0.11\%$, and the reported
$T$-count is unchanged (\Cref{fig:circuit-resource-scaling}). These
comparisons are between two quantum implementations of the same
operation; they are not comparisons with a classical solver.

\begin{figure}[!tb]
\centering
\includegraphics[width=0.92\linewidth]{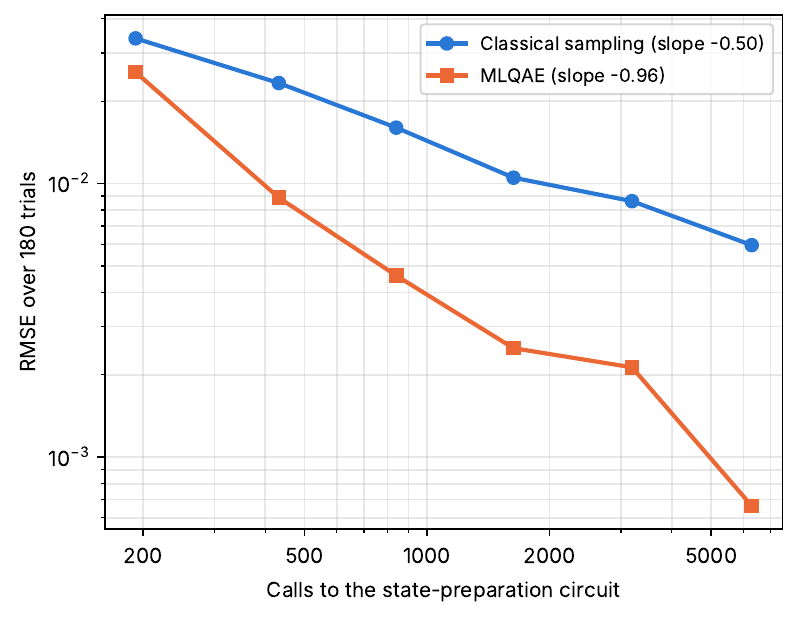}
\caption{Root-mean-square error of maximum-likelihood amplitude
estimation (MLQAE) and ordinary sampling at equal oracle-call
budgets. The model good-flag probability is $a=0.3206$; each point
is estimated from simulated measurement statistics over 180 trials
(\Cref{tab:scaling}). The figure compares sampling error, not
gate cost or runtime.}
\label{fig:scaling}
\end{figure}

We separately examine sampling error. Using simulated measurement
statistics at matched oracle-call budgets over 180 trials,
maximum-likelihood amplitude estimation (MLQAE) has an empirical
root-mean-square-error slope of approximately $-0.96$, compared
with $-0.50$ for ordinary sampling. The slope is measured against
oracle calls, not logical gates or elapsed time. A single Model I
call at $N=8$ already contains $48{,}737$ gates, so this experiment
does not demonstrate a practical runtime advantage.
\Cref{fig:scaling} shows the sampling comparison.

\subsection{Three-Dimensional Operator Checks}
\label{sec:exp-3d}

Finally, we repeat the main operator-level checks on the periodic
three-dimensional domain $\mathbb T^3$. For the $N_3=8$ test, the
constructed mean-zero divergence-free operator has a constraint
residual below $10^{-15}$. Preconditioning reduces its condition
number from about $4.3$ to $1.9$, again below the theoretical bound
$\widetilde\kappa=3$ (\Cref{fig:spectral-clustering}, right). These
are matrix-level checks, not a compiled three-dimensional frame
circuit.

We also test why the mean-flow multiplier is needed. With the
spatially varying drag $\alpha(x)=\alpha_0+0.1\sin x_1$, the
corresponding problem without the multiplier develops a nonzero
mean velocity of about $-0.035$. Enforcing zero mean with the
multiplier reduces the mean velocity to numerical roundoff; the
reported momentum residual decreases from $0.071$ to
$1.3\times10^{-15}$. This agrees with the projected Brinkman
formulation of \eqref{eq:brinkman-pde}. The detailed values appear
in \Cref{app:numerics}.

The finite-size tests agree with the conditioning bound and the
energy-encoding identities. The asymptotic gate comparison follows
under the assumptions of \Cref{thm:sobolev-advantage}.

\section{Conclusion}
\label{sec:conclusion}

For steady Brinkman flow with spatially varying drag, the polar
Riesz map gives a positive-definite system in divergence-free
coordinates. The Sobolev--Riesz preconditioner bounds its condition
number by $\alpha_{\max}/\alpha_{\min}$ independently of grid
resolution. A coherent circuit encodes the mean kinetic energy
without reconstructing the velocity field, and finite-size tests
check the conditioning bound and the implemented circuit components.

The resulting gate-complexity advantage is conditional on efficient
coherent inputs, the forcing and discretization assumptions, and
comparison with single-level classical grid-resolving solvers.
The simulations test individual components at finite sizes on
classical hardware.

Extensions to nonperiodic boundaries and weakly nonlinear flow
would broaden the range of problems covered. More efficient arithmetic
and state preparation, together with comparisons against classical
reduced-order solvers, would help assess the overall cost for
structured drag families.

\FloatBarrier

\onecolumn
\appendix

\section{Appendix Guide and Notation}
\label{app:notation}

The main text introduces each result before stating its mathematical
form. Here we provide the proof details and implementation information
needed to verify those statements. The appendices follow the same order
as the main argument: constraint reduction, continuum Brinkman model,
circuit construction, and numerical verification. \Cref{tab:appendix-guide} identifies the role of each appendix.

\begin{table}[!ht]
\centering
\caption{Contents of the appendices.}
\label{tab:appendix-guide}
\small
\begin{tblr}{width=\textwidth, colspec={Q[wd=1.4cm,l] X[l] X[l]}}
\toprule
Appendix & Contents & Supports \\
\midrule
\ref{app:proofs-early} & Proofs for constraint elimination, the coherent evaluator and the Stokes benchmark & \Cref{sec:reduction,sec:access,sec:stokes-benchmark} \\
\ref{app:matched-problem} & Well-posedness of the Brinkman problem; two forms of the Sobolev--Riesz map & \Cref{sec:sobolev-riesz} \\
\ref{app:circuit} & Block encoding, reversible arithmetic, normalization chain, gate counts & \Cref{thm:sobolev-advantage} \\
\ref{app:numerics} & Additional numerical tables & \Cref{sec:experiment} \\
\bottomrule
\end{tblr}
\end{table}

\Cref{tab:notation-physics,tab:notation-circuit} list
the symbols used in the paper; circuit quantities are listed separately.

\begin{table}[!ht]
\centering
\caption{Physical, numerical and complexity notation.}
\label{tab:notation-physics}
\small
\begin{tblr}{width=\textwidth, colspec={Q[wd=3.6cm,l] X[l]}}
\toprule
Symbol & Meaning \\
\midrule
$d,\ \mathbb T^d$ & Dimension ($d\geq 2$); periodic cell $[0,2\pi)^d$ \\
$N=2^n,\ h\asymp N^{-1}$ & Grid points per axis; mesh width \\
$M=2^{n_M},\ s$ & Number of scenarios; scenario index \\
$u_s,\ p_s,\ \lambda_s$ & Velocity; pressure; constant mean-flow multiplier \\
$\nu,\ \alpha_s(x)$ & Kinematic viscosity; porous drag field \\
$\alpha_{\min},\ \alpha_{\max},\ \alpha_0$ & Drag bounds; midpoint $\tfrac12(\alpha_{\min}+\alpha_{\max})$ \\
$\widetilde\kappa,\ r_{\max}$ & Contrast $\alpha_{\max}/\alpha_{\min}$; $r_{\max}=(\widetilde\kappa-1)/(\widetilde\kappa+1)<1$ \\
$D,\ C,\ \Ker D$ & Discrete divergence; generator with $\Ran C=\Ker D$; feasible velocities \\
$R,\ R_{\mathcal E}$ & Polar Riesz map (\Cref{sec:reduction}); Sobolev--Riesz map (\Cref{sec:sobolev-riesz}) \\
$A_s,\ A_0,\ K_s,\ \widetilde K_s$ & Brinkman operator; reference $-\nu\Delta+\alpha_0I$; reduced $R^*A_sR$; preconditioned $R_{\mathcal E}^*A_sR_{\mathcal E}$ \\
$r=r_0+\beta$ & Spatial H\"older regularity of the drag \\
\midrule
$g_s,\ \mu$ & Normalized scenario output in $[0,1]$; its ensemble mean \\
$\mathcal E_s,\ \bar{\mathcal E}$ & Physical kinetic energy $\|u_s\|_{L^2}^2$; its ensemble mean \\
$\epsilon,\ \delta$ & Target additive error; failure probability \\
$p,\ m$ & Classical convergence order; ensemble growth $M=\Theta(\epsilon^{-m})$ \\
$g_Q,\ g_C,\ B(p,m)$ & Per-sample cost exponents; sampling budget $\min\{m,1/p\}-1$ \\
$\ell,\ \sigma$ & Growth exponents of the drag expansion length and the forcing sparsity \\
$Q_C,\ Q_Q$ & Classical and quantum query counts \\
$\Gamma_Q,\ G_C$ & Quantum and classical cost of one scenario evaluation \\
\bottomrule
\end{tblr}
\end{table}

\begin{table}[!ht]
\centering
\caption{Quantum-circuit notation.}
\label{tab:notation-circuit}
\small
\begin{tblr}{width=\textwidth, colspec={Q[wd=3.6cm,l] X[l]}}
\toprule
Symbol & Meaning \\
\midrule
$r_{\rm be},\ \gamma_\alpha$ & Block-encoding scale in $(r_{\max},1)$; high-contrast choice $\min\{\frac43r_{\max},\frac{1+r_{\max}}{2}\}$ \\
$\gamma_{\rm QSVT},\ B_{\rm norm}$ & QSVT polynomial scale; factor converting flag probability to kinetic energy \\
$d_{\rm inv}$ & QSVT polynomial degree \\
$b_{\rm arith},\ b_{\rm in}$ & Reversible fixed-point width; bit width of a fixed-point drag value \\
$O_{\rm coeff}$ & Scrambled Sobol' coefficient circuit \\
$\cA,\ \mathcal G$ & Full state-preparation circuit; Grover iterate \\
$M_{\rm AE},\ n_\delta$ & Grover steps per phase-estimation run; number of runs for the median \\
\bottomrule
\end{tblr}
\end{table}

\section{Proofs for the Constraint Reduction and Stokes Benchmark}
\label{app:proofs-early}

This appendix completes three preliminary arguments used throughout
the paper. We first show that solving in divergence-free coordinates is
equivalent to solving the constrained system. We then control the bias
introduced by approximate circuit stages, and finally justify the
finite-list Stokes query benchmark.

\begin{proof}[Proof of \Cref{prop:reduced}]
\emph{Step 1 (the reduced operator is positive definite).}
For any nonzero $\xi\in\cZ_0$, \Cref{prop:riesz} gives $R\xi\in\Ker D$ and
$\|R\xi\|_2=\|\xi\|_2>0$, so coercivity of $A_s$ on $\Ker D$ implies
$\langle\xi,K_s\xi\rangle=\langle R\xi,A_sR\xi\rangle\geq\gamma_s\|R\xi\|_2^2=\gamma_s\|\xi\|_2^2>0$.
Thus $K_s$ is Hermitian positive definite on $\cZ_0$ and invertible.

\emph{Step 2 (equivalence with the weak form).}
Since $R|_{\cZ_0}:\cZ_0\to\Ker D$ is unitary, every $v\in\Ker D$ has the unique form
$v=R\eta$ with $\eta\in\cZ_0$, and \eqref{eq:weak} is equivalent to
$\langle\eta,K_s\xi_s\rangle=\langle\eta,R^*Rb_s\rangle=\langle\eta,b_s\rangle$
for all $\eta\in\cZ_0$, whose unique solution is $\xi_s=K_s^{-1}b_s$,
$u_s=R\xi_s$.

\emph{Step 3 (reconstruction of the multiplier).}
The residual $r_s := Rb_s - A_su_s\in\cH$ satisfies
$\langle v,r_s\rangle=0$ for all $v\in\Ker D$, so $r_s\in(\Ker D)^\perp=\Ran D^*$.
When $D$ has full row rank, $D^*$ is injective (and $DD^*\succ 0$), so there is
a unique $p_s=(DD^*)^{-1}Dr_s\in\cY$ satisfying $D^*p_s=r_s$, which together
with $Du_s=DR\xi_s=0$ yields \eqref{eq:saddle}. Conversely, any solution of
\eqref{eq:saddle} satisfies $u_s\in\Ker D$ and $\langle v,A_su_s-Rb_s\rangle=-\langle Dv,p_s\rangle=0$
for all $v\in\Ker D$, which gives uniqueness.
\end{proof}

\begin{proof}[Proof of \Cref{lem:bias}]
Set $y_s := E_s R F_s b_s$ and $\widetilde y_s := \widetilde E_s\widetilde R\widetilde F_s\widetilde b_s$.
Because each ideal and approximate stage has spectral norm at most $1$, we have
$\|y_s\|_2\leq 1$, and telescoping the difference in four steps yields
\begin{align*}
 \|\widetilde y_s - y_s\|_2
 &\leq
 \|(\widetilde E_s-E_s)\widetilde R\widetilde F_s\widetilde b_s\|_2
 + \|E_s(\widetilde R-R)\widetilde F_s\widetilde b_s\|_2 \\
 &\quad+ \|E_s R(\widetilde F_s-F_s)\widetilde b_s\|_2
 + \|E_s R F_s(\widetilde b_s-b_s)\|_2
 \leq
 \eta_E+\eta_R+\eta_F+\eta_b
 =
 \eta_{\rm impl}.
\end{align*}
Writing $\widetilde y_s = y_s + \Delta_s$ with $\|\Delta_s\|_2\leq\eta_{\rm impl}$
and applying the Cauchy--Schwarz and triangle inequalities gives
\[
 |\widetilde g_s - g_s|
 =
 \big|\|\widetilde y_s\|_2^2 - \|y_s\|_2^2\big|
 =
 \big|2\operatorname{Re}\langle y_s,\Delta_s\rangle + \|\Delta_s\|_2^2\big|
 \leq
 2\|y_s\|_2\|\Delta_s\|_2 + \|\Delta_s\|_2^2
 \leq
 2\eta_{\rm impl} + \eta_{\rm impl}^2.
\]
Averaging over $s\in\{0,\dots,M-1\}$ gives the same bound for
$|\widetilde\mu_{\rm cir}-\mu|$.
\end{proof}

\begin{proof}[Proof of \Cref{thm:quantum}]
For $\eta_{\rm impl}\leq\epsilon/8$ and $\epsilon\leq 1$, \Cref{lem:bias} bounds
the deterministic circuit bias by
\[
 |\widetilde\mu_{\rm cir}-\mu|
 \leq
 2\left(\frac{\epsilon}{8}\right) + \left(\frac{\epsilon}{8}\right)^2
 =
 \frac{\epsilon}{4} + \frac{\epsilon^2}{64}
 <
 \frac{\epsilon}{2}.
\]
A single run of Brassard--H\o yer--Mosca--Tapp amplitude
estimation~\cite[Theorem~12]{brassard2002} using $M_{\rm AE}=\Theta(\epsilon^{-1})$
applications of the Grover iterate $\mathcal G$ estimates the circuit probability
$\widetilde\mu_{\rm cir}\in[0,1]$ to additive error at most $\epsilon/2$ with
success probability at least $8/\pi^2>0.81$.  Taking the median of
$n_\delta = \Theta(\log(1/\delta))$ independent runs boosts the success
probability to at least $1-\delta$ via Hoeffding's inequality while multiplying
the total number of calls to $\cA$ and $\cA^\dagger$ by $n_\delta$.  On this
event, the triangle inequality gives
$|\widetilde\mu-\mu|\leq|\widetilde\mu-\widetilde\mu_{\rm cir}|+|\widetilde\mu_{\rm cir}-\mu|<\epsilon/2+\epsilon/2=\epsilon$,
and multiplying the $\mathcal O(\epsilon^{-1}\log(1/\delta))$ coherent calls by
the per-call gate cost $\Gamma(\kappa,\beta,\epsilon/8)$ yields \eqref{eq:qcost}.
\end{proof}

The Stokes benchmark rests on the following finite-list lower bound, which is
standard; we include a short proof with explicit constants.

\begin{lemma}[Classical lower bound for a finite list]
\label{thm:classical}
Let $M\geq 16$ and $0<\epsilon\leq 1/16$. Any randomized classical algorithm that
makes adaptive queries $s\mapsto g_s$ to an input $g\in[0,1]^M$ and, for every
input, outputs an estimate of $\mu=\frac1M\sum_{s=0}^{M-1}g_s$ within $\pm\epsilon$
with probability at least $2/3$, must make at least
$Q_C\geq\frac{1}{1000}\min\{M,\epsilon^{-2}\}$ queries on some input. The
constant is not optimized. The matching upper bound
$\mathcal O(\min\{M,\epsilon^{-2}\})$ follows from reading every entry when
$M\leq\epsilon^{-2}$ and from $\mathcal O(\epsilon^{-2})$ uniform samples otherwise.
\end{lemma}

\begin{proof}
We restrict to binary inputs $g\in\{0,1\}^M$. The proof has three steps: reduce
estimation to distinguishing two hidden lists, show that adaptive queries see
exactly what random sampling without replacement would see, and bound how far
apart the two resulting sampling laws are. The two hypotheses can be
distinguished only if the observed query transcripts differ sufficiently. We
therefore bound the total-variation distance between the transcript laws, which
limits the success probability of every test, and show that it remains below
$1/3$ when the number of queries is below the claimed threshold.

\emph{Step 1 (two hypotheses).} Put $a_0=\lfloor M/2\rfloor$,
$t=\lceil 2\epsilon M\rceil+1$ and $a_1=a_0+t$. Under hypothesis $H_b$, the list
$g$ is uniform over the lists with exactly $a_b$ ones. Since
$(a_1-a_0)/M>2\epsilon$, an $\epsilon$-accurate estimate falls on the correct side
of the midpoint $(a_0+a_1)/(2M)$, so an estimator that succeeds with probability
$2/3$ on every input decides between $H_0$ and $H_1$ with probability at least
$2/3$ under each hypothesis. By Yao's principle we may fix the algorithm's random
bits, so it becomes a deterministic adaptive decision tree; it can then succeed
only if the laws $L_0,L_1$ of its query transcripts satisfy
$\mathrm{TV}(L_0,L_1)\geq 1/3$. Repeated queries give no new information, so we
may assume that no position is queried twice.

\emph{Step 2 (adaptivity does not help).} Under $H_b$ the list is a uniformly
random arrangement of $a_b$ ones and $M-a_b$ zeros. Given any history of answers,
the unqueried positions still hold a uniformly random arrangement of the
remaining entries, so the next answer equals $1$ with probability (remaining
ones)/(remaining positions), whichever position is chosen. By induction, the
answers of any adaptive strategy with $q$ queries have the law of $q$ draws
without replacement from an urn with $a_b$ ones among $M$ balls. The number of
ones is a sufficient statistic, so
$\mathrm{TV}(L_0,L_1)=\mathrm{TV}\big(\mathrm{Hyp}(M,a_0,q),\mathrm{Hyp}(M,a_1,q)\big)$.

\emph{Step 3 (the two laws are close).} Let $p_b=a_b/M$ and $\Delta=p_1-p_0=t/M$.
Throughout the proof we use the total-variation convention
$\mathrm{TV}(P,Q)=\sup_A|P(A)-Q(A)|$. For an urn with two colours, the
sampling-without-replacement bound of Diaconis and
Freedman~\cite[Theorem~4]{diaconis1980} gives
\[
 \mathrm{TV}\big(\mathrm{Hyp}(M,a_b,q),\mathrm{Bin}(q,p_b)\big)\leq\frac{2q}{M}.
\]
We use the weaker bound $4q/M$ below. The triangle inequality and Pinsker's
inequality then give
\[
 \mathrm{TV}(L_0,L_1)\ \leq\ \frac{8q}{M}+\sqrt{\tfrac12\,q\,\mathrm{kl}(p_0\Vert p_1)},
 \qquad
 \mathrm{kl}(p_0\Vert p_1)\leq\frac{\Delta^2}{p_1(1-p_1)}\leq\frac{16}{3}\,\Delta^2,
\]
where $\mathrm{kl}$ is the Bernoulli relative entropy and the last step uses
$p_1\in[1/4,3/4]$, which holds because $p_1\leq\tfrac12+2\epsilon+2/M$.

If $M\leq\epsilon^{-2}$, then $\epsilon\leq M^{-1/2}$ and
$\Delta\leq 2\epsilon+2/M\leq 4M^{-1/2}$. For $q\leq M/1000$ this gives
$\mathrm{TV}(L_0,L_1)\leq 0.008+\sqrt{\tfrac12\cdot\tfrac{16}{3}\cdot\tfrac{16q}{M}}\leq 0.008+0.207<\tfrac13$.

If $M>\epsilon^{-2}$, which is the intermediate regime of the classical argument,
then $2/M<2\epsilon^2\leq\epsilon$, so $\Delta\leq 3\epsilon$. For
$q\leq\epsilon^{-2}/1000$ we also have $q<M/1000$, and
$\mathrm{TV}(L_0,L_1)\leq 0.008+\sqrt{\tfrac12\cdot\tfrac{16}{3}\cdot 9\epsilon^2q}\leq 0.008+0.155<\tfrac13$.

In both regimes, fewer than $\min\{M,\epsilon^{-2}\}/1000$ queries give
$\mathrm{TV}(L_0,L_1)<1/3$, which contradicts Step~1.
\end{proof}

\begin{proof}[Proof of \Cref{thm:optimal-stokes}]
Observe directly that $\partial_x v_1 + \partial_y v_2 = 0$,
$\int_{\mathbb T^2}v\,dx\,dy=0$, $-\Delta v = \sqrt2\,(0,\sin x)^T = v$, and
\[
 \|v\|_{L^2(\mathbb T^2)}^2
 =
 \frac{1}{(2\pi)^2}\int_0^{2\pi}\!\!\int_0^{2\pi} 2\sin^2 x\,dx\,dy
 =
 1.
\]
Hence for every $s\in\{0,\dots,M-1\}$, the unique Stokes velocity--pressure pair
is $u_s(x,y)=z_s v(x,y)$ and $p_s(x,y)=0$, whose normalized kinetic energy is
$g_s=\|u_s\|_{L^2(\mathbb T^2)}^2=z_s^2\|v\|_{L^2(\mathbb T^2)}^2=z_s\in\{0,1\}$.
Thus estimating the Stokes ensemble kinetic energy $\mu=\frac1M\sum_{s=0}^{M-1}g_s$
is isomorphic to estimating the Hamming weight $\frac1M\sum_{s=0}^{M-1}z_s$.
Classically, exhaustive evaluation of all $M$ scenarios paired with ordinary
Monte Carlo gives $Q_C=\mathcal O(\min\{M,\epsilon^{-2}\})$, matched below by
\Cref{thm:classical}.  Quantumly, evaluating all $M$ bits classically when
$M\leq\epsilon^{-1}$ and running coherent Stokes state preparation plus
amplitude estimation~\cite{brassard2002} when $M>\epsilon^{-1}$ achieves
$Q_Q=\mathcal O(\min\{M,\epsilon^{-1}\})$, while the quantum mean-estimation
lower bound of Nayak and Wu~\cite{nayak1999} gives
$Q_Q=\Omega(\min\{M,\epsilon^{-1}\})$.
\end{proof}

\section{Continuum Brinkman Model}
\label{app:matched-problem}

This appendix records the well-posedness of the continuum Brinkman problem
\eqref{eq:brinkman-pde} and explains how the pressure and the mean-flow
multiplier are recovered from the velocity. It supports
\Cref{sec:sobolev-riesz}. Throughout, $d\geq 2$, $\nu>0$ and
$0<\alpha_{\min}<\alpha_{\max}$; the inner product is
$\langle v,w\rangle_{L^2(\mathbb T^d)}:=\int_{\mathbb T^d}v^*\cdot w\,d\mu$ with
$d\mu=(2\pi)^{-d}dx$, and $\widehat w(k):=\int_{\mathbb T^d}w(x)e^{-ik\cdot x}\,d\mu(x)$.

We write
$\dot H^1_{\rm div}(\mathbb T^d):=\dot L^2_{\rm div}(\mathbb T^d)\cap H^1(\mathbb T^d)^d$
and $\dot H^1(\mathbb T^d):=\{q\in H^1(\mathbb T^d):\int_{\mathbb T^d}q\,dx=0\}$,
and let $\dot P_{\ker D}$ be the orthogonal (Leray) projector onto
$\dot L^2_{\rm div}(\mathbb T^d)$, given in Fourier coordinates by
$\widehat{(\dot P_{\ker D}F)}(0)=0$ and
$\widehat{(\dot P_{\ker D}F)}(k)=(I_d-kk^T/|k|^2)\widehat F(k)$ for $k\neq 0$. The
projected Brinkman operator is the unbounded operator
$\mathcal L_{\alpha_s}:=\dot P_{\ker D}(-\nu\Delta+\alpha_s)$ on
$\dot L^2_{\rm div}(\mathbb T^d)$ with domain
$\dot H^1_{\rm div}(\mathbb T^d)\cap H^2(\mathbb T^d)^d$ (for a general
$H^1$ field, $-\Delta u$ is not in $L^2$; on $\dot H^1_{\rm div}$ the operator is
understood weakly, as the map $\dot H^1_{\rm div}\to(\dot H^1_{\rm div})'$
induced by the form $a_s$ below). For measurable
$\alpha_s\in[\alpha_{\min},\alpha_{\max}]$ and real $f_s\in L^2(\mathbb T^d)^d$, a
triple
$(u_s,p_s,\lambda_s)\in(\dot H^1_{\rm div}(\mathbb T^d)\cap H^2(\mathbb T^d)^d)\times\dot H^1(\mathbb T^d)\times\mathbb R^d$
solves \eqref{eq:brinkman-pde} if and only if
$\mathcal L_{\alpha_s}u_s=\dot P_{\ker D}f_s$, and then
$\lambda_s=(2\pi)^{-d}\int_{\mathbb T^d}(f_s-\alpha_s u_s)\,dx$; this is part (i) of
the next lemma.

\begin{lemma}[Well-posedness and recovery of the multipliers]
\label{lem:app-matched-pde-oracle}
Let $f\in L^2(\mathbb T^d)^d$ be real and let
$\mathcal G_d:=\{\nabla p+\lambda:p\in\dot H^1(\mathbb T^d),\;\lambda\in\mathbb C^d\}$.
Then $L^2(\mathbb T^d)^d=\dot L^2_{\rm div}(\mathbb T^d)\oplus\mathcal G_d$
orthogonally in $\langle\cdot,\cdot\rangle_{L^2(\mathbb T^d)}$.  Consequently,
for every real-valued $\alpha_s\in L^\infty(\mathbb T^d)$ satisfying
$\alpha_{\min}\leq\alpha_s(x)\leq\alpha_{\max}$ a.e., there exists a unique real
triple $(u_s,p_s,\lambda_s)\in(\dot H^1_{\rm div}(\mathbb T^d)\cap H^2(\mathbb T^d)^d)\times\dot H^1(\mathbb T^d)\times\mathbb R^d$
solving the strong Brinkman system \eqref{eq:brinkman-pde} in $L^2(\mathbb T^d)^d$,
and $u_s$ coincides with the unique Lax--Milgram solution on
$\dot H^1_{\rm div}(\mathbb T^d)$ of
\begin{equation*}
 a_s(v,u_s)
 :=
 \nu\langle\nabla v,\nabla u_s\rangle_{L^2(\mathbb T^d)}
 +
 \langle v,\alpha_s u_s\rangle_{L^2(\mathbb T^d)}
 =
 \langle v,f\rangle_{L^2(\mathbb T^d)}
 \qquad\forall v\in\dot H^1_{\rm div}(\mathbb T^d).
\end{equation*}
\end{lemma}

The decomposition in part (i) is what recovers the two multipliers. The
residual $f-(-\nu\Delta+\alpha_s)u_s$ of the velocity equation is orthogonal to
every divergence-free mean-zero field, so it lies in $\mathcal G_d$; its unique
decomposition as $\nabla p_s+\lambda_s$ then defines the pressure (the gradient
part, carried by the modes $k\neq 0$) and the mean-flow multiplier (the constant
part, carried by the mode $k=0$).

\begin{proof}
\emph{Orthogonality.} For any $p\in\dot H^1(\mathbb T^d)$ ($\widehat p(0)=0$,
$\sum_{k\neq 0}|k|^2|\widehat p(k)|^2<\infty$), $\lambda\in\mathbb C^d$, and
$v\in L^2(\mathbb T^d)^d$, Plancherel's theorem gives
\[
 \langle\nabla p+\lambda,v\rangle_{L^2(\mathbb T^d)}
 =
 \lambda^*\cdot\widehat v(0)
 +
 \sum_{k\in\mathbb Z^d\setminus\{0\}}(-ik\,\widehat p(k))^*\cdot\widehat v(k).
\]
If $v\in\dot L^2_{\rm div}(\mathbb T^d)$ ($\widehat v(0)=0$ and $k\cdot\widehat v(k)=0$
for all $k\neq 0$), every term vanishes, so $\dot L^2_{\rm div}(\mathbb T^d)\perp\mathcal G_d$.
Conversely, for any $F\in L^2(\mathbb T^d)^d$, set $\lambda:=\widehat F(0)\in\mathbb C^d$,
$\widehat p(0):=0$, and $\widehat p(k):=-i|k|^{-2}(k\cdot\widehat F(k))$ for $k\neq 0$.
Since $\sum_{k\neq 0}|k|^2|\widehat p(k)|^2=\sum_{k\neq 0}|k|^{-2}|k\cdot\widehat F(k)|^2\leq\sum_{k\neq 0}|\widehat F(k)|_2^2\leq\|F\|_{L^2}^2<\infty$,
we have $p\in\dot H^1(\mathbb T^d)$ and $F-(\nabla p+\lambda)=\dot P_{\ker D}F\in\dot L^2_{\rm div}(\mathbb T^d)$,
with uniqueness of $(p,\lambda)$ because $\nabla p+\lambda=0$ implies $\lambda=\widehat 0(0)=0$
and $ik\widehat p(k)=0$ ($\forall k\neq 0$).

\emph{Existence and regularity.}
On the Hilbert space $\dot H^1_{\rm div}(\mathbb T^d)$ equipped with
$\|v\|_{\mathcal E}^2=\nu\|\nabla v\|_{L^2}^2+\alpha_0\|v\|_{L^2}^2=\sum_{k\neq 0}(\nu|k|^2+\alpha_0)|\widehat v(k)|_2^2$,
pointwise bounds $\alpha_{\min}\leq\alpha_s(x)\leq\alpha_{\max}$ yield
$\frac{\alpha_{\min}}{\alpha_0}\|u\|_{\mathcal E}^2\leq a_s(u,u)$ and
$|a_s(v,u)|\leq\frac{\alpha_{\max}}{\alpha_0}\|v\|_{\mathcal E}\|u\|_{\mathcal E}$.
By the Lax--Milgram theorem, there exists a unique $u_s\in\dot H^1_{\rm div}(\mathbb T^d)$
solving $a_s(v,u_s)=\langle v,f\rangle_{L^2}$ ($\forall v\in\dot H^1_{\rm div}(\mathbb T^d)$).
Because $\alpha_s$ and $f$ are real-valued, complex conjugation shows that
$\bar u_s\in\dot H^1_{\rm div}(\mathbb T^d)$ satisfies $a_s(v,\bar u_s)=\langle v,f\rangle_{L^2}$
for all $v$, so $\bar u_s=u_s$ by uniqueness.  Testing with $v(x)=w e^{ik\cdot x}$
($k\neq 0$, $w\in\mathbb C^d$, $k\cdot w=0$) shows that
$(I_d-kk^T/|k|^2)[(\nu|k|^2+\alpha_0)\widehat u_s(k)-\widehat F_s(k)]=0$ where
$F_s:=f-(\alpha_s-\alpha_0)u_s\in L^2(\mathbb T^d)^d$; since $k\cdot\widehat u_s(k)=0$,
this gives $(\nu|k|^2+\alpha_0)\widehat u_s(k)=(I_d-kk^T/|k|^2)\widehat F_s(k)$, whence
$\sum_{k\neq 0}|k|^4|\widehat u_s(k)|_2^2\leq\nu^{-2}\|F_s\|_{L^2}^2<\infty$ and
$u_s\in H^2(\mathbb T^d)^d$.  Setting $\mathcal R_s:=f-(-\nu\Delta u_s+\alpha_s u_s)\in L^2(\mathbb T^d)^d$,
integration by parts gives $\langle v,\mathcal R_s\rangle_{L^2}=\langle v,f\rangle_{L^2}-a_s(v,u_s)=0$
for all $v\in\dot H^1_{\rm div}(\mathbb T^d)$ (which is dense in $\dot L^2_{\rm div}(\mathbb T^d)$
via Fourier truncation), so $\mathcal R_s\in(\dot L^2_{\rm div}(\mathbb T^d))^\perp=\mathcal G_d$
decomposes uniquely as $\mathcal R_s=\nabla p_s+\lambda_s$ with real $(p_s,\lambda_s)\in\dot H^1(\mathbb T^d)\times\mathbb R^d$
and $\lambda_s=\int_{\mathbb T^d}(f-\alpha_s u_s)\,d\mu(x)$.  Conversely, any strong
solution of \eqref{eq:brinkman-pde} satisfies $a_s(v,u_s)=\langle v,f\rangle_{L^2}$
because $\langle v,\nabla p_s+\lambda_s\rangle_{L^2}=0$ for all $v\in\dot H^1_{\rm div}(\mathbb T^d)$.
\end{proof}

\begin{remark}[The variant without a mean-flow constraint]
\label{rem:unconstrained}
Suppose one sets $\lambda_s\equiv 0$ and drops the zero-mean velocity constraint
in \eqref{eq:brinkman-pde}, solving $-\nu\Delta u_s+\alpha_s(x)u_s+\nabla p_s=f_s$ with
$\nabla\cdot u_s=0$ and $\int_{\mathbb T^d}p_s\,dx=0$ on
$H^1_{\rm div}(\mathbb T^d)=\dot H^1_{\rm div}(\mathbb T^d)\oplus\mathbb R^d$, so that the
mean velocity $(2\pi)^{-d}\int_{\mathbb T^d}u_s\,dx$ may be nonzero. Because
$\alpha_s\geq\alpha_{\min}>0$, the reference operator $A_0=-\nu\Delta+\alpha_0I$ is
positive definite at $k=0$ as well ($A_0(0)=\alpha_0I_d$), unlike pure Stokes flow.
Keeping the $d$ constant modes by setting $R(0)=I_d$, $K_0(0)^{-1/2}=\alpha_0^{-1/2}I_d$
and $\beta_w(0)=0$ ($w(0)=1$) preserves
$\frac{\alpha_{\min}}{\alpha_0}I\preceq\widetilde K_s^{\rm unconstr}\preceq\frac{\alpha_{\max}}{\alpha_0}I$,
so $\kappa(\widetilde K_s^{\rm unconstr})\leq\widetilde\kappa$ and the bound of
\Cref{prop:sobolev-riesz} holds for this variant as well. The three-dimensional
check in \Cref{sec:exp-3d} compares the two variants numerically.
\end{remark}

\subsection{Equivalent Forms of the Sobolev--Riesz Map}
\label{app:generator}

The map $R_{\mathcal E}$ can also be written in terms of $C$, but only up to a
change of orthonormal coordinates in general. With the compact singular value
decomposition $C=U\Sigma V^*$ and $B:=U^*A_0U\succ 0$, one has $K_0=VBV^*$ and
$RK_0^{-1/2}=UB^{-1/2}V^*$, whereas
$C(C^*A_0C)^{\dagger/2}=U\,\Sigma(\Sigma B\Sigma)^{-1/2}V^*$. The matrix
$X:=\Sigma(\Sigma B\Sigma)^{-1/2}$ satisfies $X^*BX=I$, so $B^{1/2}X$ is unitary and
\[
 C\big(C^*A_0C\big)^{\dagger/2}=R_{\mathcal E}W_C,
 \qquad
 W_C:=V B^{1/2}X V^*\ \text{unitary on }\cZ_0 .
\]
If $\Sigma$ and $B$ commute, then $X=B^{-1/2}$, $W_C=I$, and
$C(C^*A_0C)^{\dagger/2}=R_{\mathcal E}$ exactly. This is the case for the Fourier
generators used in this paper (the transverse Fourier frame, and the curl of a
streamfunction or vector potential): on each wavenumber $k$, both $C^*C$ and
$U^*A_0U$ act as multiples of the identity on the plane perpendicular to $k$.
Since every statement below depends on $R_{\mathcal E}$ only through
$R_{\mathcal E}^*A_sR_{\mathcal E}$ up to unitary conjugation, the factor $W_C$
would not change any spectral bound.

\section{Quantum Circuit Construction and Resource Bounds}
\label{app:arith}
\label{app:circuit}

A mathematical preconditioner helps a quantum algorithm only if its
circuit implementation is efficient. This appendix therefore analyzes
the block encoding and the reversible arithmetic used to represent the
spatially varying drag. We first prove the normalization and precision
guarantees, then show how the flag probability corresponds to the
physical kinetic energy. We use these results to establish the
conditional gate-complexity theorem. Register layouts and detailed
angle-circuit specifications appear afterwards so that the implementation
can be checked without interrupting the main proof.

\subsection{Grid-Independent Block-Encoding Normalization}
\label{app:be-normalization}

To analyze multi-mode perturbations cleanly, we distinguish the \emph{Fourier
$\ell_1$ (Wiener algebra) semi-norm} from the \emph{pointwise $L^\infty$
semi-norm} of $\delta\alpha_s(x):=\alpha_s(x)-\alpha_0$:
\begin{equation}
 \Delta\alpha_{\ell_1}
 :=
 \sup_s \sum_{\ell=1}^L |c_\ell y_{s,\ell}|,
 \qquad
 \Delta\alpha_\infty
 :=
 \sup_{s,x} |\alpha_s(x)-\alpha_0|
 \leq
 \frac{\alpha_{\max}-\alpha_{\min}}{2}.
 \label{eq:l1-vs-linf}
\end{equation}
Because trigonometric sums can cancel in space, $\Delta\alpha_{\ell_1}$ can be much
larger than $\Delta\alpha_\infty$ when $L$ is large, so \Cref{lem:lcu-subnorm}
gives two constructions.

\begin{lemma}[Block encodings of $H_s$ with grid-independent normalization]
\label{lem:lcu-subnorm}
Let $W(k):=w(k)R(k)$ with $w(k)=\alpha_0^{1/2}(\nu|k|^2+\alpha_0)^{-1/2}$, so that
$\|W\|_2\leq 1$. For every grid size $N$, dimension $d$ and viscosity $\nu>0$, the
operator $H_s$ has the following two block encodings.
\begin{enumerate}[label=(\roman*),leftmargin=2.2em]
\item \textbf{Construction A (Fourier shifts).} Writing
$H_s=\sum_{\ell=1}^L\alpha_0^{-1}c_\ell y_{s,\ell}\,W^*T_{q_\ell}W$, where
$T_{q_\ell}|k\rangle=|k+q_\ell\rangle$, gives a weighted sum of operators
$W^*T_{q_\ell}W$. These terms are contractions, not unitaries (the weight
$w(k)<1$ enters through $W$), but each can be block-encoded using the
contraction $W$ and the unitary shift $T_{q_\ell}$; a linear combination of
these block encodings then has normalization
\begin{equation}
 \alpha_{\rm LCU}(H_s)=\frac{1}{\alpha_0}\sum_{\ell=1}^L|c_\ell y_{s,\ell}|
 \leq\frac{\Delta\alpha_{\ell_1}}{\alpha_0}.
\end{equation}
\item \textbf{Construction B (position space).} Writing
$H_s=W^*\mathcal F_d^\dagger\operatorname{diag}_x\!\big((\alpha_s(x)-\alpha_0)/\alpha_0\big)\mathcal F_dW$,
where $\mathcal F_d$ is the $d$-dimensional quantum Fourier transform, gives a
block encoding with normalization
\begin{equation}
 \alpha_\infty(H_s)=\frac{\Delta\alpha_\infty}{\alpha_0}
 \leq\frac{\widetilde\kappa-1}{\widetilde\kappa+1}=r_{\max}<1.
\end{equation}
\end{enumerate}
\end{lemma}

Neither normalization depends on $N$, $d$ or $\nu$. Construction~A needs the
Fourier coefficients of the drag to have a bounded sum; Construction~B needs only
the pointwise bounds \eqref{eq:alpha-bounds}, at the price of two quantum Fourier
transforms ($\mathcal O(d\log^2N)$ gates).

\begin{proof}
For every $k\neq 0$, $\|R(k)\|_2=1$ and $0<\alpha_0(\nu|k|^2+\alpha_0)^{-1}<1$,
so $\|W\|_2\leq 1$. For part (i), $W$ is a contraction, so it has a unitary
block encoding with normalization $1$, and $W^*T_{q_\ell}W$ is block-encoded with
normalization $1$ by the product of the block encodings of $W^*$, $T_{q_\ell}$
and $W$. Combining these $L$ block encodings by the standard
linear-combination-of-block-encodings construction~\cite{gilyen2019} with weights
$\alpha_0^{-1}c_\ell y_{s,\ell}$ gives normalization
$\alpha_0^{-1}\sum_\ell|c_\ell y_{s,\ell}|$. Part (ii) follows because $\mathcal F_d$ is unitary and the diagonal
operator $\operatorname{diag}_x((\alpha_s(x)-\alpha_0)/\Delta\alpha_\infty)$ is a
contraction under \eqref{eq:alpha-bounds}.
\end{proof}

\subsection{Reversible Arithmetic: Accuracy and Cost}
\label{app:arith-summary}

The drag-dependent rotation angle is a function of the scenario and
grid coordinates. Model~II evaluates it reversibly instead of storing
all angles in a lookup table. In this subsection we establish two facts
needed for the complexity claim: the computed angle is uniformly
accurate, and the number of arithmetic operations grows only with the
register widths under the stated input assumptions.

This replaces a large table of angles by a compact calculation,
provided that the scenario coefficients themselves are efficiently
computable. Write
$\widetilde{\mathcal U}_{{\rm BE},s}$ for the Hermitian block encoding of
$\widetilde X_s:=H_s/r_{\rm be}$ built in stages (2) to (6), with
$r_{\rm be}\in(r_{\max},1)$ (\Cref{app:registers}). We distinguish two
compilations of the same circuit: in Model~I the rotation angles are exact, and
in Model~II they are computed by reversible arithmetic of width $b_{\rm arith}$.
Let $X_s^{({\rm I})}$ and $X_s^{({\rm II})}$ be the corresponding signal blocks.

\begin{proposition}[Accuracy and gate cost of one block-encoding call]
\label{prop:be-cost}
\begin{enumerate}[label=(\roman*),leftmargin=2.2em]
\item \textbf{Accuracy.} If the viscous, frame and drag angles are computed with
worst-case errors $\delta_{\beta_w}$, $\delta_{\phi_{\rm rot}}$ and
$\delta_{\beta_\alpha}$, the Model~II block encoding differs from the Model~I
block encoding by at most
\begin{equation}
 \|X_s^{({\rm II})}-X_s^{({\rm I})}\|_2
 \leq
 \|\widetilde{\mathcal U}_{{\rm BE},s}^{({\rm II})}-\widetilde{\mathcal U}_{{\rm BE},s}^{({\rm I})}\|_2
 \leq
 \delta_{\beta_w} + \delta_{\phi_{\rm rot}} + \frac12\delta_{\beta_\alpha}
 =
 \mathcal O(2^{-b_{\rm arith}}).
 \label{eq:be-telescoping-bound}
\end{equation}
\item \textbf{Cost.} Suppose the $L(\epsilon)$ scenario coefficients come from a
reversible oracle with $\operatorname{polylog}(M,1/\eta_K)$ gates
(Assumption~\ref{as:A1} below) and $b_{\rm arith}=\mathcal O(\log(1/\eta_K))$.
Then one call to $\widetilde{\mathcal U}_{{\rm BE},s}^{({\rm II})}$ uses
$\mathcal O(d\log N+b_{\rm arith}^2)$ reusable clean ancillas and
\begin{equation}
 \begin{split}
 G_K(\eta_K)
 =
 \mathcal O\Big(&
 d(\log N)^2
 + d\log N\,b_{\rm arith}^2
 + L(\epsilon)\big(\operatorname{polylog}(M,1/\eta_K) + d\log N + b_{\rm arith}^3\big)
 \\
 &\qquad + b_{\rm arith}\,\log\frac{b_{\rm arith}}{\eta_K}
 \Big)
 \end{split}
 \label{eq:gk-sobolev}
\end{equation}
Clifford$+T$ gates.
\end{enumerate}
\end{proposition}

For coefficients generated from $|s\rangle$ by a scrambled Sobol' digital
net~\cite{matousek1998}, the scenario oracle costs
$\mathcal O((\log M)\,b_{\rm arith}+b_{\rm arith}^2)$ gates
(\Cref{prop:global-arith-error}(i)). Loading an unstructured table of $M$
scenarios would cost $\mathcal O(Mb_{\rm arith})$ gates instead.

\begin{proof}
The block encoding is the product of the five unitary stages
$U_{W,R},U_{\rm rot},U_{{\rm refl},\alpha},U_{\rm rot}^\dagger,U_{W,L}$. Replacing
each exact rotation angle by its computed value changes a rotation by at most
$\|R_y(\theta+\delta)-R_y(\theta)\|_2=2|\sin(\delta/4)|\leq\tfrac12|\delta|$, so
telescoping the product gives
$2(\tfrac12\delta_{\beta_w})+2(\tfrac12\delta_{\phi_{\rm rot}})+\tfrac12\delta_{\beta_\alpha}$,
which is \eqref{eq:be-telescoping-bound}; the angle errors are
$\mathcal O(2^{-b_{\rm arith}})$ by \Cref{prop:global-arith-error}. The gate
count \eqref{eq:gk-sobolev} is the sum of the costs of the arithmetic components,
which are listed in \Cref{app:be-cost-details}.
\end{proof}

The drag angle requires care at high contrast. It involves $\arccos m$ with
$|m|$ close to $1$, where the derivative is unbounded. \Cref{prop:global-arith-error}
handles this case uniformly in the contrast by evaluating the equivalent
half-angle form $2\arccos|m|=4\arcsin\sqrt{(1-|m|)/2}$ on dyadic ranges of the gap
$1-|m|$.

Write $m_s(x):=(\alpha_s(x)-\alpha_0)/(r_{\rm be}\alpha_0)$ for the normalized drag
modulation, and let $\delta_{\beta_w}$, $\delta_{\theta_E}$, $\delta_{\phi_{\rm rot}}$ and $\delta_{\beta_\alpha}$
denote the worst-case errors of the four computed angles. The next proposition
summarizes the accuracy of the reversible circuits; the full configuration
(register widths, polynomial degrees and guard bits) and the proof are in
\Cref{app:arith}.

\begin{proposition}[Accuracy of the reversible angle circuits]
\label{prop:global-arith-error}
Fix $\nu>0$, $\alpha_0>0$ and $r_{\rm be}\in(0,1)$, and let the fixed-point width
be $b_{\rm arith}\geq 8$.
\begin{enumerate}[label=(\roman*),leftmargin=2.2em]
\item \textbf{Scenario coefficients.} For every $M=2^{n_M}\geq 2$, the scrambled
Sobol' circuit $O_{\rm coeff}$ computes each coefficient
$c_{s,\ell}/(r_{\rm be}\alpha_0)$ from $|s\rangle$ to within
$\tfrac12 2^{-b_{\rm arith}}$, using
$\mathcal O((\log M)\,b_{\rm arith}+b_{\rm arith}^2)$ gates.
\item \textbf{Angles.} There is a constant $C_{\rm arith}$ such that, for all
$N=2^n\geq 4$ and $M\geq 2$, the four angle circuits return every workspace qubit
to $|0\rangle$ and, on every computational basis input, satisfy
\begin{equation}
 \max\!\big\{\delta_{\beta_w},\,\delta_{\theta_E},\,\delta_{\phi_{\rm rot}},\,\delta_{\beta_\alpha}\big\}
 \leq
 C_{\rm arith}\,2^{-b_{\rm arith}}.
 \label{eq:global-arith-bound}
\end{equation}
\item \textbf{Drag angle at any contrast.} For the drag angle, (ii) holds
(a) with a direct evaluator when $\sup_{s,x}|m_s(x)|\leq m_{\max}<0.75$, and
(b) with a half-angle adapter for every contrast $\widetilde\kappa>1$ and every
$r_{\rm be}\in(r_{\max},1)$. In case (b), $C_{\rm arith}$ does not depend on
$\widetilde\kappa$, $N$, $M$, $b_{\rm arith}$ or the input width $b_{\rm in}$, and
the adapter uses
$\mathcal O((b_{\rm in}+\log\delta_{\min}^{-1}+b_{\rm arith})b_{\rm arith}^2+b_{\rm arith}^3)$
gates, where $\delta_{\min}:=1-r_{\max}/r_{\rm be}$.
\end{enumerate}
\end{proposition}

Two details are left out of the statement. In case (a) the constant
$C_{\rm arith}$ depends on $(\nu,\alpha_0,m_{\max},L)$. In (i), the Toffoli, $T$
and ancilla counts of $O_{\rm coeff}$ do not depend on $M$ as long as the Sobol'
coordinate width satisfies $B_{\rm coord}\geq b_{\rm arith}+3$; the choice
$B_{\rm coord}=\max\{32,\log_2M\}$ guarantees this for $b_{\rm arith}\leq 29$, and
beyond that the counts grow at most linearly in $\log_2M$.

\begin{remark}[Scope of the explicit gate-level construction in dimension $d$]
\label{rem:general-d}
The four angle circuits of \Cref{prop:global-arith-error}, the bound
\eqref{eq:be-telescoping-bound}, the gate count \eqref{eq:gk-sobolev} and all
gate-level simulations are constructed explicitly for $d=2$, where the frame
stage $U_{\rm rot}$ is a single rotation by $\phi_{\rm rot}(k_x,k_y)$. The
mathematical results (\Cref{prop:riesz,prop:reduced,prop:sobolev-riesz},
\Cref{lem:lcu-subnorm,lem:norm-chain}) hold for every $d\geq 2$, and the
three-dimensional checks of \Cref{sec:exp-3d} are operator-level, not
gate-level. For $d\geq 3$ the frame $R(k)$ can be realized, for example, by the
Householder reflection $I-2vv^T/\|v\|^2$ with $v=k/|k|-e_d$, which maps $e_d$ to
$k/|k|$ and whose first $d-1$ columns span $\ker(k^T)$. Implementing it on the
$\lceil\log_2 d\rceil$-qubit component register requires reversibly computing
$k/|k|$ and the $\mathcal O(d)$ angles of a state-preparation circuit for $v$,
and then conjugating a reflection about $|0\rangle$. We do not give a complete
reversible synthesis or uniform precision analysis of this stage here.
Accordingly, for $d\geq 3$ the gate-level statements of
\Cref{prop:be-cost} and \Cref{thm:sobolev-advantage}(i) are conditional on the
following \emph{frame-synthesis hypothesis}: the frame stage can be implemented
with $\mathcal O(d\operatorname{polylog}(N,1/\eta_K))$ Clifford$+T$ gates and
operator-norm error $\mathcal O(d\,2^{-b_{\rm arith}})$. Under this hypothesis
$\delta_{\phi_{\rm rot}}$ in \eqref{eq:be-telescoping-bound} is replaced by
$\mathcal O(d\,2^{-b_{\rm arith}})$, which changes $b_{\rm arith}$ only by an
additive $\mathcal O(\log d)$ and leaves \eqref{eq:gamma-q-sobolev} unchanged.
\end{remark}

\subsection{Recovering Kinetic Energy from the Flag Probability}
\label{app:norm-chain}

\begin{lemma}[From flag probability to kinetic energy]
\label{lem:norm-chain}
Suppose every forcing $f_s$ is supported on wavenumbers $0<|k|\leq k_0$ and its
reduced form $b_s=R^*\hat f_s$ satisfies $b_{\min}\leq\|b_s\|_2\leq b_{\max}$ with
$b_{\min}>0$. Put $\widetilde b_{\max}:=(\nu+\alpha_0)^{-1/2}b_{\max}$, fix
$r_{\rm be}\in(r_{\max},1)$, and choose a polynomial scale
$\gamma_{\rm QSVT}=\tfrac12\,\mathrm{scale}$ with
$0<\mathrm{scale}\leq(1-r_{\rm be}^2)/2$.
\begin{enumerate}[label=(\roman*),leftmargin=2.2em]
\item \textbf{Source loading.} The preconditioned source
$\widetilde b_s=K_0^{-1/2}b_s$ satisfies $\|\widetilde b_s\|_2\leq\widetilde b_{\max}$,
so a unitary $U_b$ can load it without postselection:
\[
 U_b|s\rangle|0\rangle_{\rm src}|0\rangle_{\rm mode}
 =
 |s\rangle\Big(
 |0\rangle_{\rm src}\frac{\widetilde b_s}{\widetilde b_{\max}}
 +
 |1\rangle_{\rm src}|\perp_s\rangle
 \Big).
\]
\item \textbf{Bounded polynomials.} Let $P_{\rm even}$ and $P_{\rm odd}$ be the
degree-$d_{\rm inv}$ Chebyshev truncations of
$f_{\rm even}(x)=\mathrm{scale}/(1-r_{\rm be}^2x^2)$ and
$f_{\rm odd}(x)=-\mathrm{scale}\,r_{\rm be}x/(1-r_{\rm be}^2x^2)$. Then
\begin{equation}
\begin{aligned}
 \sup_{x\in[-1,1]}|P_{\rm even}(x)|
 &\leq
 \sum_{m=0}^{d_{\rm inv}/2}|a_{2m}|
 \leq
 \frac{\mathrm{scale}}{1-r_{\rm be}^2}
 \leq \frac12, \\
 \sup_{x\in[-1,1]}|P_{\rm odd}(x)|
 &\leq
 \sum_{m=0}^{(d_{\rm inv}-1)/2}|b_{2m+1}|
 \leq
 \frac{\mathrm{scale}\,r_{\rm be}}{1-r_{\rm be}^2}
 < \frac12,
\end{aligned}
 \label{eq:analytic-cheb-bound}
\end{equation}
so QSVT can implement both polynomials.
\item \textbf{Energy readout.} With the QSVT inverse
$U_{\rm QSVT}\simeq\gamma_{\rm QSVT}\widetilde K_s^{-1}$ and the observable
$E_s=\sqrt{\nu+\alpha_0}\,K_0^{-1/2}$, the flag probability of scenario $s$ is
\begin{equation}
 p_{{\rm good},s}
 =
 \Big\|
 E_s\Big(
 \frac{\gamma_{\rm QSVT}}{\widetilde b_{\max}}
 \widetilde K_s^{-1}\widetilde b_s
 \Big)
 \Big\|_2^2
 =
 \frac{\|u_s\|_{L^2}^2}{B_{\rm norm}},
 \qquad
 B_{\rm norm}:=\frac{b_{\max}^2}{\gamma_{\rm QSVT}^2(\nu+\alpha_0)^2},
 \label{eq:exact-norm-chain}
\end{equation}
and for every $N$, $d$ and $s$ it is bounded below by
\begin{equation}
 \|u_s\|_{L^2}
 \geq
 \frac{b_{\min}}{\nu k_0^2+\alpha_{\max}},
 \qquad
 p_{{\rm good},s}
 \geq
 \frac{\gamma_{\rm QSVT}^2(\nu+\alpha_0)^2 b_{\min}^2}
      {b_{\max}^2(\nu k_0^2+\alpha_{\max})^2}.
 \label{eq:uniform-prob-lower-bound}
\end{equation}
\end{enumerate}
\end{lemma}

Part (iii) is what lets amplitude estimation read out the energy. The factor
$B_{\rm norm}$ does not depend on the grid, so converting a probability into an
energy costs a fixed factor, and the lower bound keeps the flag probability away
from zero however fine the grid is.

\begin{proof}
Part (i) follows from $\min_{k\neq 0}(\nu|k|^2+\alpha_0)=\nu+\alpha_0$. For part
(ii), $(1-r_{\rm be}^2x^2)^{-1}=\sum_{n\geq 0}r_{\rm be}^{2n}x^{2n}$ has
nonnegative power-series coefficients, hence nonnegative Chebyshev coefficients,
so the Chebyshev $\ell_1$-norm of $f_{\rm even}$ equals $f_{\rm even}(1)$; the same
holds for $-f_{\rm odd}$. Since $|T_k(x)|\leq 1$ on $[-1,1]$, the truncations obey
\eqref{eq:analytic-cheb-bound}. For part (iii), $E_s$ is a contraction with
$\|E_s\|_2=1$, implemented by rotating the observable flag by
$\theta_E(|k|^2)=2\arccos\sqrt{(\nu+\alpha_0)/(\nu|k|^2+\alpha_0)}$. Moreover
$K_0^{-1/2}\widetilde K_s^{-1}\widetilde b_s=K_0^{-1/2}(K_0^{1/2}K_s^{-1}K_0^{1/2})(K_0^{-1/2}b_s)=K_s^{-1}b_s=\xi_s$
and $\|R\xi_s\|_{L^2}=\|\xi_s\|_2=\|u_s\|_{L^2}$, which gives
\eqref{eq:exact-norm-chain}. For the lower bound, test the weak form
$\langle v,A_su_s\rangle=\langle v,Rb_s\rangle$ with $v=Rb_s\in\Ker D$. Since
$b_s$ is supported on $|k|\leq k_0$, $\|-\nu\Delta Rb_s\|_{L^2}\leq\nu k_0^2\|b_s\|_2$
and $\|\alpha_sRb_s\|_{L^2}\leq\alpha_{\max}\|b_s\|_2$, so
\[
 \|b_s\|_2^2
 =
 \langle Rb_s,A_su_s\rangle
 \leq
 \|A_sRb_s\|_{L^2}\|u_s\|_{L^2}
 \leq
 (\nu k_0^2+\alpha_{\max})\|b_s\|_2\|u_s\|_{L^2},
\]
which is \eqref{eq:uniform-prob-lower-bound} after inserting it into
\eqref{eq:exact-norm-chain}.
\end{proof}

\Cref{lem:norm-chain}(ii) bounds the magnitude of the truncated polynomials.
The degree used in \Cref{thm:sobolev-advantage} also needs a uniform
approximation bound, which the next lemma supplies; the degree-14 numerics of
\Cref{tab:sobolev-brinkman} illustrate it but do not prove it.

\begin{lemma}[Uniform accuracy and degree of the QSVT inverse]
\label{lem:qsvt-degree}
Let $r:=r_{\rm be}\in(0,1)$ and $\rho:=(1+\sqrt{1-r^2})/r>1$, and let
$P_{\rm even}$, $P_{\rm odd}$ be the degree-$n$ Chebyshev truncations of
\Cref{lem:norm-chain}(ii). Then
\begin{equation}
 \sup_{x\in[-1,1]}
 \Big|P_{\rm even}(x)+P_{\rm odd}(x)-\frac{\mathrm{scale}}{1+rx}\Big|
 \leq
 \frac{2\,\mathrm{scale}\;\rho^{-(n+1)}}{\sqrt{1-r^2}\,(1-\rho^{-1})}.
 \label{eq:qsvt-tail}
\end{equation}
With $r_{\rm be}=(1+r_{\max})/2$, so that $1-r_{\rm be}=(\widetilde\kappa+1)^{-1}$,
uniform accuracy $\eta$ is reached with
\begin{equation}
 n+1\;\geq\;\sqrt{\widetilde\kappa+1}\;
 \ln\!\frac{4\,\mathrm{scale}\,(\widetilde\kappa+1)}{\eta},
 \qquad\text{so}\qquad
 d_{\rm inv}=\mathcal O\!\big(\sqrt{\widetilde\kappa}\,\log(\widetilde\kappa/\eta)\big)
 \subseteq\mathcal O\!\big(\widetilde\kappa\log(\widetilde\kappa/\eta)\big).
 \label{eq:qsvt-degree}
\end{equation}
Since $P_{\rm even}$ and $P_{\rm odd}$ are real, have definite parity and satisfy
$\sup_{[-1,1]}|P|\leq\tfrac12$ by \eqref{eq:analytic-cheb-bound}, QSVT phase
sequences of length $n$ implementing them exist and can be computed
classically~\cite{gilyen2019}; each is realized as the real part of a QSVT
circuit, and the two parities are combined on the register $|q_{\rm lcu}\rangle$.
\end{lemma}

\begin{proof}
With $a:=1/r>1$ and $\rho=a+\sqrt{a^2-1}$, the classical expansion
$(a-x)^{-1}=(a^2-1)^{-1/2}\big[1+2\sum_{k\geq 1}\rho^{-k}T_k(x)\big]$ holds
uniformly on $[-1,1]$. Writing
$f_{\rm even}=\tfrac{\mathrm{scale}}{2r}\big[(a-x)^{-1}+(a+x)^{-1}\big]$,
$f_{\rm odd}=-\tfrac{\mathrm{scale}}{2r}\big[(a-x)^{-1}-(a+x)^{-1}\big]$ and using
$T_k(-x)=(-1)^kT_k(x)$ and $r\sqrt{a^2-1}=\sqrt{1-r^2}$, the Chebyshev
coefficients of $f_{\rm even}+f_{\rm odd}=\mathrm{scale}/(1+rx)$ have modulus
$2\,\mathrm{scale}\,\rho^{-k}/\sqrt{1-r^2}$ for $k\geq 1$. Since $|T_k|\leq 1$,
the truncation error is at most the coefficient tail
$\tfrac{2\,\mathrm{scale}}{\sqrt{1-r^2}}\sum_{k>n}\rho^{-k}$, which is
\eqref{eq:qsvt-tail}. For \eqref{eq:qsvt-degree}, put
$t:=(\widetilde\kappa+1)^{-1/2}\in(0,1)$. Then $\sqrt{1-r^2}\geq\sqrt{1-r}=t$, and
$\ln\rho=\operatorname{arccosh}(1/r)\geq\operatorname{arccosh}(1+(1-r))\geq\sqrt{1-r}=t$
(because $\tfrac{d}{d\delta}\operatorname{arccosh}(1+\delta)=(\delta(2+\delta))^{-1/2}
\geq\tfrac{d}{d\delta}\sqrt\delta$ for $0<\delta\leq 2$). Hence
$1-\rho^{-1}\geq 1-e^{-t}\geq t/2$, and the right side of \eqref{eq:qsvt-tail} is
at most $4\,\mathrm{scale}\,e^{-(n+1)t}/t^2$, which is at most $\eta$ under the
stated condition on $n$. The existence and classical computability of the phase
sequences for bounded real polynomials of definite parity is the QSVT theorem
of~\cite{gilyen2019}.
\end{proof}

\subsection{Proof of the Conditional Gate-Complexity Bound}
\label{app:thm5-proof}

\begin{proof}[Proof of \Cref{thm:sobolev-advantage}]
\emph{Circuit parameters.} The circuit uses QSVT of degree
$d_{\rm inv}=\mathcal O(\widetilde\kappa\log(\widetilde\kappa/\epsilon))$, which by
\Cref{lem:qsvt-degree} (with uniform accuracy $\eta=\Theta(\epsilon/B_{\rm norm})$)
suffices for the polynomial approximation of the inverse, with
polynomial scale $\gamma_{\rm QSVT}=\Theta(\widetilde\kappa^{-1})$ and reversible
arithmetic of width $b_{\rm arith}=\mathcal O(\log(\widetilde\kappa^2 d_{\rm inv}/\epsilon))$.
Observable bias $\mathcal O(\epsilon)$ corresponds to raw probability precision
$\eta_{\rm impl}\leq\epsilon/(8B_{\rm norm})$. The controlled-gate decompositions
and register counts are given in \Cref{app:thm5-circuit}.

\emph{Cost.}
By \Cref{prop:sobolev-riesz,lem:lcu-subnorm}, both $\kappa(\widetilde K_s)\leq\widetilde\kappa$
and the block-encoding subnormalization are $\mathcal O(1)$ independently of
$N=\Theta(\epsilon^{-1/r})$.  By \Cref{lem:norm-chain}, the normalization scale
$B_{\rm norm}=\Theta(\widetilde\kappa^2)$ multiplies the required raw probability
precision and amplitude-estimation query count by $\mathcal O(\widetilde\kappa^2)$.
Telescoping the $(d_{\rm inv}+1)$ block-encoding calls via
\eqref{eq:be-telescoping-bound} together with the source and observable flag
rotations $U_{\rm src},U_{\rm obs}$ (angle error $\delta_{\theta_E}=\mathcal O(2^{-b_{\rm arith}})$)
gives
\[
 \|\cA^{({\rm II})}|0\rangle-\cA^{({\rm I})}|0\rangle\|_2
 \leq
 \delta_{\theta_E}
 +
 (d_{\rm inv}+1)\!\left(\delta_{\beta_w}+\delta_{\phi_{\rm rot}}+\frac12\delta_{\beta_\alpha}\right)
 =
 \mathcal O(d_{\rm inv}\,2^{-b_{\rm arith}}),
\]
so $b_{\rm arith}=\mathcal O(\log(\widetilde\kappa^2 d_{\rm inv}/\epsilon))$
guarantees physical observable bias $\mathcal O(\epsilon)$.  Multiplying
$d_{\rm inv}=\mathcal O(\widetilde\kappa\log(\widetilde\kappa/\epsilon))$ by
\eqref{eq:gk-sobolev} with $\log N=\mathcal O(\frac1r\log\frac1\epsilon)$, and
adding the $\mathcal O(S(\epsilon)\,d\log N)$ source preparation cost, the $\mathcal O(d(\log N)^2 + d\log N\cdot b_{\rm arith}^2 + b_{\rm arith}^3)$
cost of $U_{\rm src}$ and $U_{\rm obs}$, and the $\mathcal O(\log M+d\log N)$
linear-Toffoli Grover reflections~\cite{barenco1995}, gives \eqref{eq:gamma-q-sobolev}.
Combining this upper bound with the classical lower bound in \Cref{thm:budget}
proves sufficiency of \eqref{eq:sobolev-frontier-exact}.
\end{proof}

\subsection{Register Layout and Workspace}
\label{app:registers}

This subsection lists the registers and the structure of one block-encoding
call (\Cref{fig:circuit-architecture}). It is needed to reproduce the circuits, not to follow the argument.

\paragraph{Registers.}
$|s\rangle$ ($\log_2 M$ qubits) stores the scenario index;
$|q_{\rm lcu}\rangle$ ($2$ qubits) combines the even and odd Chebyshev QSVT
branches;
$|a_{\rm BE}\rangle=|a_{\rm WL},a_\alpha,a_{\rm WR},a_{\rm pol}\rangle$ ($4$
ancilla qubits in $d=2$; in general $d\geq 2$, three ancilla qubits
$|a_{\rm WL},a_\alpha,a_{\rm WR}\rangle$ plus the longitudinal basis state of a
$\lceil\log_2 d\rceil$-qubit velocity-component register $|c\rangle$) flags the
block-encoding signal subspace $\Pi_{\rm sig}$;
$|k\rangle$ ($d\log_2 N$ qubits) stores two's-complement Fourier wavenumbers (or
grid coordinates $|x\rangle$ after inverse QFT); and
$|f_{\rm src}\rangle,|f_{\rm obs}\rangle$ ($2$ qubits) mark source preparation
and observable extraction.
\paragraph{The block encoding.}
Fix a subnormalization scale $r_{\rm be}\in(r_{\max},1)$ so that
$\widetilde X_s := H_s/r_{\rm be}$ satisfies $\|\widetilde X_s\|_2<1$.  Define
the scalar Fourier energy weight $w(k):=\sqrt{\alpha_0/(\nu|k|^2+\alpha_0)}\in(0,1)$
for $k\neq 0$ (and $w(0):=0$).  In Construction~B, $\widetilde X_s$ is factored
symmetrically into unitary basis changes and single-qubit reflections
$R_y(\theta)Z$ (\Cref{fig:circuit-architecture}(b)):
\[
 \widetilde{\mathcal U}_{{\rm BE},s}
 :=
 S_{LR}\, U_{W,L}\, B^\dagger\, U_{{\rm refl},\alpha}\, B\, U_{W,R},
 \qquad
 B := (I_{\rm anc}\otimes\mathcal F_d^\dagger)\, U_{\rm rot}.
\]
Reading from right to left: $U_{W,R}$ rotates ancilla $|a_{\rm WR}\rangle$ by
$\beta_w(|k|^2)=2\arccos(w(k))$ ($\beta_w(0)=\pi$) to weight mode $k$ by $w(k)$;
$U_{\rm rot}$ maps transverse polarization states onto divergence-free Cartesian
velocity vectors perpendicular to $k$ (a single rotation by
$\phi_{\rm rot}(k_x,k_y)=2\arctan 2(k_x,k_y)$ in $d=2$, the case constructed and
simulated here; in general $d$ a Householder reflection, whose gate-level
synthesis is the hypothesis of \Cref{rem:general-d}); $\mathcal F_d^\dagger$ transforms $|k\rangle$ to
position space $|x\rangle$; $U_{{\rm refl},\alpha}$ rotates $|a_\alpha\rangle$
by $\beta_\alpha(s,x)=2\arccos((\alpha_s(x)-\alpha_0)/(r_{\rm be}\alpha_0))$ to
apply pointwise drag multiplication; $B^\dagger$ and $U_{W,L}$ apply the
conjugate return path onto $|a_{\rm WL}\rangle$; and
$S_{LR}=\operatorname{SWAP}(a_{\rm WL},a_{\rm WR})$ swaps the two weight
ancillas so that $\widetilde{\mathcal U}_{{\rm BE},s}$ is strictly Hermitian and
self-inverse ($\widetilde{\mathcal U}_{{\rm BE},s}^\dagger=\widetilde{\mathcal U}_{{\rm BE},s}$,
$\widetilde{\mathcal U}_{{\rm BE},s}^2=I$).
\paragraph{Two compilations.}
We use two implementations of the same rotation stages
($U_{\rm src}$, $U_{W,R}$, $U_{\rm rot}$, $U_{{\rm refl},\alpha}$, $U_{W,L}$
and $U_{\rm obs}$). \emph{Model~I} loads precomputed floating-point angles through
ancilla-free Walsh--Hadamard Gray-code multiplexors~\cite{shende2006}; its
block is the exact reference $X_s^{({\rm I})}=\widetilde X_s$ and its gate
cost is $\Theta(MN^d)$. \emph{Model~II} evaluates the angles in superposition
with $b_{\rm arith}$-bit reversible fixed-point arithmetic on reusable clean
workspace qubits; its block is $X_s^{({\rm II})}$ and its gate cost is
polylogarithmic in $N$ and $M$ under the input assumptions of
\Cref{thm:sobolev-advantage}.

\subsection{Reversible Angle-Evaluation Circuits}

This subsection proves \Cref{prop:global-arith-error}. Each angle evaluator
splits its input range into octaves, splits each octave into eight sub-bins, and
evaluates a Chebyshev polynomial on each sub-bin by Horner's rule in fixed-point
arithmetic. \Cref{tab:arith-config} lists the parameters.

\begin{table}[htbp]
\centering
\caption{Parameters of the reversible angle evaluators, as functions of the
arithmetic width $b_{\rm arith}$, the ensemble size $M$ and the contrast margin
$\delta_{\min}:=1-r_{\max}/r_{\rm be}$.}
\label{tab:arith-config}
\small
\begin{tblr}{width=\textwidth, colspec={Q[wd=5.2cm,l] X[l]}}
\toprule
Parameter & Value \\
\midrule
Output width $b_{\rm out}$ (angles in $[0,4\pi)$) & $b_{\rm arith}+1$ \\
Mantissa width $q$ & $\max\{5,b_{\rm arith}-1\}$ \\
Chebyshev degree for $\beta_w,\theta_E,\phi_{\rm rot}$ & $D=\max\{1,\lfloor(b_{\rm arith}-1)/4\rfloor\}$ \\
Chebyshev degree for the drag angle & $D_{\rm acos}=\max\{1,\lfloor(b_{\rm arith}-1)/3\rfloor\}$ \\
Horner guard bits & $g_{\rm bits}=3+\lceil\log_2(D_{\rm stab}+2)\rceil$, $D_{\rm stab}\in\{D,D_{\rm acos}\}$ \\
Exact point values & low octaves, $1\leq|k|^2\leq 15$ \\
Sobol' coordinate width & $B_{\rm coord}=\max\{32,\log_2M\}$ \\
Coefficient product width & $q_{\rm prod}=\max\{7,b_{\rm arith}+1\}$ \\
Net truncation width & $B_{\rm net}=\min\{B_{\rm coord},q_{\rm prod}+2\}$ \\
Adapter gap width (case (b)) & $q_m=\max\{2+\lceil\log_2\delta_{\min}^{-1}\rceil,\ b_{\rm arith}+4+\lceil\tfrac12\log_2\max\{4,\delta_{\min}^{-1}\}\rceil\}$ \\
\bottomrule
\end{tblr}
\end{table}

The drag angle $\beta_\alpha(s,x)=2\arccos m_s(x)$ is computed in one of two ways.
(a) When $\sup_{s,x}|m_s(x)|\leq m_{\max}<0.75$, the direct eight-sub-bin evaluator
is used. (b) At any contrast, the half-angle adapter takes the $b_{\rm in}$-bit drag
value, computes the gap $u_s(x):=1-|m_s(x)|$ in width $q_m$, finds its leading bit
(the octave $e\in\{1,\dots,q_m\}$), and evaluates
$2\arccos|m|=4\arcsin\sqrt{u/2}$ on that octave. It loads the constant $\pi$ at
$u=1$ and reflects $\theta\mapsto 2\pi-\theta$ when $m<0$. Any
$r_{\rm be}\in(r_{\max},1)$ works for (b); the circuits use
$r_{\rm be}=\gamma_\alpha=\min\{\frac43r_{\max},\frac12(1+r_{\max})\}$.

\begin{proposition}[Circuit-level accuracy and cost]
\label{prop:arith-full}
With the parameters of \Cref{tab:arith-config}, the following hold.
\begin{enumerate}[label=(\roman*),leftmargin=2.2em]
\item \textbf{Scenario oracle.} For every $M=2^{n_M}\geq 2$ and every mode $\ell$,
$O_{\rm coeff}$ computes $c_{s,\ell}/(r_{\rm be}\alpha_0)$ from $|s\rangle$ with
error at most $2^{-q_{\rm prod}}\leq\tfrac12 2^{-b_{\rm arith}}$. It uses at most
$B_{\rm net}(n_M+1)$ $\mathrm{CX}$ gates for the scrambled net and $B_{\rm net}-1$
Cuccaro additions on $q_{\rm prod}+2+\lceil\log_2(B_{\rm net}+2)\rceil$ bits. Its
Toffoli, $T$ and ancilla counts do not depend on $M$ when
$B_{\rm coord}\geq q_{\rm prod}+2$, and otherwise grow at most linearly in
$\log_2M$.
\item \textbf{Angles.} For every $N=2^n\geq 4$, $M\geq 2$ and $b_{\rm arith}\geq 8$,
the four evaluators return every workspace qubit to $|0\rangle$ and satisfy
\begin{equation}
 \max\!\big\{\delta_{\beta_w},\,\delta_{\theta_E},\,\delta_{\phi_{\rm rot}},\,\delta_{\beta_\alpha}\big\}
 \leq
 C_{\rm arith}\,2^{-b_{\rm arith}}
 \label{eq:global-arith-bound-full}
\end{equation}
on all basis inputs. The constant $C_{\rm arith}$ depends only on
$(\nu,\alpha_0,m_{\max},L)$ in case (a), and on none of $\widetilde\kappa$,
$b_{\rm in}$, $N$, $M$ and $b_{\rm arith}$ in case (b).
\item \textbf{Adapter cost.} The case (b) adapter uses
$\mathcal O((b_{\rm in}+\log\delta_{\min}^{-1}+b_{\rm arith})b_{\rm arith}^2+b_{\rm arith}^3)$
Clifford$+T$ gates.
\end{enumerate}
\Cref{prop:global-arith-error} is a direct consequence.
\end{proposition}

\begin{proof}
Part (i) analyzes the scenario oracle. For parts (ii) and (iii), Steps 1 to 3
bound the polynomial approximation error of each angle on its sub-bins, and Step 4
bounds the rounding error of the fixed-point evaluation.

\textbf{Part (i): the scenario oracle.} Let $G^{(\ell)}=S^{(\ell)}V^{(\ell)}\bmod 2\in\mathbb F_2^{B_{\rm coord}\times n_M}$
and $\sigma^{(\ell)}\in\mathbb F_2^{B_{\rm coord}}$ be the scrambled Sobol'
generator matrix and shift.  Computing $d_{s,\ell}=G^{(\ell)}_{1:B_{\rm net}}s\oplus\sigma^{(\ell)}_{1:B_{\rm net}}$
into $B_{\rm net}$ clean wires uses at most $B_{\rm net}n_M$ $\mathrm{CX}$ gates
and $B_{\rm net}$ $\mathrm{X}$ gates; extracting $\mathrm{neg}_{s,\ell}=1\oplus d_{s,\ell,1}$
and complementing $d_{s,\ell,2:B_{\rm net}}$ via $B_{\rm net}-1$ $\mathrm{CX}$
gates yields $|2x_{s,\ell}-1|=\sum_{j=2}^{B_{\rm coord}}2^{-(j-1)}(d_{s,\ell,j}\oplus\mathrm{neg}_{s,\ell})+2^{-B_{\rm coord}}$
with zero Toffoli gates.  Accumulating the top $B_{\rm net}-1$ dyadic weights
with $g_{\rm cs}=2+\lceil\log_2(B_{\rm net}+2)\rceil$ guard bits bounds the
combined tail ($<2^{-(q_{\rm prod}+2)}$), weight-rounding ($\leq\tfrac18 2^{-q_{\rm prod}}$),
and output-rounding ($\tfrac12 2^{-q_{\rm prod}}$) errors by $2^{-q_{\rm prod}}\leq\tfrac12 2^{-b_{\rm arith}}$,
with Toffoli/$T$ count determined solely by $(B_{\rm net},q_{\rm prod})$ and
hence independent of $M$ whenever $B_{\rm coord}\geq q_{\rm prod}+2$.

\textbf{Parts (ii) and (iii), approximation error.} Map each sub-bin $z\in[0,1]$ ($s\in\{0,\dots,7\}$) to $t=2z-1\in[-1,1]$
and let $E_\rho=\{t=\tfrac12(w+w^{-1}):|w|=\rho\}$ be the Bernstein ellipse of
parameter $\rho>1$.
\emph{Step 1 ($\beta_w$ and $\theta_E$ for all $N$).} $K=0$ is loaded at a single
point, and for octaves $e\in\{0,1,2,3\}$ ($2^e\leq 8$), $K=2^e(1+(s+z)/8)\in\mathbb Z$
holds for at most one point ($z=0$) per sub-bin, so exact constant loading has
zero polynomial error.  For every octave $e\geq 4$ ($2^e/8\geq 2$), on $E_{16}$
($\rho_1=16=2^4$) we have $|t|\leq\tfrac12(16+\tfrac{1}{16})=8.03125$, so
$\operatorname{Re}(z)\geq -3.515625$ and $\operatorname{Re}(8+s+z)\geq 4.484375$.
Consequently, for $W_e(z):=\nu\frac{2^e}{8}(8+s+z)+\alpha_0$,
\[
 \operatorname{Re} W_e(z)
 \geq
 2\nu(4.484375)+\alpha_0
 =
 8.96875\,\nu+\alpha_0
 >
 \nu+\alpha_0
 >
 0
\]
for $e\geq 4$, $s\in\{0,\dots,7\}$ and $t\in E_{16}$.
Because $\operatorname{Re}X>\gamma>0$ implies both $\operatorname{Re}(\gamma/X)>0$
and $\operatorname{Re}(1-\gamma/X)\geq 1-\gamma/\operatorname{Re}X>0$, both
$\zeta_w=\alpha_0/W_e(z)$ and $\zeta_E=(\nu+\alpha_0)/W_e(z)$ remain uniformly
bounded away from the branch cuts $(-\infty,0]\cup[1,\infty)$ of
$\arccos\sqrt{\zeta}$ and $\arcsin\sqrt{\zeta}$ across all $e\geq 4$ and
$t\in E_{16}$, with $\sup_{e\geq 4,s,t\in E_{16}}\{|\beta_w|,|\theta_E|\}=:M_1(\nu,\alpha_0)<\infty$
independent of $e$ (and hence of $N$).  With $D=\max\{1,\lfloor(b_{\rm arith}-1)/4\rfloor\}$,
$16^{-(D+1)}\leq 2^{-b_{\rm arith}}$, so standard Chebyshev interpolation bounds
give $\sup_{z\in[0,1]}|f(z)-P_D(z)|\leq\frac{2M_1}{15}\,2^{-b_{\rm arith}}$.
\emph{Step 2 ($\phi_{\rm rot}$ and $|\cos(\pi u)|$).} On $E_{16}$,
$|\operatorname{Im}((s+z)/8)|\leq\frac{16-1/16}{32}=0.49805<1$, avoiding the
branch points $\pm i$ of $\arctan$, while on each sub-bin $s\in\{0,\dots,3\}$
(resp.\ $s\in\{4,\dots,7\}$), $|\cos(\pi(s+z)/8)|=+\cos(\pi(s+z)/8)$ (resp.\
$-\cos(\pi(s+z)/8)$) is entire on $\mathbb C$ because the sign change at $\pi/2$
falls on the sub-bin boundary $(s=4,z=0)$.
\emph{Step 3 ($\beta_\alpha$ on $E_8$, cases (a) and (b)).}
In case~(a), $|m_s|\leq m_{\max}<0.75=6/8$, so on $E_8$ ($\rho_2=8=2^3$) for
$s\in\{0,\dots,5\}$ we have $\operatorname{Re}(z)\in[-1.53125,2.53125]$, keeping
$(s+z)/8$ at distance $>0.05$ from $(-\infty,-1]\cup[1,\infty)$.  In case~(b),
for any $0<r_{\max}<r_{\rm be}<1$, every $b_{\rm in}$-bit midpoint input
$a\in\{0,\dots,2^{b_{\rm in}}-1\}$ has exact $u_{\rm ex}(a)=1-|m_{\rm ex}(a)|>\delta_{\min}=1-r_{\max}/r_{\rm be}>0$.
Accumulating the top $B_{\rm eff}=\min\{b_{\rm in},q_m+3\}$ dyadic weights with
$g_{\rm acc}=3+\lceil\log_2(B_{\rm eff}+2)\rceil$ guard bits produces
$\tilde u=U_{\rm int}2^{-q_m}$ ($U_{\rm int}\in\{1,\dots,2^{q_m}\}$) satisfying
$|\tilde u-u_{\rm ex}|<2^{-q_m}\leq\tfrac14\delta_{\min}$, so
$\tilde u>\tfrac34\delta_{\min}\geq 3\cdot 2^{-q_m}$ never underflows and, since
$\frac{d}{du}[4\arcsin\sqrt{u/2}]=\sqrt{2/(u(2-u))}\leq\sqrt{2/u}$, perturbs
$4\arcsin\sqrt{u/2}$ by $<\sqrt{8/(3\delta_{\min})}\,2^{-q_m}<\tfrac18 2^{-b_{\rm arith}}$
independently of $\widetilde\kappa$.  Writing $\tilde u=2^{-e}(1+(s+z)/8)$ via
leading-bit priority encoding over $q_m+1$ bits ($e\in\{1,\dots,q_m\}$, exact
point values on $U_{\rm int}\leq 15$, and constant $\pi$ at $U_{\rm int}=2^{q_m}$),
on $E_8$ every $e\geq 1$ and $s\in\{0,\dots,7\}$ satisfies
$\operatorname{Re}(1+\frac{s+z}{8})\geq\frac{8-1.53125}{8}=0.80859>0$ (avoiding
$(-\infty,0]$ of $\sqrt{\cdot}$) and
$|\zeta_{e,s}(z)|:=2^{-(e+1)}|1+\frac{s+z}{8}|\leq\frac{17.53125}{32}=0.54785<1$,
whence $|\sqrt{\zeta_{e,s}(z)}|<0.741<1$ avoids $(-\infty,-1]\cup[1,\infty)$ and
$\sup_{e\geq 1,0\leq s\leq 7,t\in E_8}|4\arcsin\sqrt{\zeta_{e,s}(z)}|\leq 4\arcsin(0.741)<3.35=:M_{\rm acos}$
is a universal constant independent of $e$ and $\widetilde\kappa$.  In both
cases, $D_{\rm acos}=\max\{1,\lfloor(b_{\rm arith}-1)/3\rfloor\}$ satisfies
$8^{-(D_{\rm acos}+1)}\leq 2^{-b_{\rm arith}}$, while priority encoding and
coefficient loading across $8q_m$ pairs use $\mathcal O(q_m b_{\rm arith}^2)$
gates plus $\mathcal O(B_{\rm eff}q_m+b_{\rm arith}^3)$ for accumulation and
Horner steps.

\textbf{Part (ii), rounding error.}
\emph{Step 4 (rounding in Horner's rule).} By Cauchy's estimate inside
$E_\rho$, $\sup_{D,z\in[0,1]}|P_D'(z)|\leq L_f<\infty$ is independent of $D$, so
mantissa/division truncation $|\tilde z-z|\leq 2^{-q}\leq 2^{-(b_{\rm arith}-1)}$
perturbs $P_D(\tilde z)$ by $\leq 2L_f 2^{-b_{\rm arith}}$.  In register width
$w=b_{\rm out}+g_{\rm bits}$, each of the $D_{\rm stab}$ Horner multiply-accumulate
steps on $\tilde z\in[0,1)$ multiplies prior error by $\tilde z<1$ and injects
at most $\tfrac32\mathrm{ULP}_w$ (coefficient rounding plus single post-shift
truncation).  Because $2^{g_{\rm bits}}\geq 8(D_{\rm stab}+2)$, the total
accumulated Horner rounding error satisfies
\[
 \sum_{d=0}^{D_{\rm stab}-1}\tilde z^d\left(\frac32\,\frac{4\pi}{2^w}\right)
 \leq
 \frac{3\pi(D_{\rm stab}+1)}{2^{g_{\rm bits}}}\,2^{-b_{\rm arith}}
 <
 \frac{3\pi}{8}\,2^{-b_{\rm arith}},
\]
canceling the linear $D_{\rm stab}$ growth, and final rounding to $b_{\rm out}$
bits adds at most $\pi\,2^{-b_{\rm arith}}$.  Summing the polynomial, mantissa,
Horner, and output-rounding bounds establishes \eqref{eq:global-arith-bound},
while gate-by-gate uncomputation restores every clean ancilla wire to $|0\rangle$.
\end{proof}

\subsection{Gate Cost of a Block-Encoding Call}
\label{app:be-cost-details}

This subsection lists the arithmetic components of one block-encoding call and
their gate counts; summing them gives \eqref{eq:gk-sobolev}. The circuit computes
three angles directly from $(\nu,\alpha_0,r_{\rm be})$:
\[
\begin{aligned}
 \beta_w(|k|^2)&=2\arccos\sqrt{\frac{\alpha_0}{\nu|k|^2+\alpha_0}}\quad(\beta_w(0)=\pi),
 \qquad
 \phi_{\rm rot}(k_x,k_y)=2\arctan2(k_x,k_y)\quad(d=2),\\
 \beta_\alpha(s,x)&=2\arccos\frac{\alpha_s(x)-\alpha_0}{r_{\rm be}\alpha_0}.
\end{aligned}
\]
\begin{enumerate}[label=(\roman*),leftmargin=2.2em]
\item \textbf{Wavenumber squaring.} Two's-complement squaring
$|k|^2=\sum_{j=1}^dk_j^2$ with Cuccaro ripple-carry adders~\cite{cuccaro2004} uses
$\mathcal O(d(\log N)^2)$ Toffoli gates.
\item \textbf{Octave normalization.} A priority encoder over the $2\log_2N+1$
octaves of $|k|^2$ selects the Chebyshev coefficients of the local
approximation in $\mathcal O(d\log N\cdot b_{\rm arith}^2)$ gates.
\item \textbf{Polynomial evaluation.} On each of $8$ sub-bins per octave, a
Chebyshev polynomial of degree $D=\Theta(b_{\rm arith})$ is evaluated by signed
shift-and-add Horner steps over $\Theta(b_{\rm arith})$-bit registers, in
$\mathcal O(b_{\rm arith}^3)$ gates and without $2^{\Theta(b_{\rm arith})}$ lookup
tables~\cite{haner2018}.
\item \textbf{Frame angle.} Reversible restoring division for $\arctan2$ uses
$\mathcal O(\log N\cdot b_{\rm arith})$ gates.
\item \textbf{Drag evaluation.} Coherent evaluation of $\alpha_s(x)$ across the
$L(\epsilon)$ modes adds
\[
 \mathcal O\big(L(\epsilon)\,(\operatorname{polylog}(M,1/\eta_K)+d\log N+b_{\rm arith}^3)\big)
\]
gates.
\item \textbf{Rotation synthesis.} The $\mathcal O(b_{\rm arith})$ dyadic
single-qubit $R_y/R_z$ kickback rotations are synthesized into Clifford$+T$ gates
by the method of Ross and Selinger~\cite{rossselinger2016}, at cost
$\mathcal O(b_{\rm arith}\log(b_{\rm arith}/\eta_K))$.
\end{enumerate}
All arithmetic workspace is uncomputed to $|0\rangle$ after each kickback stage,
which gives the ancilla count $\mathcal O(d\log N+b_{\rm arith}^2)$. With a
scrambled Sobol' digital net, $O_{\rm coeff}$ computes each
$c_{s,\ell}/(r_{\rm be}\alpha_0)$ from the $\log_2M$ qubits of $|s\rangle$ in
$\mathcal O((\log M)\,b_{\rm arith}+b_{\rm arith}^2)$ gates; one-hot multiplexing of
an unstructured table would need $\mathcal O(Mb_{\rm arith})$ gates and
$\mathcal O(M)$ temporary wires.

\subsection{Controlled Operations and Source Preparation}
\label{app:thm5-circuit}

The QSVT sequence applies coherent six-qubit branch phase gates on the registers
$(q_{\rm lcu},a_{\rm BE})$. The single controlled call needed for the odd branch is
implemented by the conjugation
\[
 \Lambda_1(\widetilde{\mathcal U}_{\rm BE})
 =
 \Lambda_1(\operatorname{SWAP})\,\Lambda_1(U_{W,L})\,B^\dagger\,
 \Lambda_1(U_{{\rm refl},\alpha})\,B\,\Lambda_1(U_{W,R}),
 \qquad
 B=(I_{\rm anc}\otimes\mathcal F_d^\dagger)U_{\rm rot},
\]
so that the Fourier transforms and the frame rotation need no control. The
Grover reflections use linear-Toffoli multi-controlled gates~\cite{barenco1995}.
In $d=2$ the system register has $2\log_2N+\log_2M+8$ qubits. When $b_s$ is
supported on $|k|^2\leq 2$, the source preparation reduces to
$30+2(\log_2N-2)=\mathcal O(\log N)$ elementary $\{\mathrm{RY},\mathrm{CX}\}$
gates by two's-complement sign extension.

\subsection{Comparison with Reduced-Order Classical Solvers}
\label{app:reduced-basis}

\begin{remark}[Comparison with reduced-basis methods]
\label{rem:reduced-basis}
Finite regularity forces $N=\Theta(\epsilon^{-1/r})$ for uniform grids, but it
does not show that every classical algorithm needs $\Omega(N^d)$ operations to
estimate one scalar output. When the drag depends on a fixed or slowly growing
number of parameters, reduced-basis, polynomial-chaos and homogenization methods
can be much cheaper than a grid-resolving solver~\cite{kuo2012,dick2014}, and
\Cref{thm:sobolev-advantage} gives a comparison with the single-level grid-resolving baseline. A
point-query lower bound would concern a separate information model.
\end{remark}

\section{Detailed Numerical Verification}
\label{app:numerics}

The main text reports the numerical trends that bear on the theory.
This appendix records the test configurations and numerical values needed
to reproduce and assess those observations. We separate PDE-level
checks, gate-by-gate statevector simulations, and circuit audits so that
each result is interpreted at the level actually tested. \Cref{tab:general} validates the polar Riesz reduction,
\Cref{tab:grover,tab:scaling} concern amplitude amplification and sampling, and
\Cref{tab:model2-resources,tab:polylog-gate-audit,tab:polylog-gate-audit-b,tab:scenario-m-scaling,tab:scenario-m-scaling-b}
give the size of the emitted Model~II circuits. In the subcircuit tests of
\Cref{sec:exp-stokes}, the wavenumber-squaring circuits had 908 gates (2D $|k|^2$
at $N=4$) and 717 gates (1D $k_x^2$ at $N=8$), both on 17 qubits, and the
18-qubit compute--kickback--uncompute circuit had 1{,}836 gates; all matched the
exact action to $5.6\times 10^{-15}$ and agreed with our unitary extractor to
$4.5\times 10^{-15}$.

\subsection{Circuit Correctness Tests}

\paragraph{Constraint elimination.} Dense linear-algebra tests of
\Cref{prop:riesz,prop:reduced} for constrained spaces of dimension up to $16$
(\Cref{tab:general}) give constraint residuals, isometry defects, projector gaps
and saddle-point discrepancies at double-precision roundoff (at most
$3\times 10^{-15}$). The bias bound \eqref{eq:bias} held in all $200$ random
four-stage contraction chains.

\paragraph{Stokes benchmark.} A spectral discretization of the Stokes family of
\Cref{thm:optimal-stokes} at $N=64$ with $257$ hidden-bit scenarios reproduces
$g_s=z_s$ to $4.4\times 10^{-16}$, with zero discrete divergence and Stokes
residual $3.2\times 10^{-13}$.

\paragraph{Amplitude amplification.} For a model ensemble on the $N=4$
wavenumber grid ($M=16$), whose scenario probabilities are a degree-two dyadic
Chebyshev approximation of the inverse amplitude, the reduced statevector on the
scenario and flag registers follows the Grover law $\sin^2((2m+1)\theta)$ for
$m\in\{0,1,2,4\}$ to $10^{-15}$ (\Cref{tab:grover}). This checks the reflections
used by amplitude estimation; it is not a gate-level circuit simulation, which is
reported separately below.

\paragraph{Model~I circuit.} To check the full multi-scenario circuit without
matrix diagonalization or classical FFT shortcuts, we compiled $\cA$ (Model~I)
into elementary $\{\mathrm{H},\mathrm{Z},\mathrm{P},\mathrm{RY},\mathrm{RZ},\mathrm{CX}\}$
gates at $N=4,M=16$ (16 qubits, 14{,}475 gates) and $N=8,M=16$ (18 qubits,
48{,}737 gates). The low-wavenumber forcing ($|k|^2\leq 2$) is prepared by a
two's-complement sign-extension circuit of 29 gates at $N=4$ and 31 gates at
$N=8$, with error $3\times 10^{-16}$. In all Brinkman experiments the reduced
forcing is $b\propto\sum_{j=1}^{6}j^{-1}e_{k_j}$ with
$(k_1,\dots,k_6)=((0,1),(-1,0),(1,0),(0,-1),(-1,1),(1,-1))$ for $N\geq 8$ and
$((0,1),(0,-1),(1,0),(-1,0),(1,1),(1,-1))$ for $N=4$. These lists are fixed in
the code rather than produced by sorting $|k|^2$, whose ties would otherwise make
the choice depend on the sort implementation. Any six modes with $|k|^2\leq 2$
satisfy \ref{as:A2}. The coefficients are not symmetric under $k\mapsto -k$, so
this forcing is complex-valued; \Cref{lem:norm-chain} does not require a real
forcing. Applying $\cA$ and then $\cA^\dagger$ to the
262{,}144-amplitude statevector returns to $|0^{18}\rangle$ with residual
$1.4\times 10^{-13}$. The mean good-subspace probability is
$\bar p_{\rm good}=0.023$, and the smallest scenario value, $0.022$, exceeds the
analytic lower bound $0.0097$ of \eqref{eq:uniform-prob-lower-bound}. Multiplying
by $B_{\rm norm}=15.0$ recovers the ensemble kinetic energy $0.341$, which differs
from the classical PDE value by $1.4\times 10^{-8}$. With the good-state reflection
(338 Clifford$+T$ gates) and the zero-state reflection (1{,}371 gates at $N=8$)
built from linear-Toffoli constructions~\cite{barenco1995}, gate-by-gate execution
of $\mathcal G^m\cA|0^{18}\rangle$ for $m\in\{0,1,2,4\}$ gives probabilities $0.023$,
$0.19$, $0.47$ and $0.96$, which agree with $\sin^2((2m+1)\theta)$ to
$1.3\times 10^{-12}$.

\paragraph{Numerical agreement with Model~I.} The Gray-code circuit (Model~I) uses
only 16 to 18 qubits at $N=4,8$, but its gate count grows as $\Theta(MN^2)$ in two
dimensions. Model~II replaces the angle tables by reversible fixed-point
arithmetic that computes $|k|^2$, the scenario coefficients, and the four angles
$\beta_w,\theta_E,\phi_{\rm rot},\beta_\alpha$ with the configuration of
\Cref{prop:arith-full}. On dense superposition statevectors, the
wavenumber-squaring subcircuits and an 18-qubit compute--kickback--uncompute
circuit match the exact action to $6\times 10^{-15}$. For the complete emitted
Model~II circuit (192 to 416 qubits, $b_{\rm arith}\in\{8,\dots,16\}$, $N=4$ and $8$),
a dense statevector cannot be stored, so each compute--kickback--uncompute
subcircuit is evaluated gate by gate over all system basis states (ancilla
residual at most $4\times 10^{-16}$ at every step), and the resulting subcircuit
unitaries evolve the 16- or 18-qubit system state through all stages.
\Cref{tab:model2-accuracy} and \Cref{fig:model2-precision} show that the block error decays as
$2^{-b_{\rm arith}}$ (fitted $\log_2$ slope $-1.0$ per bit at both grid sizes),
as predicted by \eqref{eq:be-telescoping-bound} and
\Cref{prop:global-arith-error}, and that the kinetic-energy error reaches
$1.9\times 10^{-8}$ at $N=8$, $b_{\rm arith}=16$. This is finite-precision agreement
between two compilations, not an exact identity of circuits.

\begin{table}[htbp]
\centering
\caption{Accuracy of the emitted Model~II circuit against Model~I and the classical ensemble ($M=16$, $\nu=0.1$, $\alpha_0=2.0$; exact $\beta_w(1)=0.43998$ rad).}
\label{tab:model2-accuracy}
\small
\begin{tabular}{rrrrrrr}
\toprule
$N$ & $b_{\rm arith}$ ($D,D_{\rm a}$) & $\beta_w(1)$ & $\max_s\|X_s^{({\rm II})}-X_s^{({\rm I})}\|_2$ & bound \eqref{eq:be-telescoping-bound} & $|\Delta\bar p_{\rm good}|$ & energy error \\
\midrule
4 & 8 (1,2) & 0.44179 & $5.15\times10^{-3}$ & $2.46\times10^{-2}$ & $2.77\times10^{-6}$ & $4.16\times10^{-5}$ \\
4 & 10 (2,3) & 0.44179 & $1.10\times10^{-3}$ & $5.33\times10^{-3}$ & $1.12\times10^{-6}$ & $1.68\times10^{-5}$ \\
4 & 12 (2,3) & 0.44025 & $3.41\times10^{-4}$ & $2.04\times10^{-3}$ & $4.63\times10^{-7}$ & $6.98\times10^{-6}$ \\
4 & 14 (3,4) & 0.43987 & $7.32\times10^{-5}$ & $2.36\times10^{-4}$ & $7.09\times10^{-8}$ & $1.08\times10^{-6}$ \\
4 & 16 (3,5) & 0.43996 & $1.76\times10^{-5}$ & $6.92\times10^{-5}$ & $2.27\times10^{-8}$ & $3.25\times10^{-7}$ \\
\midrule
8 & 8 (1,2) & 0.44179 & $4.79\times10^{-3}$ & $3.49\times10^{-2}$ & $2.80\times10^{-6}$ & $4.21\times10^{-5}$ \\
8 & 10 (2,3) & 0.44179 & $1.25\times10^{-3}$ & $6.79\times10^{-3}$ & $4.13\times10^{-7}$ & $6.19\times10^{-6}$ \\
8 & 12 (2,3) & 0.44025 & $2.90\times10^{-4}$ & $2.11\times10^{-3}$ & $1.11\times10^{-8}$ & $1.80\times10^{-7}$ \\
8 & 14 (3,4) & 0.43987 & $7.61\times10^{-5}$ & $4.89\times10^{-4}$ & $1.15\times10^{-7}$ & $1.75\times10^{-6}$ \\
8 & 16 (3,5) & 0.43996 & $1.63\times10^{-5}$ & $1.21\times10^{-4}$ & $2.20\times10^{-9}$ & $1.91\times10^{-8}$ \\
\bottomrule
\end{tabular}
\end{table}

\subsection{Resource-Scaling Tests}

\paragraph{Sampling.} At matched total oracle calls over $180$ independent trials
for the model probability $a=0.3206$ of the amplitude-amplification test,
maximum-likelihood QAE (MLQAE)~\cite{suzuki2020} has an empirical log--log error
slope of $-0.96$, against $-0.50$ for classical sampling, and a $9.0\times$
smaller RMSE at $6{,}288$ calls (\Cref{fig:scaling} and \Cref{tab:scaling}).
The RMSE is not monotone in the call count because the likelihood can have
several nearby maxima at small sample sizes. The comparison counts calls, not
gates, and is not evidence of a runtime advantage.

\paragraph{Grid size.} We compiled and audited the complete Model~II netlist for
$N$ from $8$ to $1024$ at $b_{\rm arith}=10$ and $M=16$, with bitwise truth-table
and ancilla checks over all reachable $|k|^2$ values (up to $524{,}289$), all
wavevector pairs (up to about $10^6$), and all scenario--position triples (up to
about $1.7\times 10^7$ at $N=1024$). The Model~II gate count grows quadratically in
$\log_2N$ (fit $R^2=0.9998$), falls below Model~I at $N=256$ ($1.4\times 10^7$
against $4.6\times 10^7$ gates), and is $51$ times smaller at $N=1024$
(\Cref{fig:circuit-resource-scaling}; \Cref{tab:polylog-gate-audit,tab:polylog-gate-audit-b}). Both models are quantum
circuits; this is not a comparison with a classical solver.

\paragraph{Ensemble size.} At $N=64$ and $b_{\rm arith}=10$ we varied the ensemble
size from $M=16$ to $M=2^{20}$ and verified $O_{\rm coeff}$ on all $M$ basis
inputs, with maximum coefficient error $3.8\times 10^{-4}$ and no dirty ancillas
(\Cref{fig:circuit-resource-scaling}; \Cref{tab:scenario-m-scaling,tab:scenario-m-scaling-b}). As
\Cref{prop:global-arith-error}(i) predicts, the gate count grows linearly in
$\log_2M$, by 886 CX gates per bit (fit $R^2=0.998$), an increase of $0.11\%$ over
a $65{,}536$-fold increase in $M$, and the $T$-count stays at $5.3\times 10^6$. At
$M=2^{20}$, Model~II uses about $10^4$ times fewer gates than Model~I
($1.4\times 10^{11}$ gates). This ratio compares two of our own quantum circuits;
it is not a speedup over classical computation.

\subsection{Three-Dimensional Checks}

On $\mathbb T^3$ at $N_3=8$, the explicit $1{,}022\times 1{,}022$ mean-zero
divergence-free operator has $\|DR_{\mathcal E,3{\rm D}}\|_2=9\times 10^{-16}$, and
preconditioning lowers its condition number from $4.3$ to $1.9$, below
$\widetilde\kappa=3.0$; the $1{,}025\times 1{,}025$ operator without the mean-flow
constraint (\Cref{rem:unconstrained}) gives $2.0$ (\Cref{fig:spectral-clustering}, right). For the drag
$\alpha(x)=\alpha_0+0.1\sin x_1$ ($\nu=0.1$, $\alpha_0=2.0$, $N_3=32$), the problem
without the mean-flow multiplier ($\lambda\equiv 0$) develops a bulk flux
$\langle u_2\rangle=-0.035$, whereas the constrained problem
\eqref{eq:brinkman-pde} keeps $\|\langle u\rangle\|_2=6\times 10^{-17}$ and recovers
$\lambda=(0,-0.071,0)^T$. Omitting $\lambda$ would leave a momentum residual of
$0.071$; with it the residual is $1.3\times 10^{-15}$.

\begin{table}[H]
\centering
\caption{Numerical validation of the constrained polar Riesz reduction
(\Cref{prop:riesz,prop:reduced}).}
\label{tab:general}
\begin{tabular}{rrrrrr}
\toprule
$\dim\cH$ & $\dim\cY$ & $\|DR\|_2$ & $\|R^*R-I\|_2$
& projector gap & saddle gap \\
\midrule
8  & 2 & $1.02\times10^{-15}$ & $2.67\times10^{-15}$
& $2.67\times10^{-15}$ & $1.02\times10^{-15}$ \\
12 & 4 & $1.66\times10^{-15}$ & $2.57\times10^{-15}$
& $2.43\times10^{-15}$ & $1.14\times10^{-15}$ \\
16 & 6 & $1.41\times10^{-15}$ & $2.30\times10^{-15}$
& $2.44\times10^{-15}$ & $1.06\times10^{-15}$ \\
\bottomrule
\end{tabular}
\end{table}

\begin{table}[H]
\centering
\caption{Grover law on the reduced statevector (scenario and flag registers) for
the model ensemble at $N=4$, $M=16$ with a degree-two polynomial inverse.}
\label{tab:grover}
\begin{tabular}{rrrr}
\toprule
Grover power & statevector probability & Grover law & absolute gap \\
\midrule
0 & 0.3205827212 & 0.3205827212 & $5.6\times10^{-17}$ \\
1 & 0.9458431538 & 0.9458431538 & $2.2\times10^{-16}$ \\
2 & 0.0173620236 & 0.0173620236 & $9.0\times10^{-17}$ \\
4 & 0.5804380788 & 0.5804380788 & $1.0\times10^{-15}$ \\
\bottomrule
\end{tabular}
\end{table}

\begin{table}[H]
\centering
\caption{Equal-call RMSE comparison between MLQAE and classical sampling across
180 independent trials ($a=0.3206$).}
\label{tab:scaling}
\begin{tabular}{rrrr}
\toprule
Calls & MLQAE RMSE & sampling RMSE & sampling/MLQAE \\
\midrule
192  & $2.55289\times10^{-2}$ & $3.39353\times10^{-2}$ & 1.329 \\
432  & $8.88151\times10^{-3}$ & $2.33002\times10^{-2}$ & 2.623 \\
840  & $4.61022\times10^{-3}$ & $1.60085\times10^{-2}$ & 3.472 \\
1632 & $2.50177\times10^{-3}$ & $1.05060\times10^{-2}$ & 4.199 \\
3192 & $2.13016\times10^{-3}$ & $8.62385\times10^{-3}$ & 4.048 \\
6288 & $6.64160\times10^{-4}$ & $5.95578\times10^{-3}$ & 8.967 \\
\bottomrule
\end{tabular}
\end{table}

\begin{table}[H]
\centering
\caption{Size of the emitted Model~II circuit for the runs of \Cref{tab:model2-accuracy}.}
\label{tab:model2-resources}
\small
\begin{tabular}{rrrr}
\toprule
$N$ & $b_{\rm arith}$ & qubits (system / total) & gates in $\cA$ \\
\midrule
4 & 8 & 16 / 192 & 6{,}750{,}555 \\
4 & 10 & 16 / 244 & 11{,}279{,}307 \\
4 & 12 & 16 / 275 & 15{,}267{,}839 \\
4 & 14 & 16 / 339 & 26{,}925{,}045 \\
4 & 16 & 16 / 411 & 34{,}684{,}561 \\
\midrule
8 & 8 & 18 / 197 & 7{,}002{,}883 \\
8 & 10 & 18 / 249 & 11{,}599{,}499 \\
8 & 12 & 18 / 280 & 15{,}603{,}811 \\
8 & 14 & 18 / 344 & 27{,}300{,}485 \\
8 & 16 & 18 / 416 & 35{,}077{,}537 \\
\bottomrule
\end{tabular}
\end{table}

\begin{table}[H]
\centering
\caption{Model~II circuit size against grid size ($b_{\rm arith}=10$, $(D,D_{\rm acos})=(2,3)$, $M=16$; no dirty ancillas over all 16{,}777{,}216 states at $N=1024$).}
\label{tab:polylog-gate-audit}
\small
\begin{tabular}{rrrrr}
\toprule
$N$ & $n$ & qubits (system / total) & reversible $|k|^2$ gates (pairs checked) & block-encoding gates \\
\midrule
8 & 3 & 18 / 249 & 1{,}842 (64) & 765{,}434 \\
16 & 4 & 20 / 254 & 3{,}100 (256) & 787{,}476 \\
32 & 5 & 22 / 259 & 4{,}682 (1{,}024) & 811{,}174 \\
64 & 6 & 24 / 264 & 6{,}588 (4{,}096) & 836{,}472 \\
128 & 7 & 26 / 269 & 8{,}818 (16{,}384) & 862{,}486 \\
256 & 8 & 28 / 274 & 11{,}372 (65{,}536) & 887{,}276 \\
512 & 9 & 30 / 279 & 14{,}250 (262{,}144) & 911{,}186 \\
1024 & 10 & 32 / 284 & 17{,}452 (1{,}048{,}576) & 936{,}032 \\
\bottomrule
\end{tabular}
\end{table}

\begin{table}[H]
\centering
\caption{Full state-preparation circuit $\cA$ for the runs of \Cref{tab:polylog-gate-audit}, compared with the Gray-code circuit (Model~I).}
\label{tab:polylog-gate-audit-b}
\small
\begin{tabular}{rrrrr}
\toprule
$N$ & Model~II gates & Model~II $T$-count & Model~II depth & Model~I gates \\
\midrule
8 & 11{,}599{,}499 & 4{,}872{,}105 & 5{,}416{,}231 & 48{,}737 \\
16 & 11{,}945{,}311 & 5{,}006{,}813 & 5{,}566{,}764 & 184{,}279 \\
32 & 12{,}317{,}479 & 5{,}152{,}273 & 5{,}729{,}745 & 723{,}141 \\
64 & 12{,}715{,}123 & 5{,}308{,}485 & 5{,}901{,}282 & 2{,}875{,}283 \\
128 & 13{,}124{,}339 & 5{,}470{,}521 & 6{,}080{,}675 & 11{,}478{,}745 \\
256 & 13{,}514{,}963 & 5{,}629{,}085 & 6{,}252{,}876 & 45{,}887{,}487 \\
512 & 13{,}892{,}191 & 5{,}785{,}857 & 6{,}423{,}477 & 183{,}515{,}549 \\
1024 & 14{,}284{,}295 & 5{,}949{,}685 & 6{,}602{,}598 & 734{,}020{,}891 \\
\bottomrule
\end{tabular}
\end{table}

\begin{table}[H]
\centering
\caption{The scrambled Sobol' scenario oracle $O_{\rm coeff}$ inside Model~II at $N=64$, $b_{\rm arith}=10$.}
\label{tab:scenario-m-scaling}
\small
\begin{tabular}{rrrrrr}
\toprule
$M$ & $n_M$ & qubits (system / total) & $O_{\rm coeff}$ gates per mode & coefficient error & $U_{{\rm refl},\alpha}$ gates \\
\midrule
16 & 4 & 24 / 264 & 7{,}136 & $3.14\times10^{-4}$ & 531{,}311 \\
256 & 8 & 28 / 268 & 7{,}156 & $3.56\times10^{-4}$ & 531{,}583 \\
4{,}096 & 12 & 32 / 272 & 7{,}165 & $3.76\times10^{-4}$ & 531{,}819 \\
65{,}536 & 16 & 36 / 276 & 7{,}166 & $3.81\times10^{-4}$ & 532{,}039 \\
1{,}048{,}576 & 20 & 40 / 280 & 7{,}166 & $3.81\times10^{-4}$ & 532{,}263 \\
\bottomrule
\end{tabular}
\end{table}

\begin{table}[H]
\centering
\caption{Full circuit $\cA$ for the runs of \Cref{tab:scenario-m-scaling}. The $T$-count does not change with $M$.}
\label{tab:scenario-m-scaling-b}
\small
\begin{tabular}{rrrr}
\toprule
$M$ & Model~II gates & Model~II $T$-count & Model~I gates \\
\midrule
16 & 12{,}715{,}123 & 5{,}308{,}485 & 2{,}875{,}283 \\
256 & 12{,}719{,}207 & 5{,}308{,}485 & 34{,}332{,}567 \\
4{,}096 & 12{,}722{,}751 & 5{,}308{,}485 & 537{,}649{,}051 \\
65{,}536 & 12{,}726{,}055 & 5{,}308{,}485 & 8{,}590{,}712{,}735 \\
1{,}048{,}576 & 12{,}729{,}419 & 5{,}308{,}485 & 137{,}439{,}731{,}619 \\
\bottomrule
\end{tabular}
\end{table}


\begin{thebibliography}{41}

\bibitem{hhl2009}
A.~W. Harrow, A.~Hassidim, S.~Lloyd,
Quantum algorithm for linear systems of equations,
Phys. Rev. Lett. 103 (2009) 150502.
\doi{10.1103/PhysRevLett.103.150502}.

\bibitem{clader2013}
B.~D. Clader, B.~C. Jacobs, C.~R. Sprouse,
Preconditioned quantum linear system algorithm,
Phys. Rev. Lett. 110 (2013) 250504.
\doi{10.1103/PhysRevLett.110.250504}.

\bibitem{childs2017}
A.~M. Childs, R.~Kothari, R.~D. Somma,
Quantum algorithm for systems of linear equations with exponentially improved
dependence on precision,
SIAM J. Comput. 46 (2017) 1920--1950.
\doi{10.1137/16M1087072}.

\bibitem{gilyen2019}
A.~Gily\'en, Y.~Su, G.~H. Low, N.~Wiebe,
Quantum singular value transformation and beyond: exponential improvements for
quantum matrix arithmetics,
in: Proc. 51st Annu. ACM SIGACT Symp. Theory Comput. (STOC 2019), ACM, 2019,
pp. 193--204.
\doi{10.1145/3313276.3316366}.

\bibitem{lintong2020}
L.~Lin, Y.~Tong,
Optimal polynomial based quantum eigenstate filtering with application to
solving quantum linear systems,
Quantum 4 (2020) 361.
\doi{10.22331/q-2020-11-11-361}.

\bibitem{lowsu2026}
G.~H. Low, Y.~Su,
Quantum linear system algorithm with optimal queries to initial state
preparation,
Quantum 10 (2026) 2041.
\doi{10.22331/q-2026-03-23-2041}.

\bibitem{deiml2025}
M.~Deiml, D.~Peterseim,
Quantum realization of the finite element method,
Math. Comp. 95 (2026) 2637--2668.
\doi{10.1090/mcom/4137}.

\bibitem{fourierpde2025}
M.~Lubasch, Y.~Kikuchi, L.~Wright, C.~Mc~Keever,
Quantum circuits for partial differential equations in Fourier space,
Phys. Rev. Research 7 (2025) 043326.
\doi{10.1103/tbzc-w9x8}.

\bibitem{brassard2002}
G.~Brassard, P.~H\o yer, M.~Mosca, A.~Tapp,
Quantum amplitude amplification and estimation,
Contemp. Math. 305 (2002) 53--74.
\doi{10.1090/conm/305/05215}.

\bibitem{nayak1999}
A.~Nayak, F.~Wu,
The quantum query complexity of approximating the median and related statistics,
in: Proc. 31st Annu. ACM Symp. Theory Comput. (STOC 1999), ACM, 1999,
pp. 384--393.
\doi{10.1145/301250.301349}.

\bibitem{montanaro2015}
A.~Montanaro,
Quantum speedup of Monte Carlo methods,
Proc. R. Soc. A 471 (2015) 20150301.
\doi{10.1098/rspa.2015.0301}.

\bibitem{suzuki2020}
Y.~Suzuki, S.~Uno, R.~Raymond, T.~Tanaka, T.~Onodera, N.~Yamamoto,
Amplitude estimation without phase estimation,
Quantum Inf. Process. 19 (2020) 75.
\doi{10.1007/s11128-019-2565-2}.

\bibitem{golse2022}
F.~Golse, S.~Jin, N.~Liu,
Quantum algorithms for uncertainty quantification: Applications to partial
differential equations,
Sci. China Phys. Mech. Astron. 68 (2025) 104704.
\doi{10.1007/s11433-024-2705-7}.

\bibitem{recchia2026}
P.~Recchia, Z.~Yu, K.~Koor, P.~Rebentrost,
Quantum quasi-Monte Carlo: a window for pre-asymptotic quantum advantage,
arXiv:2609.03625 (2026).

\bibitem{gimenez2026}
J.~M. Gimenez, F.~Cirak, M.~Ortiz,
A quantum spectral solver for periodic incompressible Stokes flow,
Comput. Methods Appl. Mech. Eng. 463 (2027) 119453.
\doi{10.1016/j.cma.2026.119453}.

\bibitem{jennings2026}
D.~Jennings, K.~Korzekwa, M.~Lostaglio, R.~Ashworth, E.~Marsili, S.~Rolston,
End-to-end quantum algorithm for nonlinear fluid dynamics with bounded quantum
advantage,
PRX Quantum 7 (2026) 033060.
\doi{10.1103/xysy-q3fp}.

\bibitem{ameri2026}
A.~Ameri, J.~Carolan, A.~M. Childs, H.~Krovi,
Quantum lower bounds for simulating fluid dynamics,
arXiv:2603.12161 (2026).

\bibitem{brinkman1949}
H.~C. Brinkman,
A calculation of the viscous force exerted by a flowing fluid on a dense swarm
of particles,
Appl. Sci. Res. A 1 (1949) 27--34.
\doi{10.1007/BF02120313}.

\bibitem{jaworski2013}
J.~W. Jaworski, N.~Peake,
Aerodynamic noise from a poroelastic edge with implications for the silent
flight of owls,
J. Fluid Mech. 723 (2013) 456--479.
\doi{10.1017/jfm.2013.139}.

\bibitem{bottaro2019}
A.~Bottaro,
Flow over natural or engineered surfaces: an adjoint homogenization perspective,
J. Fluid Mech. 877 (2019) P1.
\doi{10.1017/jfm.2019.607}.

\bibitem{benzi2005}
M.~Benzi, G.~H. Golub, J.~Liesen,
Numerical solution of saddle point problems,
Acta Numer. 14 (2005) 1--137.
\doi{10.1017/S0962492904000212}.

\bibitem{mardal2011}
K.-A. Mardal, R.~Winther,
Preconditioning discretizations of systems of partial differential equations,
Numer. Linear Algebra Appl. 18 (2011) 1--40.
\doi{10.1002/nla.716}.

\bibitem{cliffe2011}
K.~A. Cliffe, M.~B. Giles, R.~Scheichl, A.~L. Teckentrup,
Multilevel Monte Carlo methods and applications to elliptic PDEs with random
coefficients,
Comput. Vis. Sci. 14 (2011) 3--15.
\doi{10.1007/s00791-011-0160-x}.

\bibitem{giles2015}
M.~B. Giles,
Multilevel Monte Carlo methods,
Acta Numer. 24 (2015) 259--328.
\doi{10.1017/S096249291500001X}.

\bibitem{kuo2012}
F.~Y. Kuo, C.~Schwab, I.~H. Sloan,
Quasi-Monte Carlo finite element methods for a class of elliptic partial
differential equations with random coefficients,
SIAM J. Numer. Anal. 50 (2012) 3351--3374.
\doi{10.1137/110845537}.

\bibitem{dick2014}
J.~Dick, Q.~T. Le~Gia, C.~Schwab,
Higher order quasi-Monte Carlo integration for holomorphic, parametric operator
equations,
SIAM/ASA J. Uncertain. Quantif. 4 (2016) 48--79.
\doi{10.1137/140985913}.

\bibitem{matousek1998}
J.~Matou\v{s}ek,
On the $L_2$-discrepancy for anchored boxes,
J. Complexity 14 (1998) 527--556.
\doi{10.1006/jcom.1998.0489}.

\bibitem{he2019}
Z.~He,
On the error rate of conditional quasi-Monte Carlo for discontinuous functions,
SIAM J. Numer. Anal. 57 (2019) 854--874.
\doi{10.1137/18M118270X}.

\bibitem{traub1988}
J.~F. Traub, G.~W. Wasilkowski, H.~Wo\'zniakowski,
Information-Based Complexity,
Academic Press, Boston, 1988.

\bibitem{novak1988}
E.~Novak,
Deterministic and Stochastic Error Bounds in Numerical Analysis,
Lecture Notes in Mathematics, vol. 1349, Springer, Berlin, 1988.
\doi{10.1007/BFb0079792}.

\bibitem{heinrich1994}
S.~Heinrich,
Random approximation in numerical analysis,
in: K.~D. Bierstedt et al. (Eds.), Functional Analysis, Marcel Dekker,
New York, 1994, pp. 123--171.

\bibitem{heinrich2006}
S.~Heinrich,
The quantum query complexity of elliptic PDE,
J. Complexity 22 (2006) 691--725.
\doi{10.1016/j.jco.2006.04.005}.

\bibitem{barenco1995}
A.~Barenco, C.~H. Bennett, R.~Cleve, D.~P. DiVincenzo, N.~Margolus, P.~Shor,
T.~Sleator, J.~A. Smolin, H.~Weinfurter,
Elementary gates for quantum computation,
Phys. Rev. A 52 (1995) 3457--3467.
\doi{10.1103/PhysRevA.52.3457}.

\bibitem{cuccaro2004}
S.~A. Cuccaro, T.~G. Draper, S.~A. Kutin, D.~P. Moulton,
A new quantum ripple-carry addition circuit,
arXiv:quant-ph/0410184 (2004).

\bibitem{shende2006}
V.~V. Shende, S.~S. Bullock, I.~L. Markov,
Synthesis of quantum-logic circuits,
IEEE Trans. Comput.-Aided Des. Integr. Circuits Syst. 25 (2006) 1000--1010.
\doi{10.1109/TCAD.2005.855930}.

\bibitem{rossselinger2016}
N.~J. Ross, P.~Selinger,
Optimal ancilla-free Clifford+$T$ approximation of $z$-rotations,
Quantum Inf. Comput. 16 (2016) 901--953.

\bibitem{haner2018}
T.~H\"aner, M.~Roetteler, K.~M. Svore,
Optimizing quantum circuits for arithmetic,
arXiv:1805.12445 (2018).

\bibitem{diaconis1980}
P.~Diaconis, D.~Freedman,
Finite exchangeable sequences,
Ann. Probab. 8 (1980) 745--764.
\doi{10.1214/aop/1176994663}.

\bibitem{montanaro2016fem}
A.~Montanaro, S.~Pallister,
Quantum algorithms and the finite element method,
Phys. Rev. A 93 (2016) 032324.
\doi{10.1103/PhysRevA.93.032324}.

\bibitem{novak2001}
E.~Novak,
Quantum complexity of integration,
J. Complexity 17 (2001) 2--16.
\doi{10.1006/jcom.2000.0566}.

\bibitem{heinrich2002}
S.~Heinrich,
Quantum summation with an application to integration,
J. Complexity 18 (2002) 1--50.
\doi{10.1006/jcom.2001.0629}.

\end{thebibliography}
\end{document}